\documentclass[12pt]{article}

\usepackage{times}
\usepackage{fancybox}
\usepackage{multirow}
\usepackage{amsmath}
\usepackage{amssymb}
\usepackage{amscd}
\usepackage{epic}
\usepackage{eepic}
\usepackage{epsfig}
\usepackage{rotating}

\newcommand{\dA}{\hbox{$A$ }}
\newcommand{\dB}{\hbox{$B$ }}
\newcommand{\dC}{\hbox{$C$ }}

\newcommand{\bi}[1]{\mathbf{#1}}
\newtheorem{conj}{Conjecture}
\newtheorem{theo}{Theorem}
\newtheorem{lemm}{Lemma}

\begin{document}
\begin{titlepage}
\begin{center}
 
\vspace*{2cm}
 
{\Large{\bf On Consistent and Calibrated Inference about \\[0.3cm]
            the Parameters of Sampling Distributions}}
 
\vskip 1.5 cm
\renewcommand{\thefootnote}{\fnsymbol{footnote}}
   Toma\v z\;Podobnik$^{1),\;2),}$\footnote{e-mail: Tomaz.Podobnik@ijs.si}
   and Tomi\;\v Zivko$^{2),}$\footnote{e-mail: Tomi.Zivko@ijs.si}
 \setcounter{footnote}{0}
\vskip 0.5cm
$^{1)}$Physics Department, University of Ljubljana, Slovenia\\
$^{2)}$''Jo\v zef Stefan'' Institute, Ljubljana, Slovenia 

\vskip 3cm

\begin{abstract}
      The theory of probability, based on very general rules referred
      to as the Cox-P\'{o}lya-Jaynes Desiderata, can be used both as
      a theory of random mass phenomena and as a quantitative theory of 
      plausible inference about the parameters of sampling distributions. 
      The existing applications of the Desiderata
      must be extended in order to allow for consistent 
      inferences
      in the limit of complete \textit{a priori} ignorance about the
      values of the parameters. Since the limits of consistent quantitative
      inference from incomplete information can clearly be 
      established, the developed theory 
      is necessarily an effective one. It is interesting
      to note that when applying the Desiderata strictly, 
      we find no contradictions between the so-called
      Bayesian and frequentist schools of inductive reasoning.
\end{abstract}
 
\vfill
Ljubljana, August 2005
 
\end{center}
\end{titlepage}

\thispagestyle{empty}
\begin{equation*}
\end{equation*}
\clearpage

\setcounter{page}{1}
\hfill \parbox{0.7\linewidth}{\small
\noindent
As for prophecies, they will pass away; as for tongues, they will
cease; as for knowledge, it will pass away. For we know in part
and we prophesy in part. 
\\ \vskip 0mm
\hfill
1 Corinthians 13, 8-9.
} \\

\section{Introduction}
\label{sec:preface}

The term \textit{inference} (\cite{dic}, p.\,436) stands for two kinds
of reasoning (\cite{jay}, p.\,xix): \textit{deductive} or 
\textit{demonstrative reasoning} whenever enough information is at hand 
to permit it, and \textit{inductive} or \textit{plausible reasoning} when
not all of the necessary information is available. ``The difference between
the two kinds of reasoning is great and manifold. Demonstrative reasoning
is safe, beyond controversy, and final. Plausible reasoning is hazardous,
controversial, and provisional. Demonstrative reasoning penetrates the science 
just as far as mathematics does, but is in itself (as mathematics
is in itself) incapable of yielding essentially new knowledge about
the world around us. Anything new that we learn about the world involves
plausible reasoning, which is the only kind of reasoning for which
we care in everyday affairs. Demonstrative reasoning has rigid standards,
codified and clarified by logic (formal or demonstrative 
logic\footnote{A branch of mathematics, also referred to as \textit{deductive
logic}.}), which is the theory of demonstrative reasoning. The standards
of plausible reasoning are fluid, and there is no theory of such reasoning
that could be compared to demonstrative logic in clarity or would
command comparable consensus.'' So George P\'{o}lya in the Preface
to his \textit{Mathematics and Plausible Reasoning} (\cite{pol}, p.\,v).

In the second volume \cite{pol1} of the work he collects patterns  of
plausible reasoning and dissects our intuitive common sense into
a set of elementary qualitative desiderata that represent basic
rules of inductive reasoning. When formulating his views in mathematical
terms (\cite{pol1}, Chapter\,XV), he recognizes his rules to be in
a close agreement with the calculus of probability as developed by Laplace
in the late 18th century \cite{lap1}, but P\'{o}lya advances a thesis
that when applying the calculus of probability to plausible reasoning,
it should be applied only \textit{qualitatively} (see, for example, 
\cite{pol1}, pp.\,136-139), i.e. numerical values should be strictly  
avoided.

In the present paper we formulate 
\textit{a quantitative theory of inductive reasoning}, in particular
\textit{a consistent theory of quantitative inference about the parameters}
of sampling distributions. In Section\,\ref{sec:intro} 
we adopt the basic 
qualitative rules of plausible reasoning, the so-called
Cox-P\'{o}lya-Jaynes Desiderata, and review some of the well known results
of their direct applications, such as Cox's and Bayes'
Theorems. In addition, we clearly establish the lack of
such applications in the limit of complete prior ignorance
about the inferred parameter, with such ignorance 
representing the natural starting point of an inference. In other
words, Bayes' Theorem that can be used for \textit{updating}
probabilities, \textit{cannot} directly be used in the step
of probability \textit{assignment}.

We carefully define the state of complete prior ignorance
about the inferred parameter in Section\,\ref{sec:ignorance}.
In particular, throughout the present paper we \textit{never}
question the specified model, i.e. the form of the sampling probability
distribution is always known beyond the required precision.
What we \textit{are} completely ignorant about at the
beginning of reasoning is the value of the inferred parameter.

In Section\,\ref{sec:parameters} we define location, dispersion
and scale parameters and briefly review some of the properties
of sampling distributions determined by these parameters.
A separate section is devoted to the invariance of such distributions,
a property that turns out to be of decisive importance 
when constructing a consistent theory of inductive reasoning.
We also show that invariance of both the form of the sampling distribution
and its domain, under a continuous (or Lie) group, is found only in problems 
of inference about parameters that can be reduced to inference about location
parameters.

In Section\,\ref{sec:subjectivity} we extend the applications
of the Desiderata in order to allow for a consistent assignment
and not just for updating of probability distributions
for the inferred parameters. The so-called Consistency Theorem is 
obtained by
making use of Bayes' Theorem and by requiring that if a conclusion
can be reasoned out in more than one way, then every possible
way must lead to the same result, in particular by requiring
logically independent pieces of information to be
commutative. The form of the Consistency Theorem is very similar to that of 
Bayes' Theorem, but there is
also a fundamental difference between the two since in the 
former \textit{a consistency factor} is used instead of \textit{the prior
probability distribution}. Hence, the consistency
factor cannot be subject to any of the requirements such as
normalization or invariance with respect to a one-to-one 
parameter transformation, that are perfectly legitimate for
well defined probability distributions.

Instead, the form of the consistency factor is determined in a way
that preserves the logical consistency of our reasoning. By consistency we
mean that, among other things, if in two problems of inference our state 
of knowledge is the same, then we must assign the same probabilities in both. 
In Section\,\ref{sec:objectivity} we find that the basic Desiderata uniquely
determine the form of consistency factors for sampling distributions
whose form and domain are invariant under a Lie group $\mathcal{G}$
of transformations. 
It is therefore only for those problems reducible
to inference about location parameters that we can give an assurance
of consistency to our parameter inference. 
The form of consistency factors for such distributions is 
then determined throughout Sections\,\ref{sec:location}-\ref{sec:locscale},
while in Section\,\ref{sec:unique}
we discuss under what circumstances the present theory is guaranteed
to be consistent in the case of pre-constrained parameters. 

In Section\,\ref{sec:calibration} we make verifiable predictions that
are based on the presented theory, thus elevating its status
above the level of a mere speculation. The predictions
are made in terms of long run relative frequencies. We show that
a consistent inference is necessarily also a calibrated one, i.e.
that consistently predicted frequencies always coincide with 
(a one-to-one function of)
actual frequencies of occurrence. This important result speaks
in favour of a complete reconciliation between the so-called
Bayesian and frequentist schools of plausible reasoning.

In counting experiments the invariance of sampling distributions
is clearly missing and so the consistency factors cannot be 
uniquely determined by following the basic Desiderata. The remedy is 
to collect enough data so that the sampling distribution approaches its 
dense limit. Until then, our reasoning is necessarily based
on some \textit{ad hoc} prescriptions that can (and very often will) 
lead to logically unacceptable results, as is demonstrated
in Section\,\ref{sec:counting}. In such cases it might 
therefore be the best to refrain from quantitative
inferences, i.e. to remain on a qualitative level.

In Section\,\ref{sec:history} we briefly review conceptual
and practical difficulties and paradoxes caused by
using Bayes' Theorem instead of the Consistency Theorem in the limit 
of complete ignorance about the inferred parameters. The problem
is contained in the self-contradicting non-informative
prior probability distributions. Long-lasting arguments
over this subject led, \textit{inter alia}, to a split in
the theory of inductive reasoning and it is in this way 
that the Bayesian and the frequentist schools emerged. In our view, 
the splitting into (at first glance) almost diametrically opposed schools
is highly artificial, provided that the two schools
strictly obey their basic rules, i.e. that they refrain from
using \textit{ad hoc} shortcuts on the course of inference. For regardless
how close to our intuitive reasoning these \textit{ad hoc}
procedures may be, how well they may have performed in some other previous
inferences, and how respectable their names may sound (e.g. the principle
of insufficient reason or its sophisticated version - the principle of maximum
entropy, the principle of group invariance, the principle of maximum 
likelihood, and the principle of reduction),
they will in general inevitably lead not only to contradictions between
the two schools of thought, but also to inferences that are neither
consistent nor calibrated.

There are also two appendices to the present paper. The first one
contains, for the sake of completeness, a proof of Cox's Theorem,
while in the second one, the so-called marginalization paradox, is
extensively discussed.

\section{Basic rules and their applications}
\label{sec:intro}

Let \textit{n hypothesis}  or \textit{an event} be an unambiguous 
\textit{proposition} \dA, i.e. a statement that can be either 
\textit{true} or \textit{false}. As we are in general not certain
about either of the two possibilities, the classical logic of 
deductive reasoning \cite{ari} is to be extended in order to allow 
for \textit{plausible} or \textit{inductive} inferences based on 
incomplete information. 

Let (\textit{a state of}) \textit{information}
$I$ summarize the information that we have
about some set $\mathcal{A}$ of propositions $A_i$, called
\textit{the basis} of $I$, and their relations to each other.
The \textit{domain} of $I$ is the logical closure of ${\cal A}$,
that is, the union of ${\cal A}$. A state of information is not 
restricted to containing only deductive information; it can also contain 
imprecise or insufficient information that says nothing with certainty, 
but still affects 
one's opinion about a certain proposition. Such kind of information
can also be updated:
we write $I'=BI$ for a state of information obtained from $I$ by adding
additional information (\textit{ evidence}) that proposition $B$ is true.

Now, let $I$ be a state of 
information of a given person and $A$ be a proposition in the
domain of $I$. Then, we introduce the (\textit{degree of}) 
\textit{plausibility}
$(A|I)$ as a \textit{degree of belief} of the person that \dA is 
\textit{true} given the information in $I$. We say that
$I$ is the \textit{knowledge base} for the assigned plausibility $(A|I)$.
In the present paper we assume all considered plausibilities 
to be subject to very general requirements
that can be listed in the following three basic 
\textit{Cox-P\'olya-Jaynes Desiderata} (\cite{jay}, \S\,1.7, pp.\,17-19):
\textit{
  \begin{description}
    \item[\textit{I.}] Degrees of plausibilities are represented by real 
                       numbers.
\end{description}}
\noindent
That is, plausibilities are numerically encoded states of knowledge
about propositions. Formally, an assigned plausibility can 
be regarded as a function:
\begin{equation*}
\begin{CD}
(\;|\;): {\cal A} \times {\cal I} @>>> \mathbb{R} \ ,
\end{CD}
\end{equation*}
where ${\cal I}$ is the set of possible states of information about 
some set ${\cal A}$ of propositions.

In addition to the first Desideratum, we adopt two natural but nonessential
conventions: 
\begin{itemize}
\item a greater degree of belief shall correspond to a greater number; 
\item the plausibility of a hypothesis that we are certain about (e.g.
      the plausibility of a tautology) equals 1.
\end{itemize}
By referring to the conventions as being nonessential we mean 
that we could have equally well adopted a convention that the plausibility
of a tautology equals a different positive constant, or that a greater 
degree of probability should correspond to a smaller number.
Nevertheless, 
according to the above Desideratum and the two conventions, the 
assigned plausibilities can range within an interval $[{\mathsf F},1]$,
where ${\mathsf F}<1$ is plausibility of the false proposition.

We say that a state of information $I$ is \textit{consistent} if
there is no proposition $A$ for which plausibilities $(A|I)$ for $A$ 
being true and $(\bar{A}|I)$ for $A$ being false can both equal unity. 
That is, based on consistent information both a proposition and its denial
\textit{cannot} be true. In order
to avoid ambiguities, we restrict ourselves to considering only 
plausibilities that are assigned upon consistent states of information.

\textit{
\begin{description}
     \item[\textit{II.}] Assignment of plausibilities must be in 
              qualitative correspondence with common sense.
\end{description}}
\noindent
In our case, the concept of common sense stands for the following conditions:
\begin{itemize}
\item Since plausible reasoning is a generalization
of deductive logic, it must be consistent with the results of
Boolean algebra \cite{boo} - the algebra of deductive logic.
\item Microscopic changes in the knowledge base should \textit{not}
cause macroscopic changes in the plausibilities assigned. In addition,
for every considered proposition $A$ there exists some set of possible
consistent states of knowledge ${\cal I}$ such that $(A|I)$, with 
$I\in {\cal I}$,
can take any of the values
within a continuous interval $(a,b)\subseteq [{\mathsf F},1]$
(continuity requirement).
\item We assume that the degree of belief $(\bar{A}|I)$ that \dA is 
\textit{false} depends in some way on the plausibility $(A|I)$ that \dA is
true. In addition, when old information $I$ is updated into $I'$ in
such a way that the plausibility of \dA is increased, $(A|I') > (A|I)$, it
must produce a decrease in the plausibility that \dA is false, 
$(\bar{A}|I') < (\bar{A}|I)$. That is, we assume that there
exists a continuous, twice differentiable, strictly decreasing
function $S$ of plausibility $(A|I)$, such that
\begin{equation*}
 (\bar{A}|I) = S\bigl[(A|I)\bigr] \ .
\end{equation*}
\item Plausibility $(AB|I)$, assigned to a hypothesis $AB$ 
that two non-contradictory hypotheses, \dA and \dB, are simultaneously true,
is assumed to be completely determined by the values of $(A|I)$, 
$(B|AI)$, $(B|I)$ and $(A|BI)$. Then it can be shown 
(see Lemma\;1 in Appendix\;\ref{sec:cox})
that evident inconsistencies are avoided only if $(AB|I)$
depends solely on $(A|I)$ and $(B|AI)$, 
i.e. if there exists a function $H$ such that
\begin{equation*}
 (AB|I) = H\bigl[(A|I),(B|AI)] \ .
\end{equation*}
We further require for the function $H$ to be strictly increasing and twice
differentiable in both of its arguments. By strictly increasing we mean
that if the knowledge base $I$ is updated to $I'$ in such a 
way that the plausibility of \dA is increased, $(A|I') > (A|I)$,
but the plausibility $(B|AI)$
remains the same, $(B|AI') = (B|AI)$, this can only produce an
increase in the plausibility that both \dA and \dB are true,
$(AB|I') \ge (AB|I)$, in which the equality can hold only if
\dB is impossible given \dA and $I$. Likewise, given information $I''$
such that $(A|I'') = (A|I)$ and $(B|AI'') > (B|AI)$, we require
that $(AB|I'') \ge (AB|I)$. 
\end{itemize}
\textit{
\begin{description}
    \item[\textit{III.}] Assignment of plausibilities must be a 
         consistent procedure:
      \begin{description}
       \item[\textit{a)}] If a conclusion can be reasoned out in more 
             than one way,
             then every possible way must lead to the same result.
       \item[\textit{b)}] When assigning plausibilities, we must always 
             take into 
             account all of the evidence we have relevant to a hypothesis.
             We do not arbitrarily ignore some of the information and base
             our conclusion only on what remains. 
       \item[\textit{c)}] Equivalent states of knowledge must be always 
             represented by
             equivalent plausibility assignments. For example, if in 
             two problems our state of knowledge is the same
             (except perhaps for the labelling of the propositions), 
             then we must assign the same plausibilities in both.
      \end{description}
\end{description}}

The requirement of consistency plays a special r\^{o}le among
various requirements which a theoretical system, or an axiomatic
system, must satisfy. It can be regarded as the first of the
requirements to be satisfied by \textit{every} theoretical system,
be it empirical or non-empirical. As for an \textit{empirical}
system, however, besides being consistent, it should satisfy
a further criterion: it must be \textit{falsifiable} (\cite{pop},
\S\,24, pp.\,91-92). According to this criterion, statements, or
systems of statements, convey information about the empirical world
only if they are capable of clashing with experience; or more
precisely, only if they can be \textit{systematically tested},
that is to say, if they can be subjected (in accordance with a
\textit{methodological decision}) to tests which \textit{might}
result in their refutation. In so far as a scientific statement
speaks about reality, it must be falsifiable: and in so far as it
is not falsifiable, it does not speak about reality \cite{pop2}.
Since every degree of belief is to be assigned on the basis
of available evidence, our aim is clearly to formulate an
empirical theory of plausible reasoning, i.e. a theory that
speaks about reality. We therefore add an additional requirement,
an operational Desideratum, to the original Cox-P\'olya-Jaynes 
Desiderata:
\textit{
  \begin{description}
    \item[\textit{IV.}] A theory of plausible inference must specify
    operations that ensure the falsifiability of every assigned degree 
    of plausibility.
\end{description}}

Richard Cox showed \cite{cox} that the following can be deduced 
when plausibilities satisfy Desiderata\;\textit{I.-III.b}: 
\begin{description}
\item[\textit{1.}] Suppose that plausibilities $(A|I)$, $(B|AI)$,
$(B|I)$, $(A|BI)$ and $(AB|I)$ can be assigned. 
Then there exists a continuous strictly increasing function $P$ of 
each of these plausibilities,
\begin{equation*}
\begin{CD}
 P: [{\mathsf F},1] @>>> [0,1] \ ,
\end{CD}
\end{equation*}
 such that
\begin{equation}
 P(AB|I) = P(A|I)\,P(B|AI) = P(B|I)\,P(A|BI) 
\label{eq:product}
\end{equation}
and
\begin{equation*}
 P({\mathsf F}) = 0 \ .
\end{equation*}
Every \textit{compositum} of function $P$ and a plausibility
assignment $(A|I)$, $P\bigl[(A|I)\bigr]$, or $P(A|I)$ in a 
simplified notation, is referred to as
\textit{the probability} for \dA to be true given available information
$I$, and the above equation \eqref{eq:product} is referred to as 
\textit{the product rule}. Note that every probability is at the same 
time also a plausibility, i.e. it is consistent with the basic Desiderata.
\item[\textit{2.}] Probabilities $P(A|I)$ and
$P(\bar{A}|I)$ sum up to the probability of a certain event, i.e. sum
up to unity:
\begin{equation}
 P(A|I)+P(\bar{A}|I) = 1 \ ,
\label{eq:sum}
\end{equation}
which is referred to as \textit{the sum rule}.
\end{description}
The above results are usually referred to as Cox's Theorem
(for a proof of the Theorem see Appendix\,A). Note that
the product and the sum rule, \textit{being only relations between
probabilities, do not of themselves assign numerical values
to any of the probabilities arising in a specific problem}. The
only numerical values, considered thus far, are those corresponding
to certainty and impossibility, one and zero, respectively,
of which the former is a 
mere consequence of a convention, adopted along with
Desideratum\,\textit{I}, rather than required by the rules of the
Theorem. Moreover, \textit{it is hardly to be supposed that 
every reasonable expectation should have a precise numerical 
value} \cite{cox}, nor is there any guarantee that every 
state of information about a particular proposition $A$ will meet 
the continuity requirement of the common sense Desideratum.

The product and the sum rule are unique in the sense that any set of rules 
for manipulating our degrees of belief, represented by real numbers,
is either
\textit{isomorphic} to \eqref{eq:product} and \eqref{eq:sum}, i.e.
different from \eqref{eq:product} and \eqref{eq:sum} only in form but
not in content, or inconsistent. 
Thus, we could have chosen any other set of plausibilities $\widetilde{P}$
that are one-to-one functions of the corresponding probabilities $P$,
and then adequately adapt the product and the sum rule.
For example, if we choose $\widetilde{P}(A|I)$ such that
\begin{equation*}
 \widetilde{P}(A|I) \equiv \sqrt[a]{P(A|I)}
\end{equation*}
with $a$ being an arbitrary positive number, the corresponding product
rule for $\widetilde{P}$ remains the same while the sum rule reads:
\begin{equation*}
 \widetilde{P}^a(A|I)+\widetilde{P}^a(\bar{A}|I) = 1 \ .
\end{equation*}
As another example, we could have chosen zero to represent the plausibility
of a proposition that we are certain about, and a plausibility $\widetilde{P}$
such that:
\begin{equation*}
 \widetilde{P}(A|I) \equiv \ln{P(A|I)} \ .
\end{equation*}
Then, the appropriate product and sum rules for $\widetilde{P}$ would
have read:
\begin{equation*}
 \widetilde{P}(AB|I) = \widetilde{P}(A|I)+\widetilde{P}(B|AI)
                     = \widetilde{P}(B|I)+\widetilde{P}(A|BI)
\end{equation*}
and
\begin{equation*}
 \exp\bigl\{ \widetilde{P}(A|I)\bigr\}
 +\exp\bigl\{ \widetilde{P}(\bar{A}|I)\bigr\} = 1 \ .
\end{equation*}

The freedom to choose an arbitrary plausibility function to 
represent our degree of belief is analogous to \textit{gauge invariance}
in field theories where \textit{potentials} (i.e. functions
that the fields are expressed by) are not rigidly fixed.
The predictions of field theories are unchanged if
the potentials are transformed according to specific rules,
i.e. if the potentials are subjects to \textit{gauge transformations}.
Then we choose one particular form of potential, i.e. \textit{we choose
a particular gauge}, not because it is more correct than any other, but  
because it is more convenient for the particular problem that we are
solving within a filed theory. For the same reason we choose
probabilities and not any other plausibilities to represent our degrees
of belief: not because they are more correct, but because it is 
for probabilities that the product and the sum rule take the simplest forms.
We comment on the choice of probability, that we adhere to throughout the 
present paper, again in Section\,\ref{sec:calibration} when we discuss
the relation between probability and frequency.

Once the probabilities are chosen from all possible plausibility
functions, i.e once \textit{the gauge is fixed}, the incompleteness
of the concept of plausibility is removed:
the product and the sum rules, \eqref{eq:product} and \eqref{eq:sum},
are the fundamental equations of probability theory, while all other
equations for manipulating probabilities follow from their repeated 
applications. For example, it is in 
this way that we obtain \textit{the general sum rule} that either 
\dA or \dB is true:
\begin{equation}
P(A+B|I) = P(A|I) + P(B|I) - P(AB|I) \ .
\label{eq:gensum}
\end{equation}

Suppose now that propositions ${A_1,A_2,...,A_n}$ form an exhaustive set
of mutually exclusive propositions. The propositions are mutually
exclusive if the evidence $I$ implies that no two of them can be true
simultaneously,
$$
P(A_iA_j|I)=0 \ {\rm for} \ i\neq j \ ,
$$
and exhaustive if one of them must be true,
$$
P\Bigl(\sum_{i=1}^{n}A_i|I\Bigr) = \sum_{i=1}^{n}P(A_i|I) = 1 \ .
$$
A classical textbook example of such a set would be six hypotheses $A_i$
arising from tossing a die, where index $i$ corresponds to the particular number 
thrown.

The hypotheses of a set are unambiguously classified by assigning one
or more numerical indices. Deciding between the hypotheses $A_i$ and
estimating the index $i$ are practically the same thing:
\begin{equation}
P(A_i|I) = p(i|I) \ ,
\label{eq:notation}
\end{equation}
with the corresponding normalization
\begin{equation}
\sum_{i=1}^{n}p(i|I) = 1 \ .
\label{eq:nor}
\end{equation}
We denote probabilities by capital $P$ when arguments are propositions,
and by small $p$ when arguments are numerical values.
By assigning probability to every possible value of the index $i$ we 
specify how our degree of belief is distributed among the hypotheses
of the set ${A_i}$, i.e. we specify the  \textit{(sampling) 
probability distribution} for $i$. 

The distribution of one's degree of belief in hypotheses labelled by
different values of index $i$ may be equivalently represented by the 
\textit{cumulative distribution function} (cdf), defined as
\begin{equation}
 F(i) \equiv \sum_{j=i_a}^{i}p(j|I) = P(j\le i|I) \ ,
\label{eq:defcdf}
\end{equation}
where permissible values of the index $i$ range from $i_a$ to $i_b$. The
probability for $i$ taking a value between $i_1$ and $i_2$ can thus
be expressed by cdf's simply as:
\begin{equation*}
 P(i_1 < i \le i_2 |I) = \sum_{i=i_1}^{i_2}p(i|I) = 
                  \sum_{i=i_a}^{i_2}p(i|I) - \sum_{i=i_a}^{i_1}p(i|I) =
                  F(i_2) - F(i_1) \ .
\end{equation*}
In addition, for an exhaustive sets of hypotheses the normalization
condition \eqref{eq:nor} implies:
\begin{equation*}
 F(b) = \sum_{i=i_a}^{i_b}p(i|I) = 1 \ .
\end{equation*}

In many cases of practical importance the hypotheses of a set become very 
numerous and dense. For example, when predicting the decay time of an unstable
particle, we start with a countable set of hypotheses $A_i$ that the decay time
of that specific particle would be, for instance, $i$ seconds. But such
a set is not an exhaustive one since the decay time could also be
$i+\frac{1}{2}$ or $i+\frac{1}{4}$ seconds. Further refinement of the 
original propositions leads to a dense set of hypotheses where neighbouring
hypotheses, i.e. hypotheses with nearly the same index value, become
barely distinguishable. In such cases there cannot be a sharply defined
hypothesis that is strongly favoured over all others. Instead, it only makes
sense to consider probabilities for index $i$ in a certain interval of its
permissible range. In this way index $i$ transforms into a continuous variable
$x_i$ with a \textit{continuous} sampling probability distribution:
\begin{equation}
p(x_i|I) = P\bigl(x\in(x_i,x_i+dx)|I\bigr) \ .
\label{eq:contprob}
\end{equation}
Every continuous distribution can be expressed by a probability density
function (pdf), $f(x|I)$:
\begin{equation}
p(x|I) \equiv f(x|I)\,dx \ .
\label{eq:pdf}
\end{equation}
Using pdf's we can now rewrite the product rule \eqref{eq:product} as
\begin{equation}
f(x_1x_2|I) = f(x_1|I)\,f(x_2|x_1I) =f(x_2|I)\,f(x_1|x_2I) \ ,
\label{eq:product1}
\end{equation}
and the normalization \eqref{eq:nor} by replacing summation over discrete
indices $i$ by integration over a dense domain $x$:
\begin{equation}
\int\limits_x f(x'|I)\,dx' = 1 \ .
\label{eq:nor1}
\end{equation}

The sum in the cdf \eqref{eq:defcdf} for a discrete variable $i$
is replaced by an integral for a continuous variable $x$:
\begin{equation}
 F(x) \equiv \int_{x_a}^{x}f(x'|I)\,dx' = P(x'\le x|I) \ ,
\label{eq:defcdf1}
\end{equation}
where $x$ ranges from $x_a$ to $x_b$. Since the probability 
$P(x_1 < x < x_2 |I)$ can be expressed by the cdf's as:
\begin{equation*}
  P(x_1 < x < x_2 |I) = \int_{x_1}^{x_2}f(x|I)\,dx = F(x_2) - F(x_1) \ ,
\end{equation*}
the normalization \eqref{eq:nor1} implies:
\begin{equation*}
 F(x_b) = \int_{x_a}^{x_b}f(x'|I)\,dx' = 1 \ .
\end{equation*}

Suppose we have a continuous variable $y$ that is related functionally
to a variable $x$ by a one-to-one relation:
$$
x = x(y) \ ; \ y = y(x) \ ,
$$
$y$ being differentiable in $x$ and vice versa. Let $y_1=y(x_1)$ and
$y_2=y(x_2)$. Since the inferences about $x$ and $y$ are based on
\textit{equivalent pieces of information} $I$ and $I'$ 
(the transformations of the variables correspond only to relabellings of 
hypotheses), Desideratum\,\textit{III.c} implies the following equality:
\begin{equation}
 P\bigl(x\in(x_1,x_2)|I\bigr) = 
\begin{cases}
 P\bigl(y\in(y_1,y_2)|I'\bigr) \ \ ; \ \ \frac{dx}{dy} > 0 \\
 P\bigl(y\in(y_2,y_1)|I'\bigr) \ \ ; \ \ \frac{dx}{dy} < 0
\end{cases}
\ ,
\label{eq:eqprobxy}
\end{equation}
where
\begin{eqnarray}
P\bigl(x\in(x_1,x_2)|I\bigr) &=& \int_{x_1}^{x_2}f(x|I)\,dx \ , \nonumber \\
P\bigl(y\in(y_1,y_2)|I'\bigr) &=& \int_{y_1}^{y_2}f(y|I')\,dy \ \ ; \ \ 
\frac{dx}{dy} > 0 \ , \\ \nonumber
P\bigl(y\in(y_2,y_1)|I'\bigr) &=& \int_{y_2}^{y_1}f(y|I')\,dy \ \ ; \ \ 
\frac{dx}{dy} < 0 \ , \nonumber
\end{eqnarray}
with $f(x|I)$ and $f(y|I')$ being the pdf's for $x$ and $y$, respectively.
That is, \textit{assigned probabilities must be invariant under
variate transformations.}
\vskip 2 true mm
\par{\noindent\narrower\small
As an example that illustrates the above reasoning, imagine two scientists, 
say Mr.\,A and Mr.\,B, measuring decay times of unstable particles.
Each time they start their clocks at the moment when a particle is produced,
but the clocks run at different speeds, so Mr.\,A measures a decay time
$t_i$ of the $i$-th particle, and Mr.\,B 
\\ \parbox{0.915\linewidth}{
\begin{equation}
 t_i'=at_i \ ,
 \label{eq:metrics1}
\end{equation}
} \\
where $a$ is an arbitrary positive constant.

Since $t$ is a continuous random variable, there is not much point
in considering probabilities $P(t=t_i|I)$ and $P(t=t_i'|I)$, since
both, $t=t_i$ and $t=t_i'$ are events with zero probability
(probability measure). Instead, we should consider probabilities
for measuring $t$ in certain intervals $(t_i,t_i+dt)$ and 
$(t_i',t_i'+dt')$, where $dt$ and $dt'$ are the
widths of the two intervals. Due to the different speeds
of the two clocks, equivalent events, i.e. equivalent time intervals,
are \textit{not labelled equally} by the two observers: the interval
$(t_i,t_i+dt)$ of Mr.\,A corresponds to the interval
$(t_i',t_i+dt')=(at_i,at_i+adt)$ of Mr.\,B. For example, in the case of $a=5$, 
the interval no. 10 of Mr.\,A is split into intervals 46-50 by Mr.\,B.
That is, the variate transformation \eqref{eq:metrics1} implies
\begin{equation*}
 dt'=adt \ .
\end{equation*}
Then, since the two propositions, $t\in(t_i,t_i+dt)$ and
$t'\in(t_i',t_i'+dt')$ differ only in labelling, 
Desideratum\,\textit{III.c} implies the two probabilities
$P\bigl(t\in(t_i,t_i+dt)|I\bigr)$ and $P\bigl(t'\in(t_i',t_i'+dt')|I'\bigr)$,
assigned by Mr.\,A and Mr.\,B, respectively, to be equal: 
\begin{equation*}
 P\bigl(t\in(t_i,t_i+dt)|I\bigr) = P\bigl(t'\in(t_i',t_i'+dt')|I'\bigr) \ .
\end{equation*}
Note that the logic behind such a reasoning is very similar to the
logic of Poincar\'{e}'s relativity principle \cite{poi} of the special
theory of relativity stating that no preferred inertial frame
(or no absolute time scale) exists.

The variate transformation \eqref{eq:metrics1} 
is linear, but it need not be so, as long as it remains one-to-one. 
Suppose that Mr.\,B considers a probability distribution of 
a variate
\\ \parbox{0.915\linewidth}{
\begin{equation}
 y\equiv \ln{t} \ .
\label{eq:metrics2}
\end{equation}
} \\
If Mr.\,A divides a range $(0,1]$ of his variate $t$ into 
$n$ intervals of equal widths
\begin{equation*}
 dt = \frac{1}{n} \ ,
\end{equation*}
Mr.\,B's $n$ corresponding intervals $dy$ of highly non-uniform widths
\begin{equation*}
 dy=\frac{1}{t}\,dt = e^{-y}\,dt 
\end{equation*}
cover the infinite corresponding range $(-\infty,0]$ of $y$. 
Despite the non-uniformity of the interval widths, the transformation 
\eqref{eq:metrics2} still represents a mere relabelling of the
hypotheses: $t$ and $y$ are equivalent variates, and
$t\in(t_i,t_i+dt)$ and $y\in(y_i,y_i+dy)$ (with $y_i=\ln{t_i}$
and $dy=e^{-y_i}\,dt$) are equivalent propositions.
Imagine that because of using the transformed variate $y$
instead of $t$, Mr.\,B considers himself inferior to Mr.\,A. But
since the transformation \eqref{eq:metrics2} is one-to-one,
the inverse transformation
\begin{equation*}
 t = e^{y}
\end{equation*}
always exists: Mr.\,B can always obtain $t_i$ from $y_i$ and then 
make his inference from $t_i$ instead of from $y_i$.

To sum up, the logic behind Desideratum\,\textit{III.c} implies
equivalence of
all variates, connected \textit{via} one-to-one transformations:
while the specified models (i.e. forms of pdf's, see below) together with 
the range of
the variates may be changed (this is indicated by using
symbol $I'$ instead of $I$), 
\textit{the probability content must be invariant} under such transformations.
\par}
\vskip 2 true mm\noindent
The equality \eqref{eq:eqprobxy} is assured for all intervals
$(x_1,x_2)$ and $\bigl(y(x_1),y(x_2)\bigr)$ only if
(\cite{ken}, pp.\,20-28)
\begin{equation}
f(y|I') = f(x|I)\,\Big|\frac{dx}{dy}\Big| \ .
\label{eq:vartr1}
\end{equation}
Indeed:
\begin{equation*}
\begin{split}
 P\bigl(y\in(y_1,y_2)|I'\bigr) &= \int_{y_1}^{y_2}f(y|I')\,dy
  = \int_{y_1}^{y_2}f(x|I)\,\Big|\frac{dx}{dy}\Big|\,dy \\
  & = \int_{x_1}^{x_2}f(x|I)\,dx = P\bigl(x\in(x_1,x_2)|I\bigr)
\end{split}
\end{equation*}
for variate transformations with positive $dx/dy$,
and

\begin{equation*}
\begin{split}
 P\bigl(y\in(y_2,y_1)|I'\bigr) &= \int_{y_2}^{y_1}f(y|I')\,dy
 =\int_{y_2}^{y_1}f(x|I)\,\Big|\frac{dx}{dy}\Big|\,dy \\
 &= -\int_{x_2}^{x_1}f(x|I)\,dx = P\bigl(x\in(x_1,x_2)|I\bigr) 
\end{split}
\end{equation*}
in the case of negative $dx/dy$.
Inversely, $f(x|I)$ is expressed in terms of $f(y|I)$ as
\begin{equation}
f(x|I) = f(y|I')\,\Big|\frac{dy}{dx}\Big| \ ,
\label{eq:vartr2}
\end{equation}
where $|dy/dx|$ is the reciprocal of $|dx/dy|$,
$$
\Big|\frac{dy}{dx}\Big| = \Big|\frac{dx}{dy}\Big|^{-1} \ ,
$$
since any one-to-one transformation from $x$ to $y$ and then
from $y$ to $x$ must restore the original distribution, and 
hence
$$
\Big|\frac{dx}{dy}\Big|\,\Big|\frac{dy}{dx}\Big| = 1 \ .
$$
Note that equal information implies equal probabilities, while
in general it does \textit{not} imply equality of pdf's.

In the bivariate case where
\begin{eqnarray}
y^{(1)}&=&y^{(1)}(x^{(1)},x^{(2)}) \ , 
\ y^{(2)}=y^{(2)}(x^{(1)},x^{(2)}) \ ; \nonumber \\
x^{(1)}&=&x^{(1)}(y^{(1)},y^{(2)}) \ , 
\ x^{(2)}=x^{(2)}(y^{(1)},y^{(2)}) \ ; \nonumber
\end{eqnarray}
the relations between the pdf's for $(x^{(1)},x^{(2)})$ and
$(y^{(1)},y^{(2)})$, $f(x^{(1)},x^{(2)}|I)$ and $f(y^{(1)},y^{(2)}|I)$,
are the following: 
\begin{equation}
f(y^{(1)},y^{(2)}|I') = f(x^{(1)},x^{(2)}|I)\,|J| ,
\label{eq:vartr3}
\end{equation}
and
\begin{equation}
f(x^{(1)},x^{(2)}|I) = f(y^{(1)},y^{(2)}|I')\,|J^*| \ ,
\label{eq:vartr4}
\end{equation}
with the absolute values of the derivatives, $|dx/dy|$ and $|dy/dx|$, being
replaced by the absolute values of the corresponding Jacobians,
\begin{equation*}
J = \frac{\partial(x^{(1)},x^{(2)})}{\partial(y^{(1)},y^{(2)})}
 \equiv 
 \begin{vmatrix}
   \partial_{y^{(1)}}x^{(1)} &
   \partial_{y^{(2)}}x^{(1)} \cr
   \partial_{y^{(1)}}x^{(2)} &
   \partial_{y^{(2)}}x^{(2)}
 \end{vmatrix}
\ ,
\end{equation*}
and
$$
J^* = J^{-1} 
= \frac{\partial(y^{(1)},y^{(2)})}{\partial(x^{(1)},x^{(2)})} \ ,
$$
respectively.

Special attention is needed if the derivatives in a univariate 
case, or the Jacobians in a multivariate case, change sign within the domain 
of the pdf since in that case the variate transformations $x\to y$
and $y \to x$ are not one-to-one any more. We will meet such a difficulty
in Section\,\ref{sec:parameters} where it will be overcome
on account of the special symmetry of the specific transformation.

In the present paper we consider sampling probability distributions,
either discrete $p(i|\theta I)$ or continuous 
$p(x|\theta I) = f(x|\theta I)\,dx$, that can be 
specified by a mathematical function, determined by
the values of its \textit{parameters} $\theta$\footnote{We 
adhere to the common and useful convention of using Greek letters $\theta$,
$\mu$, $\sigma$, $\tau$, $\nu$ and $\lambda$ for parameters throughout 
the paper.}.
In such cases assignment of probabilities to hypotheses from a given set 
reduces to estimation of the parameters $\theta$ of the distribution:
what we try to achieve is to assign probabilities to different  
values of trhe parameters, i.e. to specify the probability distribution for 
$\theta$. Note that \textit{the probability distributions for parameters
are subjects to the same Desiderata as the distributions for sampling
variates.}
In this paper we will focus on the parameters with 
dense domains, where the corresponding probability distributions are 
continuous:
\begin{equation}
p(\theta|x_1I) = f(\theta|x_1I)\,d\theta \ .
\label{eq:pdftheta}
\end{equation}
The probability for $\theta$ \eqref{eq:pdftheta} is assigned upon
information that we explicitly split in our notation into evidence
$x_1$ from the measurement of the quantity $x$ whose probability distribution
is determined by $\theta$, and the additional relevant information $I$.
The reason for such splitting of the information will become evident below
when we derive Bayes' Theorem.

The pdf for $\theta$ \eqref{eq:pdftheta} is subject to the usual normalization,
\begin{equation}
\int\limits_\theta f(\theta'|x_1I)\,d\theta' = 1 \ ,
\label{eq:nor2}
\end{equation}
where integration is performed over the complete range of $\theta$. 
In addition, in the case of assigning probabilities for two parameters,
$\theta$ and $\nu$, simultaneously, the product rule \eqref{eq:product}
can be applied:
\begin{equation}
f(\theta\nu | x_1 I) = f(\theta | x_1 I)\,f(\nu | \theta x_1 I) =
                       f(\nu| x_1 I)\,f(\theta | \nu x_1 I) \ .
\label{eq:product2}
\end{equation}
Then, with the factors $f(\nu | \theta x_1 I)$ and $f(\theta | \nu x_1 I)$
being properly normalized according to \eqref{eq:nor2}, it is easy to see
that the \textit{marginalization} procedure yields
\begin{equation}
\begin{split}
\int\limits_\theta f(\theta'\nu | x_1 I)\,d\theta' &= f(\nu| x_1 I) \ , \\
\int\limits_\nu f(\theta\nu' | x_1 I)\,d\nu'    &= f(\theta | x_1 I) \ . 
\end{split}
\label{eq:marginal}
\end{equation}

The product rule can also be applied for assigning probabilities to 
$\theta$ and $x_1$:
\begin{equation}
p(\theta x_1|I) = f(\theta|I)\,d\theta\,p(x_1|\theta I) =
                  p(x_1|I)\,f(\theta|x_1 I)\,d\theta \ .
\label{eq:product3}
\end{equation}
The above equation can be rewritten into Bayes' Theorem \cite{bay,lap},
\begin{equation}
f(\theta | x_1 I) = \frac{f(\theta | I)\,p(x_1  | \theta I)}
                         {p(x_1 | I)} \ ,
\label{eq:bayes}
\end{equation}
also referred to as \textit{the principle of inverse probability}
(see \cite{jef}, \S\,1.22, p.\,28). When the domain of $x$ is also dense, 
the theorem can be written in terms of pdf's only:
\begin{equation}
f(\theta | x_1 I) = \frac{f(\theta | I)\,f(x_1  | \theta I)}
                         {f(x_1 | I)} \ ,
\label{eq:bayespdf}
\end{equation}

We interpret the theorem in the following way (\cite{hag}, \S\,1.3, p.2).
We are interested in the probability
distribution for $\theta$ and begin with the initial or \textit{prior}  
probability, also referred to as the \textit{probability a priori},
whose pdf reads $f(\theta|I)$. It is based on any additional information
$I$ that we possess beyond the immediate data $x_1$. Thus, $p(\theta |I) = 
f(\theta|I)\,d\theta$ is the probability for $\theta$ prior to taking
evidence $x_1$ into account. The \textit{posterior} probability, also
referred to as the \textit{probability a posteriori}, 
$f(\theta|x_1 I)\,d\theta$, is the probability for $\theta$ posterior to
adding evidence $x_1$ to our previous information. 
The \textit{likelihood}, $p(x_1 | \theta I)$, tells us how likely 
$x_1$ is observed, given the value $\theta$ of the parameter that 
determines the probability distribution for $x$. According to Bayes' 
Theorem \eqref{eq:bayes}, the only consistent way of
obtaining the posterior pdf, $f(\theta|x_1 I)$, is  
by multiplying the prior pdf by the likelihood,
which is usually referred to as \textit{the likelihood principle} 
(see, for example, \cite{jay}, \S\,8.5, p.\,250). Or, 
in the words of Jeffreys (\cite{jef}, \S\,2.0,
p.\,57): ''Consequently the whole of \textit{the information contained
in the observations} that is relevant to the posterior probabilities 
of different hypotheses\footnote{i.e. of different values of the
inferred parameter(s)}, is summed up in the values that they give
to the likelihood.''
Note that with the basic Desiderata being adopted, the term
``likelihood principle'' becomes inappropriate and might even be 
misleading, since the fact that all of the information that can
be extracted from the datum $x_1$ is contained in the value
of the appropriate likelihood, is \textit{a mere consequence of
application} of the basic Desiderata, rather than an additional
principle (i.e. a Desideratum) on its own.
The denominators  $p(x_1 | I)$ and $f(x_1 | I)$ can be obtained by 
the normalization requirement \eqref{eq:nor2} as:
\begin{equation}
p(x_1|I) = \int\limits_\theta {f(\theta'|I)\,p(x_1|\theta' I)\,d\theta'} 
\label{eq:norbayes}
\end{equation} 
and
\begin{equation}
f(x_1|I) = \int\limits_\theta {f(\theta'|I)\,f(x_1|\theta' I)\,d\theta'} \ . 
\label{eq:norbayes1}
\end{equation}

Bayes' Theorem is thus a rule for updating the information that an inference
is based upon. Formally, it is just a special case of the product rule of
the Cox theorem. The latter also ensures that \eqref{eq:bayes} is the only
consistent way of updating the information and, consequently, our 
probability distribution for $\theta$.

Suppose that after 
$x_1$ we learn a new piece of information, $x_2$, that we would like to include
in our inference about $\theta$. Then, $f(\theta|x_1 I)$ serves as a prior
pdf, i.e. pdf for $\theta$ prior to taking $x_2$ into account. According to
\eqref{eq:bayes}, the posterior pdf then reads:
\begin{equation}
f(\theta | x_1 x_2 I) = \frac{f(\theta |x_1 I)\,p(x_2 |\theta x_1 I)}
                         {p(x_1 x_2 | I)} 
            = \frac{f(\theta |I)\,p(x_1 |\theta I)\,p(x_2|\theta x_1 I)}
                         {p(x_1 | I)\,p(x_1 x_2 | I)}  \ ,
\label{eq:bayes1}
\end{equation}
where $p(x_2|\theta x_1 I)$ is the likelihood for $x_2$ given
$\theta$, $x_1$ and the additional information $I$.

In the limit of our \textit{complete ignorance} about the value of a parameter
$\theta$ prior to the first evidence $x_1$, when $I$ merely stands for 
our admission that we possess \textit{no} prior information relevant to 
$\theta$ apart from the specified form of the sampling distribution, 
the complete procedure for manipulating probabilities by 
using the product rule \eqref{eq:product}, or Bayes' Theorem 
\eqref{eq:bayes}, \textit{breaks down}. This is a direct consequence of the
fact that we can only assign probabilities for hypotheses on the 
basis of available relevant information: \textit{ignorance} $I$ thus
allows for \textit{no} probability assignment $f(\theta|I)$. In this event,
both the product rule \eqref{eq:product} and its derivative, Bayes'
Theorem \eqref{eq:bayes}, lack their vital components and
cannot be used. In other words, Bayes' Theorem only allows for updating
probabilities that were already assigned prior to their updating,
and therefore need be amended for the limit of complete prior
ignorance, which is a natural starting point for every sequential
updating of information. Our goal in the following sections is
to make such an amendment in a consistent way, \textit{i.e. to establish
when and how probabilities can consistently be assigned}.
\section{Complete ignorance about parameters}
\label{sec:ignorance}

As a starting point, we would like to specify precisely what we 
\textit{mean} and what we \textit{do not mean} by the limit of 
complete ignorance about the value of a parameter of a given probability
distribution for $x$. First of all, ignorance about a distribution
parameter is \textit{not} a synonym for absolute ignorance in every
possible respect. For example, throughout the paper we assume as a working
hypothesis that the probability for $x$ is distributed in a form
that is \textit{completely known} but for the value of its parameter(s)
$\theta$, i.e. the chosen form of the probability distribution
for $x$ together with its domain - the ranges of the sampling variate(s)
and the inferred parameter(s), $(x_a,x_b)$ and $(\theta_a,\theta_b)$,
is always assumed to be appropriate beyond the required
precision. This assumption is explicitly indicated by the symbol $I$
that every probability (or probability density) is conditioned upon. 

What we \textit{are completely ignorant} about is the value of $\theta$. 
There is no information at our disposal that would enable us to
assign a probability distribution for $\theta$: it can take
any value within its permissible range $(\theta_a,\theta_b)$. 
Since the value of $\theta$ is
completely unknown, then the distribution for $x$ becomes undetermined. 
This is where we then start collecting data that
we would like to use for a consistent inference about $\theta$.

The situation, described above, is an ideal limiting case that can serve
as a reasonable approximation for many real-life situations.
For example, even before the first measurement of a decay time of an
unknown unstable particle, there is not much room for doubt 
about the form of the decay time distribution. Due to past experiences with
all other unstable particles we feel almost completely certain
that the distribution would be exponential (we come back to this point 
in Section\,\ref{sec:probhyp} where the possibility of assigning
probabilities to specified models is considered).

But before the first measurement, the parameter of the distribution, the
average decay time $\tau$, is completely unknown, so we do not know
what value of the first measure decay-time, $t_1$, to expect:
it could be anything between zero and infinity. Inversely, before the
first measurement of $t$, the parameter $\tau$ can take any value
in the same interval.
Note that a hint of a symmetry between the collected data
and the inferred parameters is present in the foregoing reasoning. 
The concept will be extensively exploited during the following sections.

With $I$ representing only knowledge about the type of sampling
distribution, different states of knowledge $I_x\equiv xI$ and
$I_{\theta}\equiv\theta I$ can be enumerated according to the values
of $x$ and $\theta$, respectively. In this way, the sets
${\cal I}_x$ and ${\cal I}_{\theta}$ of possible different states
of knowledge become subsets of real numbers, 
${\cal I}_x,{\cal I}_{\theta}\subseteq\mathbb{R}$, and the pdf's
$f(x|\theta I)$ and $f(\theta|xI)$ can both be formally expressed
as functions
\begin{equation*}
\begin{CD}
f(x|\theta I),f(\theta|xI): \mathbb{R} \times \mathbb{R} @>>> \mathbb{R} \ .
\end{CD}
\end{equation*}

Many of the derivations in the present article
automatically require dense ranges for both $x$ and $\theta$, as well as 
differentiability of pdf's $f(x|\theta I)$ and $f(\theta|x_1 I)$.
The problems accompanying discrete sets of hypotheses are 
extensively discussed in Section\;\ref{sec:counting}, where
inferences about parameters of counting experiments are considered.

\section{Location, scale and dispersion parameters}
\label{sec:parameters}

We pay special attention to the so called \textit{location},
\textit{scale}, and \textit{dispersion} parameters of probability
distributions. A parameter $\mu$ of a sampling distribution is a 
location parameter, and a parameter $\sigma$ is a dispersion parameter,
if the pdf for $x$ takes the form 
\begin{equation}
f(x|\mu\sigma I) = \frac{1}{\sigma}\,\phi\Bigl(\frac{x-\mu}{\sigma}\Bigr) \ ,
\label{eq:defpdflocscale}
\end{equation}
with the range for $x$ stretching over the whole real axis, with the
range of $\mu$ being an interval $(\mu_a,\mu_b)$ on
the real axis and with the permissible range of $\sigma$ being an 
interval $(\sigma_a,\sigma_b)$ on the positive half of the real axis.
For the time being, let the permissible range of $\mu$ coincide with 
the entire real axis, $(\mu_a,\mu_b)=(-\infty,\infty)$, and the
range of $\sigma$ with its entire positive half,
$(\sigma_a,\sigma_b)=(0,\infty)$, while we postpone a discussion
about pre-constrained parameters until Section\,\ref{sec:unique}.

A bivariate pdf for two independent variates, $x^{(1)}$ and $x^{(2)}$, 
both being subject to the same pdf of the form \eqref{eq:defpdflocscale}, 
according to the product rule \eqref{eq:product1} equals the product
of univariate pdf's, $f(x^{(1)}|\mu\sigma I)$ and 
$f(x^{(2)}|\mu\sigma I)$:
\begin{equation}
\begin{split}
f(x^{(1)}x^{(2)}|\mu\sigma I) 
&= f(x^{(1)}|\mu\sigma I)\,f(x^{(2)}|x^{(1)}\mu\sigma I) \\
&= f(x^{(1)}|\mu\sigma I)\,f(x^{(2)}|\mu\sigma I) \\
&=\frac{1}{\sigma^2}\,\phi\Bigl(\frac{x^{(1)}-\mu}{\sigma}\Bigr)\,
                        \phi\Bigl(\frac{x^{(2)}-\mu}{\sigma}\Bigr) \ .
\end{split}
\label{eq:pdfx1x2}
\end{equation}
The pdf for transformed variates $x^{(1)}$ and $x^{(2)}$, 
$\bar{x}$ and $s$, where
\begin{equation*}
\begin{split}
 \bar{x} &\equiv \frac{x^{(1)} + x^{(2)}}{2} \ ,\\
       s &\equiv \frac{|x^{(1)} - x^{(2)}|}{2} = 
  \begin{cases}
   (x^{(1)} - x^{(2)})/2 &;\; x^{(1)} > x^{(2)} \\
   (x^{(2)} - x^{(1)})/2 &;\; x^{(2)} > x^{(1)} \ ,
  \end{cases}
\end{split}
\end{equation*}
or, inversely,
\begin{equation*}
x^{(1,2)} = 
  \begin{cases}
  \bar{x}\pm s &;\; x^{(1)} > x^{(2)} \\
  \bar{x}\mp s &;\; x^{(2)} > x^{(1)} \ ,
  \end{cases}
\end{equation*}
can be calculated according to \eqref{eq:vartr3}:
\begin{equation}
\begin{split}
f(\bar{x}s|\mu\sigma I) &= 
4\,f\bigl(x^{(1)}(\bar{x},s)\,x^{(2)}(\bar{x},s)|\mu\sigma I\bigr) \\ 
&= \frac{4}{\sigma^2}
   \begin{cases}
     \phi\bigl(\frac{\bar{x}+s-\mu}{\sigma}\bigr)
     \phi\bigl(\frac{\bar{x}-s-\mu}{\sigma}\bigr) &;\; x^{(1)} > x^{(2)} \\
     \phi\bigl(\frac{\bar{x}-s-\mu}{\sigma}\bigr)
     \phi\bigl(\frac{\bar{x}+s-\mu}{\sigma}\bigr) &;\; x^{(2)} > x^{(1)} 
   \end{cases} \\
&= \frac{4}{\sigma^2}
   \begin{cases}
     \phi\bigl(\frac{\bar{x}-\mu}{\sigma}+\frac{s}{\sigma}\bigr)
     \phi\bigl(\frac{\bar{x}-\mu}{\sigma}-\frac{s}{\sigma}\bigr) 
     &;\; x^{(1)} > x^{(2)} \\
     \phi\bigl(\frac{\bar{x}-\mu}{\sigma}-\frac{s}{\sigma}\bigr)
     \phi\bigl(\frac{\bar{x}-\mu}{\sigma}+\frac{s}{\sigma}\bigr) 
     &;\; x^{(2)} > x^{(1)} 
   \end{cases} \\ 
&=\frac{4}{\sigma^2}
\phi\Bigl(\frac{\bar{x}-\mu}{\sigma}+\frac{s}{\sigma}\Bigr)
\phi\Bigl(\frac{\bar{x}-\mu}{\sigma}-\frac{s}{\sigma}\Bigr) \\
&=\frac{1}{\sigma^2}\,\widetilde{\phi}
                         \Bigl(\frac{\bar{x}-\mu}{\sigma},
                         \frac{s}{\sigma}\Bigr) \ .
\end{split}
\label{eq:pdfxbars}
\end{equation}
The factor 4 in \eqref{eq:pdfxbars} that ensures appropriate normalization,
arises as a product of the absolute value of the Jacobian,
\begin{equation}
J = \frac{\partial(x^{(1)},x^{(2)})}{\partial(\bar{x},s)}
=
 \begin{cases}
  -2 &;\; x^{(1)} > x^{(2)} \\
  +2  &;\; x^{(2)} > x^{(1)} \ ,
  \end{cases}
\label{eq:jac}
\end{equation}
and an additional factor of 2, the latter being a consequence of the 
symmetry of the pdf for $x^{(1)}$ and $x^{(2)}$ with respect to the 
change of sign of the
difference $x^{(1)}-x^{(2)}$. With the sign of the difference inverted, 
the determinant \eqref{eq:jac} inverts its sign, too, 
but its absolute value, as well as the value of the pdf 
$f\bigl(x^{(1)}(\bar{x},s)\,x^{(2)}(\bar{x},s)|\mu\sigma I\bigr)$
(see \eqref{eq:pdfxbars}), remain unchanged. Therefore both the range
with the positive and the range with the negative sign of the Jacobian
can be simultaneously taken into account simply by multiplying the pdf
$f\bigl(x^{(1)}(\bar{x},s)\,x^{(2)}(\bar{x},s)|\mu\sigma I\bigr)$
by an additional factor of 2. 

When inferring the parameters of a sampling distribution of the form
\eqref{eq:defpdflocscale} it may happen that the value of one of the 
two parameters is known to a high precision. In such cases the parameter
with the precisely determined value is \textit{fixed} and we only
make an inference about the remaining one. Let first the dispersion 
parameter $\sigma$ be fixed to $\sigma_0$, e.g. to 1, so that the 
pdf for $x$, given the possible value of $\mu$ and the fixed value $\sigma_0$,
\begin{equation}
 f(x|\mu\sigma_0 I) = 
 \frac{1}{\sigma_0}\,\phi\Bigl(\frac{x-\mu}{\sigma_0}\Bigr) =
 \phi(x-\mu) \ ,
\label{eq:loc}
\end{equation}
is a function of $x$ and $\mu$ only. The fixed 
parameter ($\sigma_0$ in the present case) is usually (though not always)
omitted from explicit expressions. 

According to \eqref{eq:loc} and \eqref{eq:defcdf1},
the cdf for $x$ reads:
\begin{equation}
 F(x,\mu,\sigma_0) = \int_{-\infty}^{x} f(x'|\mu\sigma_0 I)\,dx'
                  = \int_{-\infty}^{x} \phi(x'-\mu)\,dx'
                  = \int_{-\infty}^{x-\mu} \phi(u)\,du 
                  = \Phi(x-\mu)\ .
\label{eq:cdfloc}
\end{equation}
Note that the form $\Phi(x-\mu)$ of the above cdf implies the corresponding
pdf to be of the form \eqref{eq:loc}, i.e. implies $\mu$ to be
a location parameter of a sampling probability distribution for $x$.
Indeed:
\begin{equation*}
\hskip -2mm
 f(x|\mu I) = \frac{\partial}{\partial x}F(x,\mu) =
  \frac{\partial}{\partial x}\Phi(x-\mu) 
=\frac{d}{du}\Phi(u)\,\frac{\partial u}{\partial x}
                      = \frac{d}{du}\Phi(u) = \phi(u) \ ,
\end{equation*}
where
\begin{equation*}
 u \equiv x-\mu \ .
\end{equation*}
Since the integrand in \eqref{eq:cdfloc}, 
i.e. the pdf for $x$, is a positive function, and the
upper bounds of the integral are strictly decreasing with the
increase in the parameter, the cdf is evidently strictly decreasing in $\mu$.

With the location parameter being fixed to $\mu_0$, say to 0,
the pdf \eqref{eq:defpdflocscale} for $x$ reduces to:
\begin{equation}
f(x|\sigma\mu_0 I) = \frac{1}{\sigma}\phi\Bigl(\frac{x-\mu_0}{\sigma}\Bigr)
                   = \frac{1}{\sigma}\phi\Bigl(\frac{x}{\sigma}\Bigr) \ .
\label{eq:dis}
\end{equation}
When the range of the random variate is bound to the positive half of
the real axis,
\begin{equation*}
 t \in (0,\infty) \ ,
\end{equation*}
the corresponding parameter $\tau$ of the sampling distribution,
\begin{equation}
f(t|\tau\mu_0 I) = \frac{1}{\tau}\phi\Bigl(\frac{t}{\tau}\Bigr) \ ,
\label{eq:sca}
\end{equation}
is usually referred to as the \textit{scale} parameter.
Note that in symmetric cases when
\begin{equation*}
 f(x-\mu_0|\sigma\mu_0 I) = f\bigl(-(x-\mu_0)|\sigma\mu_0 I\bigr),
\end{equation*}
the pdf \eqref{eq:dis} for $x$ with fixed $\mu_0$ can be 
reduced to a pdf \eqref{eq:sca} without any loss of either
generality or information. Namely, the pdf for a transformed
variate
\begin{equation*}
 t=|x-\mu_0| \ ,
\end{equation*}
reads:
\begin{equation*}
f(t|\sigma\mu_0 I) = 2\,f(x-\mu_0|\sigma\mu_0 I)\,\Bigl|\frac{dx}{dt}\Bigr|
                   = \frac{2}{\sigma}\phi\Bigl(\frac{t}{\sigma}\Bigr)
             = \frac{1}{\sigma}\widetilde{\phi}\Bigl(\frac{t}{\sigma}\Bigr),
\end{equation*}
where the factor of 2 arises due to shrinkage of the sample
space. In such cases dispersion parameters are evidently equivalent to
scale parameters.

Any sampling probability distribution determined by
a scale parameter, $\tau$, and a fixed location parameter
$\mu_0$, can be further transformed
into a probability distribution determined by a location parameter 
$\nu$ (see, for example, \cite{fer}, \S\,4.4, p.\,144 or \cite{har}, 
\S\,3.2.2, pp.\,22-23):
\begin{equation*}
\nu = \ln{\tau} \ ,
\end{equation*}
and a fixed dispersion parameter $\lambda_0$, say $\lambda_0=1$. 
Namely, a substitution
\begin{equation*}
z = \ln{t} 
\end{equation*}
yields:
\begin{equation}
f(z|\nu\lambda_0 I) = f(t|\tau\mu_0 I)\,\Bigl|\frac{dt}{dz}\Bigr| \\
=e^{z-\nu}\,\phi\bigl(e^{z-\nu}\bigr) \equiv
              \widetilde{\phi}_(z-\nu) \ .
\label{eq:fpmz}
\end{equation}

As was the case with cdf \eqref{eq:cdfloc}, the cdf for $t$,
\begin{equation}
 F(t,\tau,\mu_0) = \int_{0}^{t} f(t'|\tau\mu_0 I)\,dt'
                  = \int_{0}^{t} \frac{1}{\tau}\,
                    \phi\Bigl(\frac{t'}{\tau}\Bigr)\,dt'
                  = \int_{0}^{\frac{t}{\tau}} \phi(u)\,du \ ,
\label{eq:cdfscale}
\end{equation}
is also monotonically decreasing with increasing value of the parameter.

Some of the most important continuous sampling distributions are determined
by one or more parameters of the above mentioned types (see \cite{ead},
\S\,4.2, pp.\,58-83). In addition, all distributions determined by either 
location, scale or dispersion parameters share a very important property: 
they all belong to the \textit{invariant families of distributions}.

\section{Invariant distributions}
\label{sec:invariance}

Let 
\begin{equation*}
f(x|\theta I) = \phi(x,\theta) 
\end{equation*}
be a pdf for a random variable from a dense sample space
$X$ that is determined by the value of parameter $\theta$
from the parameter space $\Theta$. Let there exist \textit{a 
group} ${\mathcal G}$ of transformations $g_a\in{\mathcal G}$ of the 
sample space into itself:
\begin{equation*}
\begin{CD}
\begin{split}
g_a&:\; X \negthickspace\negthickspace @>>> \;X \ , \\
g_a&:\; x \negthickspace\negthickspace @>>> \;g_a(x) \equiv y \ ,
\end{split}
\end{CD}
\end{equation*}
where index $a$ denotes the particular element of the group.
Since ${\mathcal G}$ is a group, it is closed under
composition of transformation, i.e. a composition $g_c$ of every pair of
transformations $g_a,\,g_b\in{\mathcal{G}}$, $g_c=g_bg_a$, such
that
\begin{equation*}
 g_c(x)=g_bg_a(x)=g_b\bigl(g_a(x)\bigr) \ ,
\end{equation*}
is also contained in ${\mathcal G}$. In addition, the group
also contains an identity
$g_e$ such that
\begin{equation*}
g_e(x) = x \ , \ \forall \ x\in X \ , 
\end{equation*}
and the inverse transformation $g_a^{-1}$ for any $g_a$ such that
\begin{equation*}
g_a^{-1}g_a=g_ag_a^{-1}=g_e \ .
\end{equation*}
As a consequence, the transformations $g_a$ are one-to-one, i.e.
$g_a(x_1)=g_a(x_2)$ implies $x_1=x_2$, and \textit{onto} $X$,
i.e. for every $x_1\in X$ there exists an $x_2\in X$ such that
$g_a(x_2) = x_1$ (see, for example, \cite{fer}, \S\,4.1, p.\,143).

Since the transformation $y=g_a(x)$ is one-to-one,
the pdf for the transformed variate according to \eqref{eq:vartr1} 
and \eqref{eq:vartr2} reads:
\begin{equation}
\begin{split}
f(y|\theta I')&= f\bigl(g_a(x)|\theta I'\bigr) 
              = f(x|\theta I)\Bigl|\frac{dy}{dx}\Bigr|^{-1} \\
              &= \phi(x,\theta)\bigl|g_a'(x)\bigr|^{-1}
              = \phi\bigl(g_a^{-1}(y),\theta\bigr)\bigl|g_a'(x)\bigr|^{-1} \ .
\end{split}
\label{eq:invar0}
\end{equation}

In addition to ${\mathcal G}$, let there exist also a set 
$\bar{\mathcal G}$ of transformations $\bar{g}_a$ of the 
parameter space $\Theta$ into itself,
\begin{equation*}
\begin{CD}
\begin{split}
 \bar{g}_a&:\ \Theta \negthickspace\negthickspace @>>>\; \Theta \ , \\
 \bar{g}_a&:\ \theta\; \negthickspace\negthickspace @>>>\; 
 \bar{g}_a(\theta) \equiv \nu \ .
\end{split}
\end{CD}
\end{equation*}
Then:
\begin{equation}
\begin{split}
f(y|\nu I')&= f\bigl(g_a(x)|\bar{g}_a(\theta)I'\bigr)
            = \phi\bigl(x(y),\theta(\nu)\bigr)\bigl|g_a'(x)\bigr|^{-1} \\
  &= \phi\bigl(g_a^{-1}(y),\bar{g}_a^{-1}(\nu)\bigr)\bigl|g_a'(x)\bigr|^{-1}
           \equiv \widetilde{\phi}\bigl(g_a(x),\bar{g}_a(\theta)\bigr) \ .
\end{split}
\label{eq:invar1}
\end{equation}
If for all $x\in X,\ \text{and}\ \theta\in\Theta$, and for every
$g_a\in{\mathcal G}$ there exists $\bar{g}_a\in\bar{\mathcal G}$
such that
\begin{equation}
 \widetilde{\phi}\bigl(g_a(x),\bar{g}_a(\theta)) =
 \phi \bigl(g_a(x),\bar{g}_a(\theta)\bigr) \ , 
\label{eq:invar2}
\end{equation}
the family of distributions $f(x|\theta I)$ is said to be
\textit{invariant under the group} ${\mathcal G}$ (\cite{fer},
\S\,4.1, p.\,144 and \cite{ken2}, \S\,23.10, pp.\,300-301).

If a family of distributions is invariant under ${\mathcal G}$, then
the set $\bar{\mathcal G}$ of transformations of
$\Theta$ into itself is also a group, usually referred to
as \textit{the induced group} (\cite{ken2}, \S\,23.10, p.\,300).
Namely, according to the definition of invariance, if the
pdf for $x$ is given by $\phi(x,\theta)$, the pdf
for $g_a(x)$ is given by $\phi\bigl(g_a(x),\bar{g}_a(\theta)\bigr)$.
Hence, the pdf for $g_b\bigl(g_a(x)\bigr)=g_bg_a(x)$
is given by both $\phi\bigl(g_b(g_a(x)),\bar{g}_b(\bar{g}_a(\theta))\bigr)$
and $\phi\bigl(g_bg_a)(x),\overline{g_bg_a}(\theta)\bigr)$.
From the equality of the two it follows that 
\begin{equation*}
\overline{g_bg_a}=\bar{g}_b\bar{g}_a \ .
\end{equation*}
This shows that $\bar{\mathcal G}$ is closed under composition. It also 
shows that $\bar{\mathcal G}$ is closed under inverses if we let
$g_b=g_a^{-1}$ and note that $\bar{g}_e$ is the identity in $\bar{\mathcal G}$.

For example, a sampling distribution for $\bar{x}$ and $s$
\eqref{eq:pdfxbars}, determined by the values
of a location parameter $\mu$ and a dispersion parameter $\sigma$,
is invariant under the group of simultaneous location
and scale transformations:
\begin{equation}
\begin{split}
 g_{a,b}&:\ 
 \begin{cases}
 \begin{CD}
 \; \bar{x}  @>>>\;  g_{a,b}(\bar{x}) = a\bar{x}+b \\
  s \negthickspace @>>> \hskip -6mm g_{a,b}(s) = as
 \end{CD}
 \end{cases} \; , \\
 \bar{g}_{a,b}&:\ 
 \begin{cases}
 \begin{CD}
 \;\mu @>>>\; \bar{g}_{a,b}(\mu) = a\mu+b \\
 \sigma \negthickspace\negthickspace @>>> \hskip -6mm 
  \bar{g}_{a,b}(\sigma) = a\sigma
 \end{CD} 
\end{cases} \, , 
\end{split}
\label{eq:invlocscale}
\end{equation}
where
\begin{equation*}
a\in(0,\infty) \ \ \text{and} \ \ b\in(-\infty,\infty) \ .
\end{equation*}
By fixing the dispersion parameter, the symmetry of the pdf with
respect to the scale transformation is broken, leaving only the
symmetry with respect to a simultaneous translation of $x$ and $\mu$
by an arbitrary real number $b$:
\begin{equation}
\begin{CD}
\begin{split}
 g_b&:\ x \ \negthickspace\negthickspace @>>>\;  g_b(x) = x + b   \ , \\
 \bar{g}_b&:\ \mu\, \negthickspace\negthickspace @>>>\;  
 \bar{g}_b(\mu) = \mu + b \ .
\end{split}
\end{CD}
\label{eq:invloc}
\end{equation}
When, on the other hand, the location parameter is fixed to
$\mu_0=0$ and the dispersion (or scale) of the distribution 
is unknown, the appropriate pdf $f(x|\sigma\mu_o I)$
\eqref{eq:sca} is still invariant under the scale
transformation:
\begin{equation}
\begin{CD}
\begin{split}
 g_a&:\ x \ \negthickspace\negthickspace @>>>\; g_a(x) = ax  \ , \\
 \bar{g}_a&:\ \sigma\, \negthickspace\negthickspace @>>>\; 
 \bar{g}_a(\sigma) = a\sigma \ .
\end{split}
\end{CD}
\label{eq:invscale}
\end{equation}

Let now $f(x|\theta I)$ be an invariant sampling distribution and
let $F(x,\theta)$ be its cdf such that
\begin{equation*}
 F(x,\theta)=\int_{x_a}^{x}f(x'|\theta I)\,dx' 
            = \int_{x_a}^{x}\phi(x',\theta)\,dx' \ ,
\end{equation*}
where $x_a$ is the lower bound of the sample space.
Then the cdf for $y=g_a(x)$, given $\nu=\bar{g}_a(\theta)$, reads:
\begin{equation*}
 F(y,\nu) = F\bigl(g_a(x),\bar{g}_a(\theta)\bigr)
          = \int_{y_a}^{y}f(y'|\nu I')\,dy' 
          = \int_{y_a}^{g_a(x)}\phi\bigl(g_a(x'),\bar{g}_a(\theta)\bigr)\,
            d\bigl(g_a(x')\bigr) \ ,
\end{equation*}
where $y_a$ is the lower bound of the range of $y$. It is easy to show
that:
%
the lower and the upper bound of the range of $x$, 
         $x_a$ and $x_b$, become transformed into the bounds of $y$, 
         $y_a$ and $y_b$:
\begin{equation*}
 y_a =
 \begin{cases}
 g_a(x_a) \ \ \ ; \ \ \ g'_a(x) > 0 \\
 g_a(x_b) \ \ \ ; \ \ \ g'_a(x) < 0 
 \end{cases}
 \ \  \text{and} \ \ \ \ 
 y_b =
 \begin{cases}
 g_a(x_b) \ \ \ ; \ \ \ g'_a(x) > 0 \\
 g_a(x_a) \ \ \ ; \ \ \ g'_a(x) < 0 
 \end{cases}
\ ,
\end{equation*}
%
and that the cdf for $y$, given $\nu$, is related to the cdf
for $x$, given $\theta$, as:
\begin{equation*}
\hskip -5mm
 F(y,\nu) = F\bigl(g_a(x),\bar{g}_a(\theta)\bigr) =
 \begin{cases}
  \int_{g_a(x_a)}^{g_a(x)}\phi\bigl(y',\bar{g}_a(\theta)\bigr)\,dy'
 = \ \ \ \ F(x,\theta) \ \ \ \ ; \ g'_a(x) > 0 \\
  \int_{g_a(x_b)}^{g_a(x)}\phi\bigl(y',\bar{g}_a(\theta)\bigr)\,dy'
 = 1-F(x,\theta) \ ; \ g'_a(x) < 0 
 \end{cases}
 .
\end{equation*}
%
%
Indeed:
\begin{equation*}
\begin{split}
\hskip -5mm 
F\bigl(g_a(x),\bar{g}_a(\theta)\bigr)-F\bigl(g_a(x_a),\bar{g}_a(\theta)\bigr)
           &=\int_{g_a(x_a)}^{g_a(x)}f\bigl(y'|\bar{g}_a(\theta) 
           I'\bigr)\,dy' \\
         &=\int_{g_a(x_a)}^{g_a(x)}\phi\bigr(x',\theta)\,
           \bigr|g_a'(x')\bigr|^{-1}\,d\bigr(g_a(x')\bigr) \\
         &=\pm \int_{x_a}^{x}\phi(x',\theta)\,dx' \\
         &=\pm F(x,\theta) \ ,
\end{split}
\end{equation*}
where the positive and the negative sign correspond to $g'_a(x) > 0$
and to $g'_a(x) < 0$, respectively.
Setting $x$ to the upper bound $x_b$ of its range, the above equation
reads:
\begin{equation*}
  F\bigl(g_a(x_b),\bar{g}_a(\theta)\bigr)
 -F\bigl(g_a(x_a),\bar{g}_a(\theta)\bigr)
 =\pm F(x_b,\theta)=\pm 1 \ .
\end{equation*}
Since the cdf's are limited within $[0,1]$, this completes the
proof by implying
\begin{equation*}
 F\bigl(g_a(x_a),\bar{g}_a(\theta)\bigr) =
 \begin{cases}
 0 \ \ \ ; \ \ \ g'_a(x) > 0 \\
 1 \ \ \ ; \ \ \ g'_a(x) < 0
 \end{cases}
\ \  \text{and} \ \ \ \ \ 
 F\bigl(g_a(x_b),\bar{g}_a(\theta)\bigr) =
 \begin{cases}
 1 \ \ \ ; \ \ \ g'_a(x) > 0 \\
 0 \ \ \ ; \ \ \ g'_a(x) < 0
 \end{cases}
\end{equation*}
for every $g_a\in{\mathcal G}$ and $\bar{g}_a\in\bar{\mathcal G}$, 
and for all $\theta\in\Theta$.

A very important corollary - the Existence Theorem - can be deduced from 
the above relations. Let a probability distribution for $x$ be
invariant under $\mathcal{G}$, and let $\mathcal{G}$ and $\bar{\mathcal{G}}$ 
be continuous groups such that partial derivatives
\begin{equation*}
 \frac{\partial}{\partial a}g_a(x) \ \ \ \ \text{and} 
\ \ \ \ \frac{\partial}{\partial a}\bar{g}_a(\theta)
\end{equation*}
exist for every $g_a\in\mathcal{G}$ and $\bar{g}_a\in\bar{\mathcal{G}}$,
with both derivatives always being different from zero and finite.
In other words, $\mathcal{G}$ and $\bar{\mathcal{G}}$ are to be
\textit{Lie groups} (see, for example, \cite{ell}, \S\,7.1-7.2, 
pp.\,126-130). In addition, let the range $(x_a,x_b)$ of the sampling 
variate also be invariant under $\mathcal{G}$, i.e.
\begin{equation*}
 g_a(x_a) =
 \begin{cases}
 x_a \ \ \ ; \ \ \ g'_a(x) > 0 \\
 x_b \ \ \ ; \ \ \ g'_a(x) < 0 
 \end{cases}
 \ \  \text{and} \ \ \ \ 
 g_a(x_b) =
 \begin{cases}
 x_b \ \ \ ; \ \ \ g'_a(x) > 0 \\
 x_a \ \ \ ; \ \ \ g'_a(x) < 0 
 \end{cases}
\ .
\end{equation*}
Then the cdf for $y=g_a(x)$, given $\nu=\bar{g}_a(\theta)$, can be rewritten
as:
\begin{equation*}
\hskip -5mm
 F\bigl(g_a(x),\bar{g}_a(\theta)\bigr) =
 \begin{cases}
  \int_{g_a(x_a)}^{g_a(x)}\phi\bigl(y',\bar{g}_a(\theta)\bigr)\,dy'
 = \ \ \ \ F(x,\theta) \ \ \ \ ; \ g'_a(x) > 0 \\
  \int_{g_a(x_b)}^{g_a(x)}\phi\bigl(y',\bar{g}_a(\theta)\bigr)\,dy'
 = 1-F(x,\theta) \ ; \ g'_a(x) < 0 
 \end{cases}
 ,
\end{equation*}
which permits the following conclusions:
%
the cdf $F\bigl(g_a(x),\bar{g}_a(\theta)\bigr)$ is independent
          of the parameter $a$ of the transformations, i.e.
\begin{equation*}
 \frac{d}{da}F\bigl(g_a(x),\bar{g}_a(\theta)\bigr) = 0 \ ,
\end{equation*}
%
and the parameter $a$ enters $F\bigl(g_a(x),\bar{g}_a(\theta)\bigr)$
          only through $g_a(x)$ and $\bar{g}_a(\theta)$, i.e.
\begin{equation*}
 \frac{d}{da}F\bigl(g_a(x),\bar{g}_a(\theta)\bigr) = 
 F_1\bigl(g_a(x),\bar{g}_a(\theta)\bigr)\,\frac{\partial}{\partial a}g_a(x)+
 F_2\bigl(g_a(x),\bar{g}_a(\theta)\bigr)\,\frac{\partial}{\partial a}
 \bar{g}_a(\theta) \ ,
\end{equation*}
where $F_i$ denotes differentiation with respect to the $i$-th argument
of $F$ (we adhere to this notation throughout the present paper,
whatever the function and the arguments may be). 
%
%
Then, by combining the two conclusions and by setting $a=e$ we obtain:
\begin{equation}
F_1(x,\theta)\,k'(\theta) + F_2(x,\theta)\,h'(x) = 0 \ ,
\label{eq:invcdf1}
\end{equation}
where the derivatives $h'(x)$ and $k'(\theta)$ of functions 
$h(x)$ and $k(\theta)$ are defined as reciprocals of the corresponding 
infinitesimal operators of the Lie groups $\mathcal{G}$ and 
$\bar{\mathcal{G}}$:
\begin{equation}
 h'(x) \equiv \frac{dh(x)}{dx} \equiv 
         \biggl[\frac{\partial}{\partial a}g_a(x)\Bigl|_{a=e}\biggr]^{-1} \;\;
\label{eq:defh}
\end{equation}
and
\begin{equation}
 k'(\theta) \equiv \frac{dk(\theta)}{d\theta} \equiv 
         \biggl[\frac{\partial}{\partial a}
          \bar{g}_a(\theta)\Bigl|_{a=e}\biggr]^{-1} \ .
\label{eq:defk}
\end{equation}
By defining a function $G(x,\theta)$,
\begin{equation}
 G(x,\theta)\equiv h(x)-k(\theta) \ ,
\label{eq:defG}
\end{equation}
\eqref{eq:invcdf1} can be further reduced to
\begin{equation}
 F_1(x,\theta)\,G_2(x,\theta) - F_2(x,\theta)\,G_1(x,\theta) = 0 \ ,
\label{eq:eqF1G2}
\end{equation}
or to a functional determinant (see \cite{acz}, \S\,7.2.1, p.\,325),
\begin{equation*}
 \begin{vmatrix}
   F_1(x,\theta) & F_2(x,\theta) \\
   G_1(x,\theta) & G_2(x,\theta)
 \end{vmatrix}
= 0 \ .
\end{equation*}
With $G_1(x,\theta)=h'(x)$ being different from zero, $F_1(x,\theta)$ and 
$F_2(x,\theta)$ can be expressed as
\begin{equation}
 F_1(x,\theta) = \alpha(x,\theta)\,G_1(x,\theta)
\label{eq:F1G1}
\end{equation}
and
\begin{equation}
 F_2(x,\theta) = \beta(x,\theta)\,G_2(x,\theta) \ ,
\label{eq:F2G2}
\end{equation}
which, inserted in \eqref{eq:eqF1G2}, yield:
\begin{equation*}
 G_1(x,\theta)\,G_2(x,\theta)\,
 \bigl(\alpha(x,\theta)-\beta(x,\theta)\bigr) = 0 \ .
\end{equation*}
Since this is to be true for any $\theta$ and $x$, 
$\alpha(x,\theta)$ and $\beta(x,\theta)$ must be the same
functions:
\begin{equation*}
 \alpha(x,\theta)=\beta(x,\theta) \ .
\end{equation*}
Taking this into account, we multiply equations \eqref{eq:F1G1} 
and \eqref{eq:F2G2} by $dx$ and $d\theta$, respectively, so that their
sum reads:
\begin{equation*}
 dF(x,\theta)=\alpha(x,\theta)\,dG(x,\theta) \ ,
\end{equation*}
implying the distribution function $F(x,\theta)$ to be a function
of a single variable $G(x,\theta)$ \eqref{eq:defG},
\begin{equation}
 F(x,\theta) = \Phi\bigl(G(x,\theta)\bigr)= 
                 \Phi\bigl(h(x)- k(\theta)\bigr) \ .
\label{eq:cdfxth}
\end{equation}

By choosing
\begin{equation*}
 z\equiv h(x) \ \ \text{and} \ \ \mu\equiv k(\theta) \ ,
\end{equation*}
the cdf $F(x,\theta)$ of an invariant sampling distribution
thus reduces to
\begin{equation*}
 F(x,\theta) = \Phi(z-\mu) \ ,
\end{equation*}
which is a cdf of the variate $z$ and a location parameter $\mu$
(c.f. eq.\,\eqref{eq:cdfloc}). The above reasoning can be summarized as
\begin{theo}
A sampling distribution for a continuous variate $x$, 
given a continuous parameter $\theta$, with both its form and its domain 
being invariant under a Lie group $\mathcal{G}$, is necessarily reducible 
{\rm (by separate transformations $x\to z$ and $\theta\to\mu$)} to a 
sampling distribution for $z$ with the parameter $\mu$ being a location 
parameter. 
\end{theo}

For example, a pdf $f(t|\tau\mu_0 I)$ for $t$ \eqref{eq:sca}
with $\tau$ being a scale parameter and with the location parameter
$\mu_0$ being fixed to zero, is invariant under the 
group of scale transformations
\eqref{eq:invscale}. Then, according to \eqref{eq:defh} and
\eqref{eq:defk}, $h'(t)$ and $k'(\tau)$ read:
\begin{equation*}
h'(t) = \frac{1}{t} \; \; \text{and}\; \; k'(\tau)=\frac{1}{\tau} \ ,
\end{equation*}
which, in order to reduce the example to the problem inference
about a location parameter $\mu$, implies the appropriate 
transformations of the variate $t$ and the parameter $\tau$:
\begin{equation*}
 z=h(t)=\ln{t} \; \; \text{and} \; \; \mu=k(\tau)=\ln{\tau} \ .
\end{equation*}
Indeed, this is in perfect agreement with equation \eqref{eq:fpmz}
of Section\,\ref{sec:parameters}.

In the following two sections we will see that the invariance of
sampling distributions is indispensable when constructing a logically
consistent theory of inference about parameters.

\section{Consistency Theorem}
\label{sec:subjectivity}

Suppose that before we received the first evidence, $x_1$, we had
been completely ignorant about the value of the parameter $\theta$
that determines the probability distribution for $x$. 
We had only known 
the type of sampling distribution for $x$ and the permissible range
$(\theta_a,\theta_b)$ of the parameter.
Let the probability for $x$ taking the value $x_1$ in a discrete case,
or taking the value in the interval $(x_1,x_1+dx)$ in a continuous case, 
be denoted by $\Phi(x_1,\theta)$:
\begin{equation}
p(x_1|\theta I) = \Phi(x_1,\theta) \ .
\label{eq:lklx}
\end{equation}
In this section we prove the following proposition, henceforth referred 
to as the Consistency Theorem:
\begin{theo}
In order to meet the consistency Desideratum, the pdf 
for $\theta$ based on $x_1$ only,
must be directly proportional to the likelihood \eqref{eq:lklx}.
\end{theo}
{\bf Proof.} \;After having made the first observation, 
we know the type of sampling distribution and the values of $x_1$. 
Therefore, the pdf for $\theta$ given evidence 
$x_1$ will be proportional
to a function $\widetilde{\Phi}(x_1,\theta)$, 
\begin{equation}
f(\theta|x_1I) = 
\frac{\widetilde{\Phi}(x_1,\theta)}
{\zeta(x_1)} \ ,
\label{eq:pdft}
\end{equation}
whose form we would like to determine. 
The denominator $\zeta(x_1)$
is just a normalization constant due to \eqref{eq:nor2},
\begin{equation}
\zeta(x_1) =
\int\limits_\theta
\widetilde{\Phi}(x_1,\theta')\,d\theta' \ ,
\label{eq:norpdft}
\end{equation}
and contains no information about $\theta$.

Let now $x_2$ be another piece of evidence, independent of $x_1$, 
that we would 
like to include in our inference about $\theta$. Since $x_2$ is independent
of $x_1$, and subject to the same probability distribution \eqref{eq:lklx}
as $x_1$, its likelihood reads: 
\begin{equation*}
p(x_2|\theta x_1 I)=p(x_2|\theta I) = \Phi(x_2,\theta) \ .
\end{equation*}
In the first section we saw that the only consistent way of updating
pdf for $\theta$ is the one in accordance with 
Bayes' Theorem \eqref{eq:bayes1}.
With $f(\theta| x_1 I)$ 
taking the role of the prior pdf for $\theta$,
the pdf posterior to including $x_2$ into our reasoning about $\theta$
is written as:
\begin{equation}
f(\theta|x_1x_2I) 
= \frac{\widetilde{\Phi}(x_1,\theta)\,\Phi(x_2,\theta)}
       {\zeta(x_1,x_2)} \ ,
\label{eq:apbay1}
\end{equation}
with the normalization constant $\zeta(x_1,x_2)$ being
\begin{equation*}
\zeta(x_1,x_2)=
\int\limits_\theta
\widetilde{\Phi}(x_1,\theta')\,\Phi(x_2,\theta')\,d\theta' \ .
\end{equation*}

Nothing prevents us from reversing the order of taking the two pieces
of information, $x_1$ and $x_2$, into account, which results in the
following pdf for $\theta$:
\begin{equation}
f(\theta|x_2x_1I) = 
 \frac{\widetilde{\Phi}(x_2,\theta)\,\Phi(x_1,\theta)}
                         {\zeta(x_2,x_1)} \ ,
\label{eq:apbay2}
\end{equation}
with the appropriate normalization constant $\zeta(x_2,x_1)$,
\begin{equation*}
\zeta(x_2,x_1)
=\int\limits_\theta
\widetilde{\Phi}(x_2,\theta)\,\Phi(x_1,\theta)\,d\theta \ .
\end{equation*}
Moreover, the consistency Desideratum \textit{III.a} requires equality 
of the two results, \eqref{eq:apbay1} and \eqref{eq:apbay2}:
\begin{equation*}
f(\theta|x_1x_2I) = f(\theta|x_2x_1I) \ ,
\end{equation*}
or
\begin{equation}
\frac{\widetilde{\Phi}(x_1,\theta)\,\Phi(x_2,\theta)}
{\zeta(x_1,x_2)} =
\frac{\widetilde{\Phi}(x_2,\theta)\,\Phi(x_1,\theta)}
{\zeta(x_2,x_1)} \ .
\label{eq:cons2}
\end{equation}
The ratio of eq. \eqref{eq:cons2} and its derivative with respect to
$\theta$ yields
\begin{equation}
\frac{\widetilde{\Phi}'(x_1,\theta)}
     {\widetilde{\Phi}(x_1,\theta)} -
\frac{\Phi'(x_1,\theta)}{\Phi(x_1,\theta)} =
\frac{\widetilde{\Phi}'(x_2,\theta)}
     {\widetilde{\Phi}(x_2,\theta)} -
\frac{\Phi'(x_2,\theta)}{\Phi(x_2,\theta)} \ ,
\label{eq:difft1}
\end{equation}
where we use the notation
$$
\Phi'(x,\theta)\equiv \frac{d}{d\theta}\Phi(x,\theta) \ .
$$

Evidently, in order to ensure equality in \eqref{eq:difft1} for all
possible values of $x_1$ and $x_2$, the left and the right side of
the equation must be independent of $x_1$ and $x_2$, respectively,
but can depend on the value of the parameter $\theta$.
\vskip 2 true mm
\par{\noindent\narrower\small
Note that at this point, the two sides of equation 
\eqref{eq:difft1} can, in principle, also depend on the values
$x_a$, $x_b$, $\theta_a$ and $\theta_b$, determining the admissible
ranges of the sampling variate and of the parameter. However, in the 
following sections we will see that in all problems of parameter inference
that can be consistently solved, there is no such explicit dependence.
\par}
\vskip 2 true mm\noindent
Taking this dependence into account by introducing a function 
$h(\theta)$ we obtain
\begin{equation*}
\frac{d}{d\theta}\ln{\widetilde{\Phi}(x,\theta)}=
\frac{d}{d\theta}\ln{\Phi(x,\theta)}+h(\theta) \ ,
\end{equation*}
and, after integration of the latter,
\begin{equation}
\widetilde{\Phi}(x,\theta)=
k\,\pi(\theta)\,\Phi(x,\theta) \ ,
\label{eq:phitild}
\end{equation}
where $\pi(\theta)$ is a \textit{consistency factor},
\begin{equation}
\pi(\theta)\equiv \exp{\biggl\{\int h(\theta)\,d\theta\biggr\} } \ ,
\label{eq:consf}
\end{equation}
and $k$ an arbitrary integration constant. That is, \textit{the consistency
factor is determined only up to an arbitrary constant factor}.

The consistency factor is differentiable by construction.
From the form of 
$\pi(\theta)$ \eqref{eq:consf} it is also obvious that if $k$ is chosen
to be positive, $k\,\pi(\theta)$ is positive for every
$\theta$ defined. Consequently, the normalization factor
$\zeta(x)$, 
\begin{equation*}
\zeta(x) = 
k\int\limits_\theta
\pi(\theta')\,\Phi(x,\theta')\,d\theta' \ ,
\end{equation*}
being an integral of a product of positive factors \eqref{eq:phitild},
is also a strictly positive quantity for every $x$ defined.

By inserting the solution \eqref{eq:phitild} into \eqref{eq:pdft},
we can finally write:
\begin{equation}
f(\theta|xI) = \frac{k\,\pi(\theta)}
{\zeta(x)}\Phi(x,\theta)
               = \frac{k\,\pi(\theta)}
                 {\zeta(x)}p(x|\theta I)
\label{eq:bayprime}
\end{equation}
which completes the proof of the Consistency Theorem.

\vskip 3mm \noindent
The result is valid for $p(x|\theta I)$
being the likelihood of either a discrete or dense variable $x$.
For the latter, the Consistency Theorem can be rewritten by replacing
the likelihood with the appropriate \textit{likelihood density},
$f(x|\theta I)$,
\begin{equation}
f(\theta|xI)= \frac{k\,\pi(\theta)}
                 {\zeta(x)}\,p(x|\theta I) 
= \frac{k\,\pi(\theta)}
{k\,\eta(x)}\,f(x|\theta I) 
= \frac{\pi(\theta)}
{\eta(x)}\,f(x|\theta I) \ ,
\label{eq:bayprime1}
\end{equation}
where $\eta(x)$ is the corresponding normalization
factor,
\begin{equation} 
 \eta(x)\equiv \int\limits_\theta
 \pi(\theta')\,f(x_1|\theta' I)\,d\theta' \ .
\label{eq:normf}
\end{equation}
Evidently, the normalization constant is also determined 
up to a constant factor $k$, i.e., \textit{the consistency factor
$\pi(\theta)$ and the normalization factor $\eta(x)$ are
determined up to the same factor}.

Similarly, in terms of posterior probabilities and likelihoods
instead of the corresponding densities, the Theorem
reads:
\begin{equation}
p(\theta|xI)= \frac{k\,\pi(\theta)\,d\theta}{k\,\eta(x)\,dx}\,p(x|\theta I) 
            = \frac{\pi(\theta)\,d\theta}{\eta(x)\,dx}\,p(x|\theta I) \ .
\label{eq:bayprime2}
\end{equation}

The form of the Consistency Theorem \eqref{eq:bayprime2}
reminds very much of that of Bayes' Theorem \eqref{eq:bayes}.
In both cases, within a specified model, the complete information
about the inferred parameter $\theta$ of the model that can be extracted
from a measurement $x$, is contained in the value
of the appropriate likelihood, $p(x|\theta I)$.  
But there is also a fundamental and
\textit{very important difference} between the two Theorems: while
$f(\theta|I)$ in Bayes' Theorem represents the pdf for $\theta$
prior to including evidence $x$ in our inference about $\theta$,
the consistency factor $\pi(\theta)$ in the 
Consistency Theorem
is just a proportionality coefficient between the pdf for $\theta$
and the likelihood function. \textit{The form of the factor depends on 
the only relevant information $I$ that we possess before
the first datum $x_1$ is collected, i.e. it depends on the
specified sampling model}.

In Section\,\ref{sec:history} we comment on
how overlooking this difference led to a long-lasting confusion
in plausible reasoning. Before that we show under what conditions
the factors $\pi(\theta)$ can be consistently
determined, and uniquely determine them for such cases by following
the basic Desiderata.

\section{Objective inference and equivalence of information}
\label{sec:objectivity}

According to the definition of probability adopted in the
first section, every assigned probability is necessarily
\textit{subjective}: no probability distribution can be assigned
independently of the experience of the person who is expressing
his or her degree of belief. In the words of Bruno de\,Finetti 
(\cite{dfi}, Preface, p.\,x): ``Probability does not exist'', meaning
that there is no probability \textit{per se}. For example, when there 
is not enough relevant information at our disposal, we are simply not 
in a position to make
any probabilistic inferences. That is, even when lacking, information
should never be confused with our hopes, fears, value judgments, etc. 
Since no matter how carefully these are specified, they still represent 
mere personal biases, prejudices and speculations.

The general Desiderata represent the rules that we have to obey in
order to preserve consistency of inference, so it is evident
that the adjective \textit{subjective} does \textit{not} stand
for \textit{arbitrary}. In fact, in accordance with Desideratum
\textit{III.c}, our goal is that inferences are 
to be completely \textit{objective} in the sense that
\textit{if in two problems the state of knowledge of a person making the 
inference is the same, then he or she must assign the same probabilities 
in both}. The goal of the present and the following sections is
to show when and how information $I$ about the specified model
and its domain, within the framework of our basic Desiderata
(i.e without any additional \textit{ad hoc} assumptions), uniquely determine
the form of consistency factors, the latter being indispensable
at the starting point of any parameter inference. Within the Desiderata,
only \textit{III.a} and \textit{III.c} directly consider \textit{equalities}
of probabilities and can thus provide equations that could determine
the form of consistency factors. Since the requirements of 
Desideratum\;\textit{III.a} were extensively exploited already
throughout the previous section when the Consistency Theorem was derived,
it is only \textit{III.c} that is left at our disposal to
obtain the desired functional equation for $\pi(\theta)$.

Let 
\begin{equation*}
 f(x|\theta I) = \phi(x,\theta)
\end{equation*}
be a sampling pdf for $x$ whose parameter $\theta$ we would like to infer.
We saw in the preceding section that in the case when this can be done 
in a consistent way, the pdf for $\theta$ must take the form:
\begin{equation}
 f(\theta|xI) = \frac{\pi(\theta)}{\eta(x)}\,f(x|\theta I) 
              = \frac{\pi(\theta)}{\eta(x)}\,\phi(x,\theta ) \ ,
\label{eq:constr1}
\end{equation}
where $\eta(x)$ is the usual normalization factor
\begin{equation}
 \eta(x)= \int\limits_\theta \pi(\theta')\,f(x|\theta' I)\,d\theta' 
        = \int\limits_\theta \pi(\theta')\,\phi(x,\theta')\,d\theta' \ . 
\label{eq:norf1}
\end{equation}
Equation \eqref{eq:constr1} with the unknown pdf $f(\theta|xI)$
clearly does \textit{not} determine uniquely the form 
of the consistency factors: as long as the normalization integral 
\eqref{eq:norf1} exists, $\pi(\theta)$ can be any positive and differentiable
function of $\theta$. Additional constraints (functional equations)
are therefore needed to reduce all these functions to a single
consistent function, i.e. to the only function that is consistent
with our Desiderata. 

In the case there exists a group $\mathcal{G}$ of transformations $g_a$ 
of the sample space such that $y=g_a(x)$, the above pdf for $\theta$
can be expressed as  
\begin{equation*}
 f\bigl(\theta|g_a(x)I\bigr) = \frac{\bigl|g_a'(x)\bigr|}{\eta(x)}
                \,\pi(\theta)\,\phi(x,\theta)\,\bigl|g_a'(x)\bigr|^{-1} 
             = \frac{\bigl|g_a'(x)\bigr|}{\eta(x)}
               \,\pi(\theta)\,f(y|\theta I) \ ,
\end{equation*}
where
\begin{equation*}
 f\bigl(y|\theta I'\bigr)\equiv
 \phi(x,\theta)\,\bigl|g_a'(x)\bigr|^{-1} \ .
\end{equation*}

Let there also exist a group $\bar{\mathcal{G}}$ of transformations 
$\bar{g}_a$ of the parameter space, $\nu=\bar{g}_a(\theta)$. 
We saw already in Section\,\ref{sec:intro}
that, according to objectivity Desideratum\,\textit{III.c}, the assigned
probabilities must be invariant under variate transformations.
This is assured if the pdf's of the original and the transformed
variate, $\theta$ and $\nu$, are related \textit{via} \eqref{eq:vartr1} 
and \eqref{eq:vartr2}, so that the pdf for $\nu$, given a measured $x$, reads:
\begin{equation*}
 f(\nu|xI') = 
 f(\theta|xI)\,\bigl|\bar{g}_a'(\theta)\bigr|^{-1} \ .
\end{equation*}
That is, in order to assign equal probabilities in states of
equal knowledge, the pdf for $\bar{g}_a(\theta)$ must take the form:
\begin{equation}
\begin{split}
 f\bigl(\bar{g}_a(\theta)|x(y)I'\bigr)&=
 \frac{\bigl|g_a'(x)\bigr|}{\eta(x)}\,
 \frac{\pi(\theta)}{\bigl|\bar{g}_a'(\theta)\bigr|}
                \,\phi(x,\theta)\,\bigl|g_a'(x)\bigr|^{-1} \\
 &=\frac{\bigl|g_a'(x)\bigr|}{\eta(x)}\,
 \frac{\pi(\theta)}{\bigl|\bar{g}_a'(\theta)\bigr|}\,
               \widetilde{\phi}\bigl(g_a(x),\bar{g}_a(\theta)\bigr) \\ 
 &=\frac{\widetilde{\pi}\bigl(\bar{g}_a(\theta)\bigr)}
        {\widetilde{\eta}\bigl(g_a(x)\bigr)}\,
       \widetilde{\phi}\bigl(g_a(x),\bar{g}_a(\theta)\bigr) \ ,
\end{split}
\label{eq:constr2}
\end{equation}
where:
\begin{equation*}
 f(y|\nu I') =
 \widetilde{\phi}\bigl(g_a(x),\bar{g}_a(\theta)\bigr)\equiv
 \phi(x,\theta)\,\bigl|g_a'(x)\bigr|^{-1} \ ,
\end{equation*}
\begin{equation}
 \widetilde{\pi}\bigl(\bar{g}_a(\theta)\bigr) \equiv
 k(a)\,\frac{\pi(\theta)}{\bigl|\bar{g}_a'(\theta)\bigr|} 
\label{eq:pitilde}
\end{equation}
and, for $\bar{g}_a'(\theta)>0$,
\begin{equation}
 \widetilde{\eta}\bigl(g_a(x)\bigr) \equiv
 k(a)\,\frac{\eta(x)}{\bigl|g_a'(x)\bigr|} =
 \int_{\bar{g}_a(\theta_a)}^{\bar{g}_a(\theta_b)}
 \widetilde{\pi}\bigl(\bar{g}_a(\theta')\bigr)\,
 f\bigl(g_a(x)|\bar{g}_a(\theta') I\bigr)\,d\bigl(\bar{g}(\theta')\bigr) \ ,
\label{eq:etatilde}
\end{equation}
while for $\bar{g}_a'(\theta)<0$ the limits of the above integral are to be
interchanged\footnote{Index $a$ in $g_a$ and $\bar{g}_a$ denotes
particular elements of transformation groups, while in $x_a$ and $\theta_a$
it indicates the lower bounds of the sample and parameter space, 
respectively.}. Note that in general the value of the multiplication 
constant $k$, up to which the 
consistency and the normalization factors, \eqref{eq:consf} and
\eqref{eq:normf}, are to be uniquely determined 
(recall the preceding section), may depend on the value of the
transformation parameter $a$. 

Equation \eqref{eq:constr2} represents a constraint
on consistency factors that is additional to \eqref{eq:constr1},
but it also introduces an additional unknown
variable, function $\widetilde{\pi}\bigl(\bar{g}_a(\theta)\bigr)$.
For invariant sampling distributions, however, it is easy to demonstrate
that the form of the consistency factor must also be invariant
under the induced group $\bar{\mathcal G}$, i.e. that $\pi$
and $\widetilde{\pi}$ must be the same functions. Namely,
\textit{the forms of consistency factors $\pi(\theta)$ and 
$\widetilde{\pi}\bigl(\bar{g}_a(\theta)\bigr)$ depend on information
$I$ and $I'$ that we possess about $\theta$ and $\bar{g}_a(\theta)$,
respectively, prior to collecting datum $x$: on the forms and domains of
sampling distributions, $\phi(x,\theta)$ and 
$\widetilde{\phi}\bigl(g_a(x),\bar{g}_a(\theta)\bigr)$. 
In the case where all of these are invariant under
particular transformations $g_a$ and $\bar{g}_a$:
\begin{description}
\item[a)] 
\begin{equation*}
 \widetilde{\phi}\bigl(g_a(x),\bar{g}_a(\theta)\bigr)=
 \phi\bigl(g_a(x),\bar{g}_a(\theta)\bigr) \ ,
\end{equation*}
\item[b)]
\begin{equation*}
 g_a(x_a) =
 \begin{cases}
 x_a \ \ \ ; \ \ \ g'_a(x) > 0 \\
 x_b \ \ \ ; \ \ \ g'_a(x) < 0 
 \end{cases}
 \ \  , \ \ \ \ 
 g_a(x_b) =
 \begin{cases}
 x_b \ \ \ ; \ \ \ g'_a(x) > 0 \\
 x_a \ \ \ ; \ \ \ g'_a(x) < 0 
 \end{cases}
\ ,
\end{equation*}
and
\item[c)]
\begin{equation*}
 \bar{g}_a(\theta_a) =
 \begin{cases}
 \theta_a \ \ \ ; \ \ \ \bar{g}'_a(\theta) > 0 \\
 \theta_b \ \ \ ; \ \ \ \bar{g}'_a(\theta) < 0 
 \end{cases}
 \ \  , \ \ \ \ 
 \bar{g}_a(\theta_b) =
 \begin{cases}
 \theta_b \ \ \ ; \ \ \ \bar{g}'_a(\theta) > 0 \\
 \theta_a \ \ \ ; \ \ \ \bar{g}'_a(\theta) < 0 
 \end{cases}
\ ,
\end{equation*}
\end{description}
the information $I$ equals information $I'$ and the 
two consistency factors, $\pi(\theta)$ and 
$\widetilde{\pi}\bigl(\bar{g}_a(\theta)\bigr)$, must be 
the same functions:
\begin{equation}
 \widetilde{\pi}\bigl(\bar{g}_a(\theta)\bigr) = 
 \pi\bigl(\bar{g}_a(\theta)\bigr) \ .
 \label{eq:invconsf}
\end{equation}
}
When combined with \eqref{eq:pitilde}, this implies:
\begin{equation}
 k(a)\,\pi(\theta) = 
 \pi\bigl(\bar{g}_a(\theta)\bigr)\,\bigl|\bar{g}_a'(\theta)\bigr| \ .
\label{eq:funeqcon}
\end{equation}
The above functional equation for $\pi(\theta)$ is the cornerstone
of the entire theory of consistent \textit{assignment} of probabilities
to parameters of sampling distributions: 
\begin{conj}
 Equation \eqref{eq:funeqcon} is the only functional equation  
 within the basic Desiderata that can be used for determination of 
 consistency factors.
\end{conj}
The proof of this conjecture represents an open problem that is still
to be solved. It is a serious problem, though, since 
any additional functional equations for $\pi(\theta)$, 
independent of \eqref{eq:funeqcon}, with solutions different from solutions
of \eqref{eq:funeqcon}, would most seriously jeopardize the
consistency of the entire probabilistic approach 
to inferences about the parameters of sampling distributions.

Be that as it may, equation \eqref{eq:funeqcon} is the only functional 
equation \textit{that we know of} that can be used for determination of 
consistency factors if we want to rely exclusively on our
basic Desiderata. All other procedures for
determination of $\pi(\theta)$ (of non-informative prior
'probability' distributions; see Section\,\ref{sec:history})
\textit{that we are aware of} involve applications of
some additional \textit{ad hoc} assumptions so that there is \textit{absolutely
no guarantee} that reasonings of such a kind be consistent.
Further arguments and examples, supporting the above conjecture
by exhibiting the decisive role of the invariance of sampling
distributions under group transformations in the process of
determination of the consistency factors, are presented
in Sections\;\ref{sec:unique}, \ref{sec:calibration}, \ref{sec:counting}
and \ref{sec:history}, and in Appendix\;\ref{sec:dsz}.
 
In case of a two-parametric induced group $\bar{\mathcal{G}}$ 
of parameter transformations $\bar{g}_{a,b}$, the functional equation 
for the consistency factor $\pi(\theta^{(1)},\theta^{(2)})$ for the
inferred parameters $\theta^{(1)}$ and $\theta^{(2)}$ reads:
\begin{equation}
 k(a,b)\,\pi(\theta^{(1)},\theta^{(2)}) = 
 \pi\bigl(\bar{g}_{a,b}(\theta^{(1)}),\bar{g}_{a,b}(\theta^{(2)})\bigr)
\,|J|^{-1} \ ,
\label{eq:funeqcon1}
\end{equation}
where $J$ stands for the appropriate Jacobian
\begin{equation*}
 J\equiv \frac{\partial\bigl(\bar{g}_{a,b}(\theta^{(1)}),
                             \bar{g}_{a,b}(\theta^{(2)})\bigr)}
              {\partial\bigl(\theta^{(1)},\theta^{(2)}\bigr)} \ .
\end{equation*}
We will come across functional equation \eqref{eq:funeqcon1} in 
Section\,\ref{sec:locscale}, during a simultaneous inference about a location
and a scale parameter.

When equation \eqref{eq:invconsf} holds the normalization factor
$\widetilde{\eta}\bigl(g_a(x)\bigr)$ equals $\eta\bigl(g_a(x)\bigr)$,
\begin{equation*}
 \widetilde{\eta}\bigl(g_a(x)\bigr)=
 \int_{\theta_a}^{\theta_b}\pi\bigl(\bar{g}_a(\theta')\bigr)\,
 \phi\bigl(g_a(x),\bar{g}_a(\theta')\bigr)\,d\bigl(\bar{g}_a(\theta')\bigr) 
 =\eta\bigl(g_a(x)\bigr) \ ,
\end{equation*}
so that
\begin{equation}
 k(a)\,\eta(x) = 
 \eta\bigl(g_a(x)\bigr)\,\bigl|g_a'(x)\bigr| \ .
\label{eq:funeqnor}
\end{equation}

Under what circumstances does a unique solution of the 
functional equation \eqref{eq:funeqcon} exist?
Let us consider the problem with a sampling distribution
being invariant under a discrete group of transformations
\begin{equation}
\begin{CD}
\begin{split}
 g_a&:\ x \ \negthickspace\negthickspace @>>>\; g_a(x) = ax  \ , \\
 \bar{g}_a&:\ \theta\, \negthickspace\negthickspace @>>>\; 
 \bar{g}_a(\theta) = a\theta \ ,
\end{split}
\end{CD}
\label{eq:invparity}
\end{equation}
where $a$ can only take two values,
\begin{equation*}
 a=\{1,-1\}
\end{equation*}
for both groups, $\mathcal{G}$ and $\bar{\mathcal{G}}$. 
That is, the considered distribution possesses parity under 
simultaneous inversion of sampling and parameter space coordinates.
Then, for $a=-1$, functional equation \eqref{eq:funeqcon} reads:
\begin{equation*}
 \pi(-\theta)=k(a=-1)\,\pi(\theta) \ ,
\end{equation*}
or, after an inversion $\theta \longleftrightarrow (-\theta)$,
\begin{equation*}
 \pi(\theta)=k(a=-1)\,\pi(-\theta) \ .
\end{equation*}
Multiplying the two equations yields:
\begin{equation*}
 k^2(a=-1) = 1 \ ,
\end{equation*}
which, when the convention about $\pi$ being positive is invoked, further
implies
\begin{equation}
 \pi(-\theta)=\pi(\theta) \ .
\label{eq:pipar}
\end{equation}
That is, the consistency factor that corresponds to a sampling distribution
being invariant under simultaneous inversions of sampling and parameter 
space coordinates, must itself possess parity under inversion of
parameter space coordinates. But apart from this it can take any
form and so in this case the solution of \eqref{eq:funeqcon}
is clearly \textit{not} unique. It is not difficult to understand that
this is a common feature of all solutions based on invariance
of the sampling distributions under \textit{discrete} groups. 
If the symmetry group is discrete, the sample and the parameter
spaces break up in intervals with no connections in terms
of group transformations within the points of the same interval.
We are then free to choose the form of $\pi(\theta)$ in one
of these intervals (e.g. we can choose $\pi(\theta)$ for the
positive values of $\theta$ in the above example), so it is
evident that \textit{it is impossible to determine uniquely 
the form of consistency factors for problems that are invariant 
only under discrete groups of transformations}.

It turns out, however, that \textit{for sampling distributions that 
are invariant under Lie groups, functional equation \eqref{eq:funeqcon} 
uniquely determines the form of the corresponding consistency factors}.
But according to the Existence Theorem of Section\,\ref{sec:invariance},
the invariance of a sampling distribution under a Lie group 
is found only when the parameter of the distribution is reducible to 
a location parameter by 
one-to-one transformations of both the parameter and the sampling variate.
It is therefore sufficient to determine the form
of consistency factors for location parameters, which is accomplished in the
following three sections.
\section{Location parameters}
\label{sec:location}

We saw in Section\,\ref{sec:invariance} that a sampling distribution, 
parameterized by a location parameter $\mu$ and by a fixed dispersion
parameter $\sigma_0$, is invariant under the
group of translations \eqref{eq:invloc}. In such a case the functional
equation \eqref{eq:funeqcon} for $\pi(\mu)$ reads:
\begin{equation}
 \pi(\mu+b) = k(b)\,\pi(\mu) \ .
\label{eq:pil1}
\end{equation}
After its differentiation with respect to $b$,
\begin{equation*}
 \pi'(\mu+b) = k'(b)\,\pi(\mu) \ ,
\end{equation*}
we set $b = 0$ and obtain a simple differential equation with
separable variables $\pi$ and $\mu$,
\begin{equation}
\frac{d\pi}{\pi}=-q\,d\mu \ ,
\label{eq:pil2}
\end{equation}
with the constant $q$ being defined as 
\begin{equation*}
 q\equiv - k'(0) \ .
\end{equation*}
The general solution of \eqref{eq:pil2} reads
\begin{equation*}
 \pi(\mu) = C_{\pi}\,
  \exp\bigl\{-q\,\mu\bigr\} \ ,
\end{equation*}
where $C_{\pi}$ is an integration constant.
Since all multiplication constants can be put  into $\eta$, we can 
assume without any loss of generality that $C_{\pi} = 1$, 
obtaining in this way the general form of the consistency factor for
location parameters:
\begin{equation}
  \pi(\mu) = \exp{\bigl\{-q\,\mu\bigr\}} \ .
\label{eq:pimu}
\end{equation}
\vskip 2 true mm
\par{\noindent\narrower\small
For sampling distributions, symmetric under simultaneous inversions of 
the sampling and the parameter space, equation \eqref{eq:pipar} implies
$q=0$, i.e. implies uniform  consistency factors for 
location parameters. By invoking the symmetry of the problems of 
simultaneous inference about a location and a scale parameter 
in Section\,\ref{sec:locscale}, we show that further applications of
the basic Desiderata and their direct implications also require
$q=0$ in the case of problems without space-inversion symmetry.
\par}
\vskip 2 true mm\noindent
Based on a measured value $x_1$, the pdf for a location parameter $\mu$ 
therefore reads:
\begin{equation}
f(\mu|x_1\sigma_0 I)=\frac{e^{-q\mu}}
{\eta(x_1)}\,f(x_1|\mu\sigma_0I) \\
=\frac{e^{-q\mu}}{\eta(x_1)}\,\frac{1}{\sigma_0}
                     \phi\Bigl(\frac{x_1-\mu}{\sigma_0}\Bigr) \ .
\label{eq:givs}
\end{equation}
Now, as an example, we want to update our inference about the 
parameter $\mu$ by including additional information $x_2$ in our 
inference, where $x_2$ is a result of a measurement of $x$ that
is also subject to the same sampling distribution and 
independent of $x_1$. We can write the likelihood density for $x_2$,
\begin{equation}
f(x_2|\mu\sigma_0x_1I) = f(x_2|\mu\sigma_0 I)=\frac{1}{\sigma_0}
                     \phi\Bigl(\frac{x_2-\mu}{\sigma_0}\Bigr) \ ,
\label{eq:indpx}
\end{equation}
and the updated pdf for $\mu$,
\begin{equation}
\begin{split}
f(\mu|x_1x_2\sigma_0I) &= 
\frac{f(\mu|x_1\sigma_0I)\,f(x_2|\mu\sigma_0 I)}
{\eta(x_1,x_2)} \\
&=\frac{\pi(\mu)}{\eta(x_1,x_2)}\,
f(x_1|\mu\sigma_0 I)\,f(x_2|\mu\sigma_0 I) \\
&=\frac{e^{-q\mu}}{\eta(x_1,x_2)}\,\frac{1}{\sigma_0^2}\,
                        \phi\Bigl(\frac{x_1-\mu}{\sigma_0}\Bigr)\,
                        \phi\Bigl(\frac{x_2-\mu}{\sigma_0}\Bigr) \\
&=\frac{e^{-q\mu}}{\eta(x_1,x_2)}
\,f(x_1x_2|\mu\sigma_0 I) \ ,
\end{split}
\label{eq:mugivsigma1}
\end{equation}
with the appropriate normalization constant, $\eta(x_1,x_2)$,
being:
\begin{equation}
\eta(x_1,x_2) = 
\int_{-\infty}^{\infty}e^{-q\mu'}\,f(x_1x_2|\mu'\sigma_0 I)\,d\mu' \ .
\label{eq:etax1x2k}
\end{equation}
The update \eqref{eq:mugivsigma1} is made in accordance with Bayes' 
theorem \eqref{eq:bayes1} with the purpose of ensuring our reasoning  be
consistent. 

The product of the likelihood densities $f(x_1|\mu\sigma_0 I)$
and $f(x_2|\mu\sigma_0 I)$ in \eqref{eq:mugivsigma1} is equal to 
the combined likelihood density for the two independent events, $x_1$
and $x_2$, due to the product rule \eqref{eq:product1}. 
According to \eqref{eq:pdfxbars},
the likelihood density $f(x_1x_2|\mu\sigma_0 I)$ can equivalently
be represented by the density for $\bar{x}$ and $s$,
$f(\bar{x}s|\mu\sigma_0 I)$, where
\begin{equation}
 \bar{x} \equiv \frac{x_1 + x_2}{2} \ , \ 
s \equiv \frac{|x_1 - x_2|}{2} \ .
\label{eq:defxbars1}
\end{equation}
Written in terms of $f(\mu|\sigma_0 \bar{x}sI)$
the pdf for $\mu$ \eqref{eq:mugivsigma1} thus reads: 
\begin{equation}
\begin{split}
&f(\mu|\sigma_0 \bar{x}sI) = 
\frac{e^{-q\mu}}{\eta(\bar{x},s)}
\,f(\bar{x}s|\mu\sigma_0 I)= \\
&\frac{e^{-q\mu}}{\eta(\bar{x},s)}
\,\frac{4}{\sigma_0^2}\,
 \phi\Bigl(\frac{\bar{x}-\mu}{\sigma_0}+\frac{s}{\sigma_0}\Bigr)\,
 \phi\Bigl(\frac{\bar{x}-\mu}{\sigma_0}-\frac{s}{\sigma_0}\Bigr) \ ,
\end{split}
\label{eq:mugivsigma2}
\end{equation}
with the appropriate normalization constant 
$\eta(\bar{x},s)$,
\begin{equation}
\eta(\bar{x},s) = 
\int_{-\infty}^{\infty}e^{-q\mu'}
\,f(\bar{x}s|\mu'\sigma_0 I)\,d\mu' \ .
\label{eq:etaxbarsk}
\end{equation}

The findings of the present example
will become of particular importance in Section\,\ref{sec:locscale}
where we determine the form of the consistency factor
$\pi(\mu,\sigma)$ for simultaneous estimation of a location and a 
dispersion parameter.

\section{Inference about scale and dispersion parameters}
\label{sec:scale}

When, contrary to the preceding section, the inferred dispersion 
(or scale) parameter is unknown and the location
parameter is fixed to $\mu=\mu_0$, the problem is invariant under the group
\eqref{eq:invscale} of scale transformations. In such a case the functional
equation \eqref{eq:funeqcon} for $\pi(\sigma)$ reads:
\begin{equation}
 \pi(a\sigma)=h(a)\,\pi(\sigma) \ ,
\label{eq:funeqsca}
\end{equation}
where
\begin{equation*}
 h(a)\equiv \frac{k(a)}{a} \ .
\end{equation*}
Equation \eqref{eq:funeqsca} determines the form of the consistency
factor for dispersion and scale parameters to be
\begin{equation}
 \pi(\sigma)= \sigma^{-r} 
\label{eq:cfsig}
\end{equation}
and
\begin{equation}
 \pi(\tau)= \sigma^{-r} \ ,
\label{eq:cftau}
\end{equation}
where the value of the constant $r$,
\begin{equation*}
 r\equiv -h'(1) \ ,
\end{equation*} 
is yet to be determined in Section\;\ref{sec:locscale}.

We stressed in Section\,\ref{sec:parameters} (c.f. 
equation\,\eqref{eq:fpmz}) that an assignment of a pdf 
to a scale parameter $\tau$ (or, equivalently, to a dispersion parameter
$\sigma$ of a symmetric distribution) can 
be reduced to an assignment of a pdf, $f(\nu|z\lambda_0I)$, 
to a location parameter, $\nu$:
\begin{equation*}
f(\nu|z\lambda_0I) 
= \frac{\widetilde{\pi}(\nu)}{\widetilde{\eta}(z)}
f(z|\nu\lambda_0I) 
\end{equation*}
with
\begin{equation*}
z \equiv \ln{t} \ \ \ \ \text{and} \ \ \ \ \ \nu \equiv \ln\tau \ \ , 
\end{equation*}
and with $\lambda_0$ being a fixed dispersion parameter.
According to the findings of the previous section (see eq.\,\eqref{eq:pimu}), 
we can immediately write the appropriate consistency factor: 
\begin{equation*}
\widetilde{\pi}(\nu)= e^{-q\nu} \ .
\end{equation*}
Making use of eq.\,\eqref{eq:vartr2}, the pdf for $\nu$ can be transformed
into the pdf for $\tau$:
\begin{equation}
f(\tau|t\mu_0I) = f(\nu|z\lambda_0I)\,
\Bigl|\frac{d\nu}{d\tau}\Bigr| \ ,
\label{eq:fs}
\end{equation}
where
\begin{equation*}
\Bigl|\frac{d\nu}{d\tau}\Bigr| = e^{-\nu} 
= \tau^{-1} \ .
\end{equation*}
On the other hand, in order to make a consistent inference, the  
pdf for $\tau$, $f(\tau|t\mu_0I)$, 
based on $t$ only, must be proportional to the likelihood density 
$f(t|\tau\mu_0 I)$ (see eq. \eqref{eq:bayprime1}):
\begin{equation}
f(\tau|t\mu_0I) = 
\frac{\pi(\tau)}{\eta(t)}f(t|\tau\mu_0 I) \ .
\label{eq:fs1}
\end{equation}
Then, due to Desideratum \textit{III.a}, the two pdf's,
\eqref{eq:fs} and \eqref{eq:fs1}, must be equal, which implies the form
of the consistency factors for scale parameters,
\begin{equation}
\pi(\tau)= \tau^{-(q+1)} \ ,
\label{eq:cftau1}
\end{equation}
as well as for dispersion parameters,
\begin{equation}
\pi(\sigma)= \sigma^{-(q+1)} \ .
\label{eq:cfsig1}
\end{equation}
The same Desideratum implies equality of the factors \eqref{eq:cfsig}
and \eqref{eq:cfsig1}, as well as equality of the factors 
\eqref{eq:cftau} and \eqref{eq:cftau1}, i.e. implies the relation
\begin{equation}
r = q + 1 
\label{eq:qr}
\end{equation}
between the parameters $q$ and $r$ of the consistency factors of the
location and scale parameters. Evidently, if $q$ is determined to 
be zero, this would immediately imply $r=1$.

In the limit of complete prior ignorance about its value,
the pdf for $\sigma$ given a measured value
$x_1$ and the fixed value of $\mu=\mu_0$ therefore reads:
\begin{equation*}
f(\sigma|\mu_0x_1I)=
\frac{\pi(\sigma)}
     {\eta(x_1)}\,f(x_1|\mu_0\sigma I) \ .
\end{equation*}
Following the steps of the example of the preceding section,
we update the pdf for $\sigma$ by including result $x_2$ of an additional
measurement in our inference. The updated value of the pdf reads:
\begin{equation*}
\begin{split}
f(\sigma|\mu_0 x_1x_2I) &= 
\frac{\pi(\sigma)}{\eta(x_1,x_2)}
\,f(x_1x_2|\mu_0\sigma I) \\
&= \frac{\sigma^{-r}}{\eta(x_1,x_2)}
f(x_1x_2|\mu_0\sigma I) \\
&= \frac{\sigma^{-r}}{\eta(x_1,x_2)}
\,\frac{1}{\sigma^2}\,
                        \phi\Bigl(\frac{x_1-\mu_0}{\sigma}\Bigr)\,
                        \phi\Bigl(\frac{x_2-\mu_0}{\sigma}\Bigr)\ ,
\end{split}
\end{equation*}
with the normalization constant $\eta(x_1,x_2)$,
\begin{equation}
\eta(x_1,x_2) = \int_{0}^{\infty}
(\sigma')^{-r}\,f(x_1x_2|\mu_0\sigma' I)\,d\sigma' \ .
\label{eq:etax1x2q}
\end{equation}
In an analogy with \eqref{eq:mugivsigma2} and \eqref{eq:etaxbarsk},
both the updated pdf for $\sigma$ 
and the corresponding normalization constant can be expressed in terms of
$\bar{x}$ and $s$ \eqref{eq:defxbars1}, instead of $x_1$ and $x_2$:
\begin{equation}
\begin{split}
&f(\sigma|\mu_0 \bar{x}sI) = 
\frac{\sigma^{-r}}{\eta(\bar{x},s)}
\,f(\bar{x}s|\mu_0\sigma I)= \\
&\frac{\sigma^{-r}}{\eta(\bar{x},s)}
\,\frac{4}{\sigma^2}\,
 \phi\Bigl(\frac{\bar{x}-\mu_0}{\sigma}+\frac{s}{\sigma}\Bigr)\,
 \phi\Bigl(\frac{\bar{x}-\mu_0}{\sigma}-\frac{s}{\sigma}\Bigr) \ ,
\end{split}
\label{eq:sigmagivmu2}
\end{equation}
and
\begin{equation}
\eta(\bar{x},s) = 
\int_{0}^{\infty}(\sigma')^{-r}
\,f(\bar{x}s|\mu_0\sigma' I)\,d\sigma' \ .
\label{eq:etaxbarsq}
\end{equation}

\section{Simultaneous inference about a location and a dispersion parameter}
\label{sec:locscale}

By fixing neither the location nor the dispersion parameter,
an inference about the two parameters is invariant
under a simultaneous location and scale transformation
\eqref{eq:invlocscale}. The symmetry of the problem implies
the following form of the functional equation \eqref{eq:funeqcon1} for 
the appropriate consistency and normalization factors:
\begin{equation}
 \pi(a\mu+b,a\sigma) = h(a,b)\,\pi(\mu,\sigma) \ , 
\label{eq:ratcfmusig}
\end{equation}
where
\begin{equation*}
 h(a,b)\equiv k(a,b)\,\biggl|\frac{\partial(a\mu+b,a\sigma)}
                                 {\partial(\mu,\sigma)}\biggr|^{-1} =
       \frac{k(a,b)}{a^2} \ .
\end{equation*}

In order to solve it, we differentiate equation \eqref{eq:ratcfmusig}
separately with respect to $a$ and $b$, set afterward $a=1$ and $b=0$,
and obtain:
\begin{eqnarray}
\mu\,\pi_1(\mu,\sigma)+
\sigma\,\pi_2(\mu,\sigma)\hskip -1mm &=& \hskip -1mm -\tilde{r}\,\pi(\mu,\sigma) \ ,
\label{eq:ratcfmusig1} \\
\pi_1(\mu,\sigma)\hskip -1mm &=& \hskip -1mm -\tilde{q}\,\pi(\mu,\sigma) \ ,
\label{eq:ratcfmusig2}
\end{eqnarray}
with the constants $\tilde{q}$ and $\tilde{r}$ being defined as
\begin{equation*}
 \tilde{q} \equiv - h_2(1,0) \ \ \ \ \text{and} \ \ \ \ \tilde{r} \equiv - h_1(1,0) \ .
\end{equation*}

The general solution of differential functional equation \eqref{eq:ratcfmusig2}
is a function $\pi(\mu,\sigma)$ of the form
\begin{equation}
\pi(\mu,\sigma) = 
\Omega(\sigma)\,\exp{\bigl\{-\tilde{q}\mu\bigr\}} \ ,
\label{eq:solcfmusig1}
\end{equation}
where $\Omega(\sigma)$ is a 
non-negative function of $\sigma$. 
When inserted in \eqref{eq:ratcfmusig1}, \eqref{eq:solcfmusig1} yields:
\begin{equation*}
 \tilde{q}\mu = \tilde{r} + \sigma\,\frac{\Omega'(\sigma)}
                         {\Omega(\sigma)} \ ,
\end{equation*}
which, if it is to be true for all $\mu$ and $\sigma$, further implies
$\tilde{q}=0$ and 
\begin{equation*}
 \Omega(\sigma)\propto \sigma^{-\tilde{r}} \ ,
\end{equation*}
so that the general form of the consistency factor 
$\pi(\mu,\sigma)$ reads:
\begin{equation}
\pi(\mu,\sigma) = \sigma^{-\tilde{r}} \ ,
\label{eq:formpi}
\end{equation}
where we put all possible multiplication constants in the normalization
factor:
\begin{equation}
 \eta(\bar{x},s)=
 \int_{-\infty}^{\infty}d\mu'
\int_{0}^{\infty}d\sigma'\,(\sigma')^{-\tilde{r}}f(\bar{x}s|\mu'\sigma' I) \ .
\label{eq:norpdfmusigma}
\end{equation}

Now there is only one step dividing us from a complete determination of
consistency factors for location, scale and dispersion parameters.
Having established the form \eqref{eq:formpi} of the consistency 
factor $\pi(\mu,\sigma)$, 
we can write down a pdf for $\mu$ and 
$\sigma$ given $\bar{x}$ and $s$:
\begin{equation}
 f(\mu\sigma|\bar{x}sI) = 
 \frac{\pi(\mu,\sigma)}
     {\eta(\bar{x},s)}\,
     f(\bar{x}s|\mu\sigma I) 
 = \frac{\sigma^{-\tilde{r}}}{\eta(\bar{x},s)}\,
 f(\bar{x}s|\mu\sigma I) \ .
\label{eq:pdfmusigma1}
\end{equation}
According to the product rule \eqref{eq:product1}, the pdf can also be 
written as
\begin{equation}
f(\mu\sigma|\bar{x}sI) =
f(\mu|\sigma\bar{x}sI)\,
f(\sigma|\bar{x}sI) \ ,
\label{eq:product12}
\end{equation}
where $f(\sigma|\bar{x}sI)$ is a marginal 
pdf (see equation \eqref{eq:marginal}):
\begin{equation}
 f(\sigma|\bar{x}sI)=
 \int_{-\infty}^{\infty} f(\mu'\sigma|\bar{x}sI)\,d\mu' 
 =\,\frac{\sigma^{-\tilde{r}}}{\eta(\bar{x},s)}\,
\int_{-\infty}^{\infty}f(\bar{x}s|\mu'\sigma I)\,d\mu' \ .
\label{eq:marginalsig}
\end{equation}

Then, combination of equations (\ref{eq:mugivsigma2}-\ref{eq:etaxbarsk})
and (\ref{eq:pdfmusigma1}-\ref{eq:marginalsig}) leads to:
\begin{equation*}
 e^{q\mu} = \frac{\int_{-\infty}^{\infty}f(\bar{x}s|\mu'\sigma I)\, d\mu'}
 {\int_{-\infty}^{\infty}e^{-q\mu'}\,f(\bar{x}s|\mu'\sigma I)\, d\mu'} \ ,
\end{equation*}
solvable for any value of $\mu$ if and only if 
\begin{equation}
q = 0 \ ,
\label{eq:keq0}
\end{equation}
where $q$ is the constant of the consistency factor for location parameters
\eqref{eq:pimu}. Due to the simple relation \eqref{eq:qr} between the 
constant $q$ of the
consistency factor for the location parameters and the constant $r$ of the
factors for the scale and dispersion parameters, the above solution also
implies
 \begin{equation}
 r = 1 \ .
\label{eq:qeq1}
\end{equation}
These are highly nontrivial results since they \textit{uniquely} 
determine the consistency factors for the location, dispersion and scale 
parameters (recall eqns. \eqref{eq:pimu}, \eqref{eq:cfsig} and 
\eqref{eq:cftau}):
\begin{eqnarray}
\label{eq:cflsd}
      \pi(\mu) &=& 1 \ , \nonumber \\
      \pi(\sigma) &=& \sigma^{-1} \ , \\
      \pi(\tau)&=& \tau^{-1} \ . \nonumber
\end{eqnarray}

In addition, in order to determine also the value of the 
constant $\tilde{r}$ of the
consistency factor \eqref{eq:formpi}, we recall that
apart from \eqref{eq:product12}, the product rule \eqref{eq:product1}
also allows for the pdf \eqref{eq:pdfmusigma1} to be written as:
\begin{equation}
f(\mu\sigma|\bar{x}sI) =
f(\sigma|\mu\bar{x}sI)\,
f(\mu|\bar{x}sI) \ ,
\label{eq:product13}
\end{equation}
where the marginal distribution $f(\mu|\bar{x}sI)$ 
now stands for
\begin{equation}
  f(\mu|\bar{x}sI)=
  \int_{0}^{\infty} 
  f(\mu\sigma'|\bar{x}sI)\,d\sigma' 
  =\frac{1}{\eta(\bar{x},s)}\,
  \int_{0}^{\infty}(\sigma')^{-\tilde{r}}f(\bar{x}s|\mu\sigma' I)\,d\sigma' \ ,
\label{eq:marginal2}
\end{equation}
while the pdf for $\sigma$ given $\mu$, 
$f(\sigma|\mu\bar{x}sI)$, equals the
pdf \eqref{eq:sigmagivmu2}:
\begin{equation}
 f(\sigma|\mu\bar{x}sI) 
 =\frac{\sigma^{-r}}
  {\int_{0}^{\infty}(\sigma')^{-r}f(\bar{x}s|\mu\sigma' I)\,d\sigma'}
  \,f(\bar{x}s|\mu\sigma I) \ .
\label{eq:sigmagivmu3}
\end{equation}
Equations \eqref{eq:pdfmusigma1} and
(\ref{eq:product13}-\ref{eq:sigmagivmu3}) combined yield:
\begin{equation*}
\sigma^{r-\tilde{r}} = 
\frac{\int_{0}^{\infty}(\sigma')^{-\tilde{r}}
     \,f(\bar{x}s|\mu\sigma' I)\,d\sigma'}
     {\int_{0}^{\infty}(\sigma')^{-r}\,f(\bar{x}s|\mu\sigma' I)\,d\sigma'}
 \ ,
\end{equation*}
with $r$ being determined \eqref{eq:qeq1} to be 1. Evidently, the solution
of the above equation reads:
\begin{equation*}
\tilde{r}=r=1 \ ,
\end{equation*}
which finally determines the consistency factor 
$\pi(\mu,\sigma)$ for a symmetric sampling
distribution (see eq.\,\eqref{eq:formpi}),
\begin{equation}
\pi(\mu,\sigma) = \sigma^{-1} \ .
\label{eq:cfmusigma}
\end{equation}
\vskip 2 true mm
\par{\noindent\narrower\small
Now we would like to make use of results obtained in this and 
preceding sections and address the so-called  problem of two
means (i.e. two location parameters), also referred to as the 
Fisher-Behrens problem (see
refs. \cite{beh} and \cite{fis}, and \S\,19.47, pp.\,160-162, 
\S\,19.48, p.\,164 and \S\,26.28-26.29, pp.\,441-442 in
ref. \cite{ken2}). Imagine $x$ and $y$ being independent quantities,
both being subject to Gaussian sampling distributions,
\begin{equation*}
 f(x|\mu_1\sigma_1I) = \frac{1}{\sqrt{2\pi}\sigma_1}\,
                       \exp\biggl\{-\frac{(x-\mu_1)^2}{2\sigma_1^2}\biggr\}
\end{equation*}
and
\begin{equation*}
 f(y|\mu_2\sigma_2I) = \frac{1}{\sqrt{2\pi}\sigma_2}\,
                \exp\biggl\{-\frac{(y-\mu_2)^2}{2\sigma_2^2}\biggr\} \ ,
\end{equation*}
with parameters of both distributions being unknown and unconstrained. 
After collecting two events, $(x_1,x_2)$ and $(y_1,y_2)$, from
each of the two samples, we are in a position to make a probabilistic
inference about the unknown parameters. The pdf's for $(\mu_1,\sigma_1)$
and $(\mu_2,\sigma_2)$ read:
\\ \parbox{0.915\linewidth}{
\begin{equation}
\begin{split}
 f(\mu_1\sigma_1|\bar{x}s_xI) &= 
 \frac{\pi(\mu_1,\sigma_1)}{\eta(\bar{x},s_x)}
 \,f(\bar{x}s_x|\mu_1\sigma_1I)\\
 &= \frac{4}{2\pi}\frac{s_x}{\sigma_1^3}
  \,\exp\biggl\{-\frac{(\bar{x}-\mu_1)^2}{\sigma_1^2}
  -\frac{s_x^2}{\sigma_1^2}\biggr\} \\ 
 &= \frac{4}{2\pi}\frac{s_x}{\sigma_1^3}
  \,\exp\biggl\{-\frac{s_x^2}{\sigma_1^2}
  \,\biggl[\frac{(\bar{x}-\mu_1)^2}{s_x^2}
  +1\biggr]\biggr\}
\end{split}  
\label{eq:twm1}
\end{equation}
} \\
and
\\ \parbox{0.915\linewidth}{
\begin{equation}
\begin{split}
 f(\mu_2\sigma_2|\bar{y}s_yI) &= 
 \frac{\pi(\mu_2,\sigma_2)}{\eta(\bar{y},s_y)}
 \,f(\bar{y}s_y|\mu_2\sigma_2I) \\ 
 &=\frac{4}{2\pi}\frac{s_y}{\sigma_2^3}
 \,\exp\biggl\{-\frac{(\bar{y}-\mu_2)^2}{\sigma_2^2}
 -\frac{s_y^2}{\sigma_2^2}\biggr\} \\
&= \frac{4}{2\pi}\frac{s_y}{\sigma_2^3}
  \,\exp\biggl\{-\frac{s_y^2}{\sigma_2^2}
  \,\biggl[\frac{(\bar{y}-\mu_2)^2}{s_y^2}
  +1\biggr]\biggr\} \ ,
\end{split} 
\label{eq:twm2}
\end{equation}
} \\
where
\begin{equation*}
 \bar{x} = \frac{x_1+x_2}{2} \ \ , \ \ s_x = \frac{|x_1-x_2|}{2} \ ,
\end{equation*}
and
\begin{equation*}
 \bar{y} = \frac{y_1+y_2}{2} \ \ , \ \ s_y = \frac{|y_1-y_2|}{2} \ .
\end{equation*}

Since the two pdf's are independent, we apply the product rule 
\eqref{eq:product1} and write down a pdf for $\mu_1$, $\mu_2$, $\sigma_1$
and $\sigma_2$ as a product of pdf's \eqref{eq:twm1} and \eqref{eq:twm2}:
\begin{equation*}
\begin{split}
 &f(\mu_1\mu_2\sigma_1\sigma_2|\bar{x}\bar{y}s_xs_yI) =
 f(\mu_1\sigma_1|\bar{x}s_xI)\,f(\mu_2\sigma_2|\bar{y}s_yI)= \\
 &\frac{4}{\pi^2}\,\frac{s_x}{\sigma_1^3}\,\frac{s_y}{\sigma_2^3}\,
 \exp\biggl\{-\frac{(\bar{x}-\mu_1)^2}{\sigma_1^2}
            -\frac{(\bar{y}-\mu_2)^2}{\sigma_2^2}\biggr\}
 \,\exp\biggl\{-\frac{s_x^2}{\sigma_1^2}-\frac{s_y^2}{\sigma_2^2}\biggr\} = \\
 &\frac{4}{\pi^2}\,\frac{s_x}{\sigma_1^3}\,\frac{s_y}{\sigma_2^3}
  \,\exp\biggl\{-\frac{s_x^2}{\sigma_1^2}
  \,\biggl[\frac{(\bar{x}-\mu_1)^2}{s_x^2}
  +1\biggr]\biggr\}
  \,\exp\biggl\{-\frac{s_y^2}{\sigma_2^2}
  \,\biggl[\frac{(\bar{y}-\mu_2)^2}{s_y^2}
  +1\biggr]\biggr\} \ .
\end{split}
\end{equation*}
By integrating out parameters $\sigma_1$ and $\sigma_2$ we find the
marginal pdf for $\mu_1$ and $\mu_2$ to be of the form:
\begin{equation*}
\begin{split}
 f(\mu_1\mu_2|\bar{x}\bar{y}s_xs_yI) &=
 \int_{0}^{\infty}d\sigma_1\int_{0}^{\infty}d\sigma_2\,
 f(\mu_1\mu_2\sigma_1\sigma_2|\bar{x}\bar{y}s_xs_yI) \\ 
 &= \frac{1}{\pi^2s_xs_y}\,\biggl(\frac{(\bar{x}-\mu_1)^2}{s_x^2}
  +1\biggr)^{-1}\,\biggl(\frac{(\bar{y}-\mu_2)^2}{s_y^2}
  +1\biggr)^{-1} \ .
\end{split}
\end{equation*}
Note that the above results were all obtained simply by using some
of the applications of basic Desiderata without making any additional
assumptions or requiring any new postulates (compare to references \cite{yat}
and \cite{ken2}, \S\,26.29, p.\,442).
We refer to the Fisher-Behrens problem again in Section\,\ref{sec:history}
when we comment on difficulties with inferences about parameters
outside the framework of probability.
\par}
\section{On uniqueness of consistency factors and on
         consistency of basic rules}
\label{sec:unique}

In previous sections we saw that the general rules for plausible
reasoning - the Cox-P\'{o}lya-Jaynes Desiderata - uniquely determine the 
consistency factors for location, scale and dispersion parameters.
In other words, in the limit of complete prior ignorance, there
is only one possible way of making a consistent inference about the three
types of parameters. According to \eqref{eq:bayprime}, the pdf, 
assigned to one or to several of these parameters simultaneously, 
is to be proportional to the likelihood function, containing the
information from one or several measured events $x_i$ that are
subject to the distribution determined by the inferred parameter(s),
and to the appropriate consistency factor as determined at the end
of the preceding section.

Anyone who possesses the same information, but assigns a different
probability distribution to a given parameter , e.g. by choosing
a 'consistency factor' of a different form, thus necessarily violates
at least one of our basic Desiderata. It is certainly true that 
nobody has the authority to forbid such violations, but, at the same
time, it is also true that anyone coming to the conclusions
by violating such basic and general rules being 
would surely have difficulties in persuading anyone else, 
who was aware of these violations, to accept his conclusions.

In Section\,\ref{sec:objectivity} we stressed that a prior information
$I$ in a problem of inference about a parameter $\theta$ of a
sampling distribution, $f(x|\theta I)$, is equal to corresponding information
$I'$ about an inferred parameter $\nu=\bar{g}_a(\theta)$ of the sampling
distribution for $y=g_a(x)$, $f\bigl(g_a(x)|\bar{g}_a(\theta) I'\bigr)$,
only if three requirements are simultaneously met: a) if the sampling 
distribution is invariant under transformations $g_a$ and $\bar{g}_a$,
b) if the permissible range of sampling variate $x$, $(x_a,x_b)$ is
invariant under $g_a$, and c) if the permissible range of the inferred
parameter $\theta$, $(\theta_a,\theta_b)$ is invariant under
transformation $\bar{g}_a$. Evidently, for the ranges of parameters
$(\mu_a,\mu_b)=(-\infty,\infty)$ and $(\sigma_a,\sigma_b)=(\tau_a,\tau_b)=
(0,\infty)$, and the corresponding transformations 
(\ref{eq:invlocscale}-\ref{eq:invscale}), the condition c) is well
fulfilled for all $a\in(0,\infty)$ and $b\in(-\infty,\infty)$.

But, in practice, we usually face problems with pre-constrained
inferred parameters: we possess some additional information that
narrows the admissible range. As a 
simple example, when we are estimating the average lifetime
of a newly discovered particle, produced in an experiment
with highly energetic protons from an accelerator hitting
a fixed target, it is easy to imagine that $\tau$ cannot really be
infinite, for in that case there should be many of these
particles around as remnants of the Big Bang. In fact, it
might well be reasoned that $\tau$ is necessarily even much 
smaller than billions
of years, since in case of $\tau$ being sufficiently large,
e.g. of the order of a millisecond, we should have noticed
the particles as products of cosmic protons hitting nuclei
in the upper layers of the Earth's atmosphere. The listed two
arguments, as well
as any possible additional ones, thus lead to a finite
interval $(\tau_a,\tau_b)$ in the positive half of the
real axis. But such an interval is clearly \textit{not} invariant
under parameter transformation $\bar{g}_a(\tau)=a\tau$ and so the above
condition c) for the equality of information, $I=I'$, is \textit{not}
fulfilled. How shall we proceed in such cases?

Suppose for a moment that we ignore the fact that the condition c) is
not fulfilled. Effectively this is equivalent to a prescription 
that can often be found in textbooks on the so-called Bayesian inference: 
to use the same function $\pi(\tau)$ as in the case of no constraints
on the parameter range, and afterwards
to chop-off the unconstrained pdf for the inferred parameter, $\tau$,
outside the interval $(\tau_a,\tau_b)$ and renormalize the truncated
distribution (see, for example, \cite{hag}, \S\,3.17, pp.\,72-73).
It is easy to show that in general such an \textit{ad hoc} prescription
inevitably leads to inconsistencies.

Namely, without the constraints on $\tau$, $\widetilde{\pi}(a\tau)$ 
in equation\,\eqref{eq:pitilde} is to equal $\pi(a\tau)$, and consequently
$\pi(\tau)$ is to equal $1/\tau$. Then, \eqref{eq:pitilde} implies
\begin{equation*}
 k(a) = 1 \ \ , \ \ \forall \  a\in(0,\infty) \ ,
\end{equation*}
so the functional equation\,\eqref{eq:funeqnor} for the normalization
factor $\eta(t_1)$ reads:
\begin{equation}
 \widetilde{\eta}(at_1) = \frac{\eta(t_1)}{a} \ .
\label{eq:eqeta}
\end{equation}
By setting $a=1$ we realize that $\widetilde{\eta}(t)$ and $\eta(t)$ 
are to be the same functions,
\begin{equation}
 \widetilde{\eta}(t_1) = \eta(t_1) \ ,
\label{eq:eqeta1}
\end{equation}
or, equivalently,
\begin{equation}
 \widetilde{\eta}(at_1) = \eta(at_1) \ .
\label{eq:eqeta2}
\end{equation}
When inserted in \eqref{eq:eqeta}, \eqref{eq:eqeta2} yields
\begin{equation}
 \eta(at_1) = \frac{\eta(t_1)}{a} \ ,
\label{eq:eqeta3}
\end{equation}
which must be true for any $a\in(0,\infty)$. For $a=t_1^{-1}$ 
\eqref{eq:eqeta3} thus reads:
\begin{equation}
 \eta(t_1)=\eta(1)\,\frac{1}{t_1} \ .
\label{eq:etat1}
\end{equation}
On the other hand, by definition \eqref{eq:etatilde}, $\widetilde{\eta}(t_1)$
should ensure normalization of the pdf for $a\tau$:
\begin{equation}
 \widetilde{\eta}(t_1) = \int_{\tau_a}^{\tau_b}\hskip -2mm
 \widetilde{\pi}(\tau')\,f(t_1|\tau'\, I)\,d\tau' 
 = \int_{\tau_a}^{\tau_b}\hskip -1mm 
 \frac{1}{\tau'^2}\,e^{-\frac{t_1}{\tau'}}\,d\tau' 
 = \frac{1}{t_1}\,\Bigl(e^{-t_1/\tau_b}-e^{-t_1/\tau_a}\bigr) \ .
\label{eq:etat2}
\end{equation}
Evidently, the equality \eqref{eq:eqeta1} is assured for all $t_1\in(0,\infty)$
only if
\begin{equation*}
 \tau_a = 0 \ \ \ \ \text{and} \ \ \ \ \tau_b = \infty \ ,
\end{equation*}
i.e. only if the invariance of the parameter range under the particular
group $\bar{\mathcal{G}}$ of transformations $\bar{g}_a$ is exact. 

In practice, our reasoning would still be \textit{sufficiently consistent} if
\begin{equation}
 \frac{t_1}{\tau_a} \gg 1 \ \ \ \ {\rm and} \ \ \ \ 
 \frac{t_1}{\tau_b} \ll 1 \ , 
\label{eq:ggll}
\end{equation}
i.e. if the integrals $\epsilon_1$ and $\epsilon_2$ of the unconstrained
pdf for $\tau$ outside the allowed region (see Figure 1),
\begin{equation}
\begin{split}
 \epsilon_1&=\int_{0}^{\tau_a}f(\tau'|t_1 I)\,d\tau' = 
            \int_{0}^{\tau_a}\frac{t_1}{\tau'^2}\,e^{-t_1/\tau'}\,d\tau' \ , \\
 \epsilon_2&=\int_{\tau_b}^{\infty}f(\tau'|t_1 I)\,d\tau' = 
             \int_{\tau_b}^{\infty}\frac{t_1}{\tau'^2}\,e^{-t_1/\tau'}\,
 d\tau' \ , 
\end{split}
\label{eq:epstau}
\end{equation}
are sufficiently small. 
\begin{figure}
\vskip -48mm
\hskip -33mm
\epsfig{file=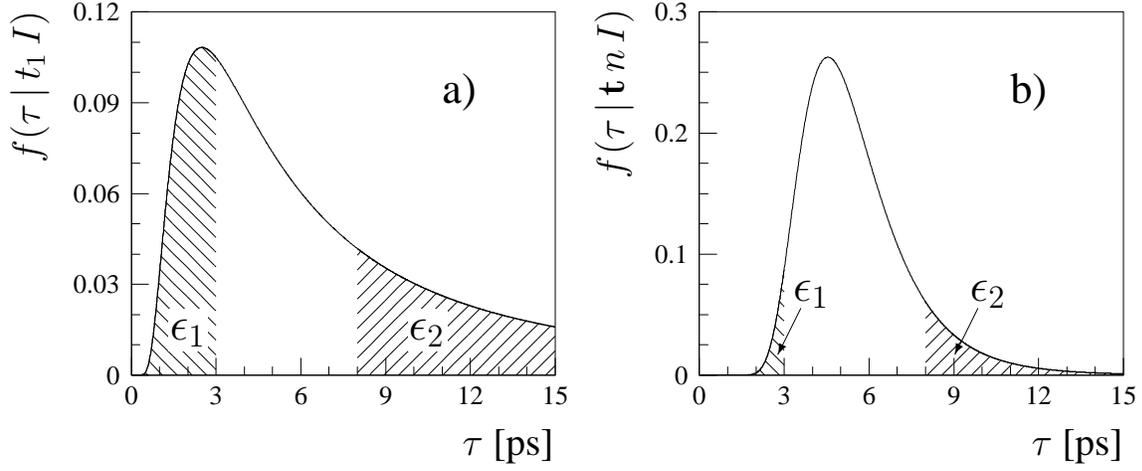,width=215mm}
\vskip -194mm
\label{fig:intpdf}
\caption{a) Unconstrained probability density function for 
           $\tau$ based on a recorded decay time $t_1=5\;{\rm ps}$.
           The hatched areas represent integrals, 
           $\epsilon_1=0.189$
           and $\epsilon_2=0.465$, of the pdf in the intervals 
           $(0,\tau_a=3\;{\rm ps})$ and $(\tau_b=8\;{\rm ps},\infty)$,
           respectively. b) Pdf for $\tau$ based on an average
           $\bar{t}=5\;{\rm ps}$ of $n=10$ recorded decay times
           with the integrals of the unconstrained pdf 
           outside the allowed region for $\tau$
           being reduced to $\epsilon_1=0.031$ and $\epsilon_2=0.102$.}
\end{figure}
That is, if the conditions \eqref{eq:ggll} 
are fulfilled, normalization factors \eqref{eq:etat1} and \eqref{eq:etat2}
are equal beyond the precision required for the particular inference. 
As admitted above, the invariance of the admissible parameter range,
and consequently also the consistency of our reasoning, is not exact
in such a case since for extremely large or extremely small values
of measured decay times $t$ the equality \eqref{eq:eqeta} would 
\textit{not} hold. But once
$t_1$ has been recorded, it is very likely that the value of the
inferred $\tau$ would also be of the same order of magnitude, i.e. it would
be extremely unlikely for $\tau$ to be much smaller or much larger
than $t_1$. Then, with $\tau\sim t_1$, it is also \textit{very unlikely} 
that we would ever observe an event $t$ orders of magnitude different
from $t_1$.

On the other hand, if the conditions \eqref{eq:ggll}
are \textit{not} fulfilled (see, for example, Figure\,1.a), 
our reasoning is clearly \textit{not consistent}:
by applying the two equally valid normalization factors, \eqref{eq:etat1} 
and \eqref{eq:etat2}, we are able to arrive at significantly different 
probabilities for the same proposition, e.g. 
$P\bigl(a\tau\in(\tau_1,\tau_2)|at_1\,I\bigr)$,
which is in direct contradiction with the consistency Desideratum
\textit{III.a}. The additional information that narrows the interval
$(\tau_a,\tau_b)$ may still be very useful, but it is just that consistent
probabilistic reasoning is impossible in such a situation.

In order to avoid such an inconsistency, we could have ignored 
the information, additional to $t_1$, and stretched
$(\tau_a,\tau_b)$ to the whole positive half of the real axis.
But this is not a solution either, since in this way we would have been 
arbitrarily  ignoring some of the available information and basing our
conclusions on what remains. Acting in explicit contradiction
with Desideratum\;\textit{III.b}, we would have allowed ideology
to break into our inference, which is inadmissible for any scientifically
respectable reasoning.

Thus, the only consistent solution to our problem of inference about the
pre-con\-strai\-ned parameter $\tau$ would be provided by recording 
additional independent decay times,\linebreak $t_2,t_3,...,t_n$, 
of particles of the same type. Then,
the unconstrained probability distribution for $\tau$, based 
on the recorded data, is described by the following pdf:
\begin{equation*}
 f(\tau|t_1t_2...t_nI)=f(\tau|\bi{t}nI)
 =\frac{(n\bar{t})^n}{(n-1)!}\,\frac{1}{\tau^{n+1}}\,e^{-n\bar{t}/\tau} \ ,
\end{equation*}
where $\bar{t}$ is an average of the recorded decay times:
\begin{equation*}
 \bar{t} = \frac{1}{n}\,\sum_{i=1}^{n}t_i \ .
\end{equation*}
The distribution narrows as $n$ increases (see Figure 1.b).
Therefore, by collecting enough data, the inconsistency is diminished
beyond the required level: by diminishing the integrals $\epsilon_1$ 
and $\epsilon_2$ of the unconstrained pdf for $\tau$ outside the
constrained domain, the limits $\tau_a$
and $\tau_b$ that caused the inconsistencies become irrelevant.

Our theory of consistent inference about parameters is therefore
valid only if certain conditions are fulfilled, i.e. in the limit
$\epsilon_{\pm} \to 0$. Such a theory is referred to as an
\textit{effective theory}, with the term coming from physics where
all theories are effective. The precision of predictions of an
effective theory is estimated by the proximity of the actual 
conditions to the ideal limit. The values of integrals $\epsilon_{\pm}$
\eqref{eq:epstau}, compared to zero, can thus serve as an estimate of 
the precision of our probabilistic inference about the pre-constrained 
parameters.

\section{Calibration}
\label{sec:calibration}
\hfill \parbox{0.7\linewidth}{\small
\noindent
The most striking achievement of the physical sciences is prediction.
\\ \vskip -2mm
\hfill
Georg P\'{o}lya (\cite{pol1}, Chapter\,XIV, \S\,4, p.\,64)
} \\ \vskip 2mm

Thus far our theory of plausible inference about parameters has been
developed by following Desiderata\;\textit{I.-III.} only, while
the implications of the operational Desideratum have \text{not} yet 
been considered. 
According to the latter Desideratum,
in order to exceed the level of a mere speculation, our theory of
inference about parameters must be exposed, i.e. must be able to
make predictions that can be verified by experiments. The magnitudes
considered by physicists such as mass, electric charge or 
reaction velocity have an \textit{operational} definition: the 
physicist knows very well which operations he or she has to perform
if he or she wishes to ascertain the magnitude of an electric
charge, for example (\cite{pol1}, Chapter\,XV, \S\,4, p.\,117). 
Is there a way for probability as a measure of a personal degree of belief 
to become operational, too?

Imagine we
were given several numbers $x_i$, all produced by a random number
generator according to a distribution $f(x_i|\theta_i I)$.
While the form of the distribution is known to us, we do not know
much about the values $\theta_i$ of the parameter: in
general they can be different for each of the numbers $x_i$ generated
and can be any number in the range $(\theta_a,\theta_b)$.

In fact, what we are asked for is to make probabilistic inferences
about the unknown values $\theta_i$ on the basis of each datum $x_i$ 
separately. Based on our
probability distribution for $\theta_i$, 
$p(\theta_i|x_i I)=f(\theta_i|x_iI)$, we should specify our 
\textit{confidence intervals} $(\theta_{i,1},\theta_{i,2})$ such that
\begin{equation}
 P\bigl(\theta_i\in(\theta_{i,1},\theta_{i,2})|x_i I\bigr) =
 \int_{\theta_{i,1}}^{\theta_{i,2}} f(\theta_i|x_i I)\,d\theta_i = 
 \delta \ ,
\label{eq:prconfint}
\end{equation}
with $\delta$ being the same for all inferences. Note that the interval
for the inference of a particular value $\theta_i$ is not unique: it can 
be the shortest of all possible intervals, the central interval with
$P(\theta_i\leq\theta_{i,1}|x_i I)=P(\theta_i > \theta_{i,2}|x_i I)
=(1-\delta)/2$, the lower-most interval with $\theta_{i,1}=\theta_a$,
the upper-most interval with $\theta_{i,2}=\theta_b$, or any other
interval as long as the probability \eqref{eq:prconfint} equals $\delta$.

After the inference has been made, we learn the values $\theta_i$
of the parameter used in the random number generator. Then, our
probability judgments are said to be \textit{calibrated}
if they agree with the actual frequencies of occurrence (\cite{hag}, \S\,6.4,
p.\,142), i.e. if the fraction of inferences with the specified intervals 
containing the actual value of the parameter for the particular example, 
coincides with $\delta$. 

The definition of long range relative frequency, although in some
way less distinct than that of an electric charge, is still operational:
it suggests definite operations that we can undertake to obtain
an approximate numerical value of such a frequency (\cite{pol1}, 
Chapter\,XV, \S\,4,p.\,117).
Our theory thus cannot be expected to give a correct prediction \textit{each}
time, but it can be verifiable from its long range consequences, i.e. it can
reasonably be expected to give the right answer in an assignable
percentage of cases in the long run.

\begin{figure}
\vskip -3.8cm
\hskip -1.8cm
\psfig{file=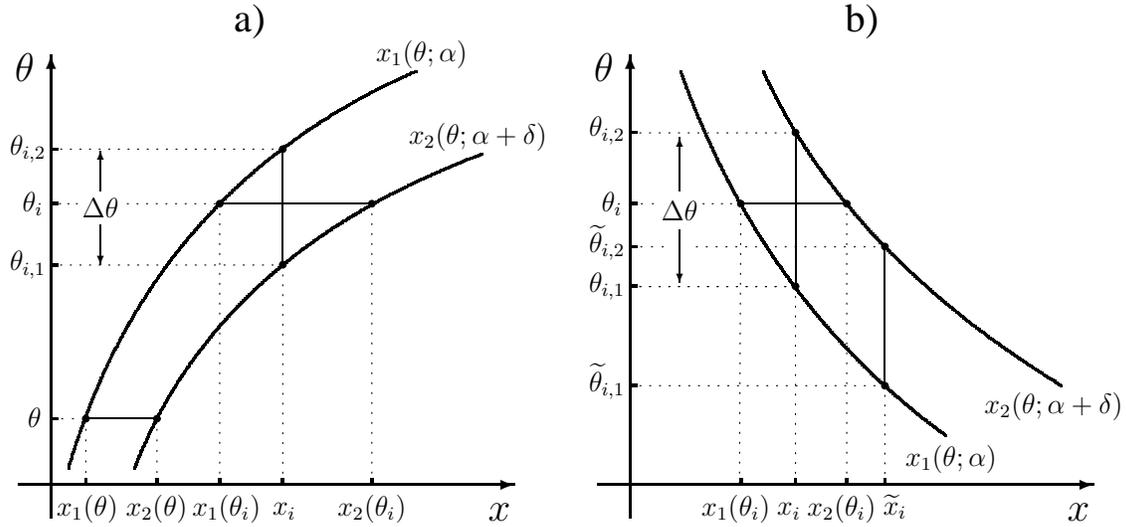,width=185mm}
\vskip -15.3cm
\caption{Confidence belts in the case of distribution function $F(x,\theta)$
         being a) strictly decreasing  and b) strictly increasing 
         in $\theta$.}
\label{fig:confint}
\end{figure}

Under what conditions is a calibrated inference achieved? To answer this
question we refer to the construction of classical confidence intervals  
(see \cite{ney}, \S\,9.2.1, pp.\,200-201 in \cite{ead}, and Chapter\,19 in
\cite{ken2}) and to fiducial theory (see refs. \cite{fis} and 
\cite{ken2}, \S\,19.44-19-47, pp.\,156-162, \S\,26.26-26.29, pp.\,440-442). 
Let $\alpha$ and $\alpha+\delta$ be probabilities 
for $x$ to take values less or equal to $x_1$ and $x_2$, respectively, 
given the value $\theta$ of the parameter of the (continuous) 
sampling distribution for $x$:
\begin{equation}
\begin{split}
 \alpha &= P\bigl(x\leq x_1(\theta)|\theta I\bigr) = 
 F\bigl(x_1(\theta),\theta\bigr) \ , \\
 \alpha+\delta &= P\bigl(x\leq x_2(\theta)|\theta I\bigr) = 
 F\bigl(x_2(\theta),\theta\bigr) \ .
\end{split}
\label{eq:alpde}
\end{equation}
Then, the probability for $x$ to take a value in the interval
$\bigl(x_1(\theta),x_2(\theta)\bigr)$, equals $\delta$:
\begin{equation}
 P\bigl(x_1(\theta)\leq x < x_2(\theta)\bigr) = 
F\bigl(x_2(\theta),\theta\bigr)- F\bigl(x_1(\theta),\theta\bigr)
= \delta \ ,
\label{eq:delta}
\end{equation}
regardless of the value of $\alpha$.
Let unique values $x_1(\theta)$ and $x_2(\theta)$, satisfying \eqref{eq:alpde},
exist for every $\theta$ in the range $(\theta_a,\theta_b)$,
and for every $\alpha\in (0,1-\delta)$. It is easy to see that 
unconstrained location and scale parameters with 
$\mu\in(-\infty,\infty)$ and $\tau\in(0,\infty)$ meet such a requirement .
The curves $x_1(\theta;\alpha)$ and $x_2(\theta;\alpha+\delta)$
are formed by varying $\theta$ in $x_1(\theta)$ and $x_2(\theta)$ but
at fixed $\alpha$ and $\delta$ (see Figure \ref{fig:confint}) and the region
between the two curves is known as a \textit{confidence belt}.

Having the confidence belts defined, we return to the estimation
of unknown values $\theta_i$ of the parameter. Given the measured
value $x_i$, the required confidence interval $(\theta_{i,1},\theta_{i,2})$
is obtained by the intersection of the vertical line  $x=x_i$ with
the curves $x_1(\theta;\alpha)$ and $x_2(\theta;\alpha+\delta)$.
By inspecting Figure \ref{fig:confint} it becomes evident that the
proposition 
\begin{equation}
 \theta_i\in(\theta_{i,1},\theta_{i,2})
\label{eq:propthi}
\end{equation}
is true if $x_i\in\bigl(x_1(\theta_i),x_2(\theta_i)\bigr)$ and is 
wrong otherwise (see, for example, $\widetilde{x}_i$ and the corresponding
confidence interval $(\widetilde{\theta}_{i,1},\widetilde{\theta}_{i,1})$
in Figure \ref{fig:confint}.b). This means that, according to
\eqref{eq:delta}, the probability for \eqref{eq:propthi} to be true
is exactly $\delta$ for \textit{every} $\theta_i$, i.e. in the long
run our inferences will be correct in $\delta$ per cent \textit{regardless}
the distribution of the values $\theta_i$.

In the case there exist unique values $\theta_{1,2}\in (\theta_a,\theta_b)$
for any $\alpha\in (0,1)$, any $\delta\in (0,1-\alpha)$ and for any 
$x_{1,2}\in (x_a,x_b)$, that solve equations \eqref{eq:alpde},
the cdf for $x$ is either \textit{strictly decreasing} or 
\textit{strictly increasing} in $\theta$. Then, by construction of 
confidence belts, the relations
\begin{equation*}
\begin{split}
 F(x_i,\theta_{i,1})&=\alpha+\delta \ , \\
 F(x_i,\theta_{i,2})&=\alpha \ , 
\end{split}
\end{equation*}
are always true for cdf's strictly decreasing in $\theta$ 
(Figure \ref{fig:confint}.a), as is the case for 
functions with $\theta$ being a location or a scale parameter 
(see Section\,\ref{sec:parameters}), while for
strictly increasing distribution functions (Figure \ref{fig:confint}.b), 
the relations read:
\begin{equation*}
\begin{split}
 F(x_i,\theta_{i,1})&=\alpha \ , \\
 F(x_i,\theta_{i,2})&=\alpha+\delta \ . 
\end{split}
\end{equation*}
Accordingly, the fraction $\delta$ of correct inferences can be expressed
as:
\begin{equation}
 \delta=\pm F(x_i,\theta_{i,1}) \mp F(x_i,\theta_{i,2}) =
        \pm F(x_i,\theta_{i,1}) \mp F(x_i,\theta_{i,1}+\Delta\theta) \ ,
\label{eq:deltaF}
\end{equation}
where the upper (lower) sign corresponds to cumulative distribution
functions strictly decreasing (increasing) in $\theta$, while
\begin{equation*}
 \Delta\theta = \theta_{i,2} - \theta_{i,1} \ .
\end{equation*}
In the limit of infinitesimally small fractions $\delta$ and differences
$\Delta\theta = d\theta$, equation \eqref{eq:deltaF} can be rewritten in terms
of a differential of the distribution function with respect to $\theta$:
\begin{equation}
 \delta = \mp \Bigl(\frac{\partial}{\partial\theta}F(x_i,\theta)\Bigr)
          \,d\theta = \mp F_2(x_i,\theta)\,d\theta \ .
\label{eq:infdelta}
\end{equation}

But recall now our probabilistic inference about $\theta_i$: given the pdf
$f(\theta|x_i I)$, the probability for $\theta_i$ being in the interval
$(\theta,\theta+d\theta)$ reads:
\begin{equation*}
 p(\theta|x_i I) = f(\theta|x_i I)\,d\theta \ .
\end{equation*}
Thus, we finally came to the point when we are able to answer: 
\textit{our inference will be calibrated}, i.e. the assigned 
probability will coincide with the fraction $\delta$ of confidence intervals
containing the true values of the parameter, \textit{if and only if}
\begin{equation}
 f(\theta|x_i I) = \mp F_2(x_i,\theta) \ .
\label{eq:answer}
\end{equation}

For location parameters with the cumulative distribution function
\eqref{eq:cdfloc} for the sampling variable $x$, 
the condition \eqref{eq:answer} implies the
calibrated pdf for $\theta$ be:
\begin{equation*}
 f(\mu|x_i I) = -\frac{\partial}{\partial\mu}
                 \int_{-\infty}^{x_i-\mu}\phi(u)\,du
              = \phi(x_i-\mu) = f(x_i|\mu I) \ .
\end{equation*}
The above pdf for $\theta$ coincides with the one obtained by
following the basic Desiderata of consistent inference, implying 
the consistency factor $\pi(\mu)$ to be constant \eqref{eq:cflsd}.
Therefore, the only consistent way of inference about location
parameters is also the only one that is calibrated. 

The latter is also true for inferences about scale parameters with the
appropriate distribution function \eqref{eq:cdfscale}. Namely,
\begin{equation*}
 f(\tau|t_i I) = -\frac{\partial}{\partial\tau} 
                 \int_{0}^{\frac{t_i}{\tau}} \phi(u)\,du
               = \frac{t_i}{\tau^2}\,\phi\Bigl(\frac{t_i}{\tau}\Bigr)
               = \frac{t_i}{\tau}\,f(t_i|\tau I)
\end{equation*}
corresponds to $\pi(\tau)=\tau^{-1}$ as
already determined in \eqref{eq:cflsd}.

The probability distributions for location, scale or dispersion 
parameters that were assigned in consistent way, passed an important 
test: they are all calibrated.
The question can be raised whether there are any other types of parameters 
that are also in accordance with the calibration requirement \eqref{eq:answer}?
We restrict the answer only to parameters whose pdf can be 
written in the form of the Consistency Theorem \eqref{eq:bayprime1}, 
obtained by requiring
logically independent pieces of information to be commutative. By combining
the two equations we obtain:
\begin{equation}
 \pi(\theta)\,F_1(x,\theta)\pm\eta(x)\,F_2(x,\theta) = 0 \ ,
\label{eq:F1F2}
\end{equation}
where the upper (lower) sign stands for cdf's
which are strictly decreasing (increasing) in $\theta$. By defining
function $G(x,\theta)$ as a difference (sum),
\begin{equation*}
 G(x,\theta) \equiv h(x) \mp k(\theta) \ ,
\end{equation*}
with $h(x)$ and $k(\theta)$ being related to $\pi(\theta)$ and 
$\eta(x_1)$ as
\begin{equation*}
 h'(x) = \eta(x) \ \ \text{and} \ \ k'(\theta) = \pi(\theta) \ ,
\end{equation*}
equation \eqref{eq:F1F2} can be rewritten as
\begin{equation*}
 F_1(x,\theta)\,G_2(x,\theta) - F_2(x,\theta)\,G_1(x,\theta) = 0 \ ,
\end{equation*}
with $G_1(x,\theta)=\eta(x)$ and $G_2(x,\theta)=\pi(\theta)$
being strictly positive functions (see Section\,\ref{sec:subjectivity}).
But as we saw in Section\,\ref{sec:invariance}, the general
solution of such a differential functional equation is
a distribution function $F(x,\theta)$ of the form
\begin{equation*}
 F(x,\theta) = \Phi(z-\mu) \ , 
\end{equation*}
where
\begin{equation*}
 z\equiv h(x) \ \ \text{and} \ \ \mu\equiv \pm k(\theta) \ ,
\end{equation*}
i.e. a distribution function that corresponds to $\mu$ being
a location parameter \eqref{eq:cdfloc}. 
Therefore, in the limit of complete prior ignorance,
\textit{an inference about a parameter $\theta$ that is 
subject to the calibration condition \eqref{eq:answer}, is 
necessarily reducible to an inference about a location parameter}. 
Note that this result was first obtained by Dennis Lindley \cite{lin}
by combining the calibration condition \eqref{eq:answer} and the
Bayes' theorem with a prior pdf $f(\theta|I)$ which is independent 
of data $x$.

It is interesting to note that the above result is identical to 
the result obtained at the end of Section\,\ref{sec:invariance}
and discussed in Section\,\ref{sec:objectivity}, despite the
fact that the requirements of logical consistency and that of
calibration may appear, at least at first sight, 
as almost diametrically opposed starting points.
As a very important consequence, \textit{every probabilistic 
inference about a parameter of a sampling distribution
that we are sure is consistent
will thus at the same time also be calibrated} and, \textit{vice versa},
\textit{every calibrated inference, based on a posterior probability
distribution that is factorized according to the Consistency Theorem, will 
simultaneously be logically consistent, too}.

Recall Section\,\ref{sec:intro} where we chose a
set of functions, called probabilities, out of all possible
plausibility functions, suitable for representing
our degree of belief according to the basic Desiderata.
The main reason for such a choice was that 
for the probabilities the basic equations for manipulating
plausibilities - the product and the sum rule, 
\eqref{eq:product} and \eqref{eq:sum}- are of especially
simple forms. 
But now
we realize another advantage of probabilities over other
plausibility functions: \textit{it is only for probabilities
that the assigned degrees of belief exactly coincide with
the long term relative frequencies}. Other plausibilities
are one-to-one functions of probabilities, so their values
correspond to (the same) one-to-one functions of
the relative frequencies. Again, this does \textit{not} imply that
predictions in terms of probabilities are more reliable
than predictions in terms of any other set of plausibility
functions: they are just the easiest to interpret.

The predictions of the systems of plausible reasoning,
however, that are \textit{not} isomorphisms of the probability
system, are, apart from being inconsistent, also necessarily
uncalibrated. This means that in general their predictions
in terms of long-range frequencies are \textit{not} correct.
\vskip 2 true mm
\par{\noindent\narrower\small
Take, as an example, the power-law distribution with the pdf
\\ \parbox{0.915\linewidth}{
\begin{equation}
 f(x|\theta I) = (\theta+1)\,x^{\theta} \ , 
\label{eq:pdfpotx} 
\end{equation}
} \\
the ranges of $x$ and $\theta$
\\ \parbox{0.915\linewidth}{
\begin{equation}
\label{eq:rangext}
 x\in(0,1) \ \ \text{and} \ \ \theta\in(-1,\infty) \ ,
\end{equation}
} \\
and the corresponding cdf
\begin{equation*}
 F(x,\theta) = \int_{0}^{x}(1+\theta)\,x'^{\,\theta}\,dx' = x^{\theta+1} \ .
\end{equation*}
Due to the range of $x$, the cdf is strictly decreasing in $\theta$,
therefore the calibration condition for the inference about the
parameter reads:
\\ \parbox{0.915\linewidth}{
\begin{equation}
  f(\theta|x I) = -F_2(x,\theta) = (\theta+1)\,x^{\theta}\;
  \frac{(-x\ln{x})}{\theta+1} \ .
\label{eq:pdfpotth}
\end{equation}
} \\
The ratio of the pdf's for $\theta$ and $x$, \eqref{eq:pdfpotx} and
\eqref{eq:pdfpotth}, 
\begin{equation*}
 \frac{f(\theta|x I)}{f(x|\theta I)} = \frac{-x\ln{x}}{\theta+1} \ ,
\end{equation*}
implies consistency and normalization factors of the form
\begin{equation*}
 \pi(\theta) = \frac{1}{\theta+1} 
\end{equation*}
and
\begin{equation*}
 \eta(x) = -\frac{1}{x\ln{x}} \ .
\end{equation*}
Their integrals, 
\begin{equation*}
 z\equiv h(x)=\int\eta(x)\,dx = -\ln{(-\ln{x})}
\end{equation*}
and
\begin{equation*}
 \mu\equiv k(\theta) = \int\pi(\theta)\,d\theta = \ln{(1+\theta)} \ ,
\end{equation*}
with the domains of both being the complete real
axis (c.f. equation \eqref{eq:rangext}),
\begin{equation*}
 z,\mu\in(-\infty,\infty) \ ,
\end{equation*}
allow for a reduction of the inference about parameter $\theta$
to an inference about a location parameter $\mu$ of the sampling
distribution of the variate $z$ with the corresponding cdf:
\begin{equation*}
 F(z,\mu) = \Phi\bigl(-\ln{(-\ln{x})}-\ln{(1+\theta)}\bigr) = \Phi(z-\mu) \ .
\end{equation*}
Indeed:
\begin{equation*}
 f(z|\mu I) 
 = f\bigl(x(z)|\theta(\mu) I\bigr)\,\Bigl|\frac{dx}{dz}\Bigr|
 = e^{-(z-\mu)}\exp{\bigl\{-e^{-(z-\mu)}\bigr\}} \ .
\end{equation*}

As a counter-example, i.e. as an example with the cdf of the sampling
distribution being neither strictly increasing nor decreasing with
respect to all of its parameters, we refer to the Weibull distribution
(\cite{ken}, \S\,5.33, pp.\,189-190) with the pdf of the form:
\begin{equation*}
 f(t|\theta\,\tau I) = \frac{\theta}{\tau}
 \Bigl(\frac{t}{\tau}\Bigr)^{\theta-1}
 \exp{\biggl\{-\Bigl(\frac{t}{\tau}\Bigr)^{\theta}\biggr\}} \ .
\end{equation*}
The range of the variate $t$, as well as the ranges of both 
parameters $\theta$ and $\tau$, coincide with the positive half of 
the real axis:
\begin{equation*}
 t,\tau,\theta\in(0,\infty) \ .
\end{equation*}
Let the value of the scale parameter $\tau$ of the distribution
be known and let $x$ be a normalized variate:
\begin{equation*}
x = \frac{t}{\tau} \ \ , \ \ x\in(0,\infty) \ ,
\end{equation*}
with the appropriate pdf,
\\ \parbox{0.915\linewidth}{
\begin{equation}
 f(x|\theta I) = f\bigl(t(x)|\theta\,\tau I\bigr)
 \,\Bigl|\frac{dt}{dx}\Bigr| =
 \theta x^{\theta-1}\exp\bigl(-x^{\theta}\bigr) \ ,
\label{eq:wbpdf}
\end{equation}
} \\
and cdf:
\begin{equation*}
 F(x,\theta) = 1-\exp\bigl(-x^{\theta}\bigr) \ . 
\end{equation*}
Clearly, the cdf is decreasing in $\theta$ for $x<1$, but increasing
in $\theta$ for $x>1$.
\par}

\section{Reduction to inference about location parameters under more
         general conditions}
\label{sec:reduction}

We can provide a consistent and calibrated parameter 
inference only when
the problem is reducible to estimation of a location parameter. 
But, as we saw in the case of the Weibull distribution,
reduction to a location parameter estimation is not always possible.
Nevertheless, under very general conditions, there is a neat way out
of such difficult situations.

\begin{figure}
\vskip -51mm
\hskip -35mm
\psfig{file=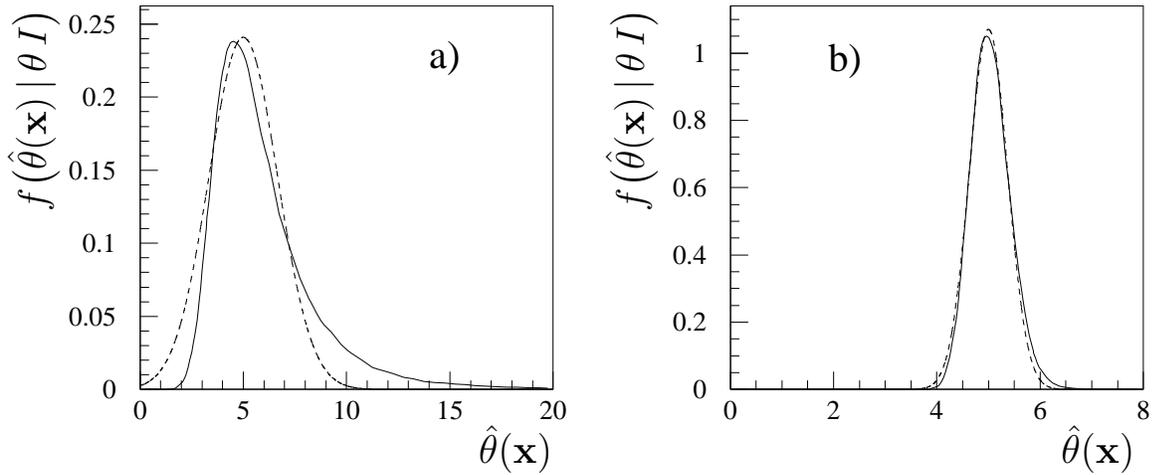,width=220mm}
\vskip -197mm
\label{fig:clt}
\caption{Sampling distribution for $\hat{\theta}(\bi{x})$ resulting from
         maximizations of the likelihood density $f(\bi{x}|\theta I)=
         \prod_{i=1}^{n}f(x_i|\theta I)$ that is a product of Weibull
         pdf's \eqref{eq:wbpdf} (continuous line) and the
         limiting Gaussian or normal distribution 
         $N\bigl(\mu=\theta,\sigma= (R(\theta)\sqrt{n})^{-1}\bigr)$ 
         (dashed line) for $\theta=5$ and for $n$ being a) 5 and b) 100.}
\end{figure}

When lacking any additional prior knowledge, all information about 
an inferred parameter $\theta$ of a specified sampling distribution 
that can be extracted from the measured events,
is contained in the value of the likelihood. After
collecting $n$ independent events $x_i$, the likelihood density
for $\bi{x}$ reads:
\begin{equation*}
 f\bigl(\bi{x}=(x_1,x_2,...,x_n)|\theta I) = 
 \prod_{i=1}^{n}f(x_i|\theta I) \ .
\end{equation*}
Let $\hat{\theta}$ be the value of the parameter that maximizes
the value of the likelihood density, given a particular set
of collected events $\bi{x}$:
\begin{equation*}
 \hat{\theta} = \hat{\theta}(x_1,x_2,...,x_n) = \hat{\theta}(\bi{x}) \ \ \ :
\ \ \ f(\bi{x}|\theta I)\big|_{\theta=\hat{\theta}} = \text{max.}
\end{equation*}
Suppose also that the integral
\begin{equation*}
 R^2(\theta) = \int\limits_x 
 \Bigl(\frac{\partial}{\partial\theta}\ln{f(x'|\theta I)}\Bigr)^2
 f(x'|\theta I)\,dx'
\end{equation*}
exists for every $\theta$ in its permissible range. Then, as the 
number $n$ of measurements $x_i$ increases, the sampling distribution
for $\hat{\theta}(\bi{x})$ converges to a Gaussian distribution with
$\mu=\theta$ and $\sigma(\theta)=\bigl(R(\theta)\sqrt{n}\bigr)^{-1}$
(see, for example, \cite{ken2}, \S\,18.10, pp.\,52-53 and 
\S\,18.16, pp.\,57-58). The width of the distribution decreases
with increasing $n$. For a sufficiently large number of measurements 
the width
can be approximated by $\sigma(\theta)\simeq\sigma(\hat{\theta})$,
\begin{equation*}
 \sigma(\hat{\theta}) = \frac{1}{R(\hat{\theta})\sqrt{n}} \ ,
\end{equation*}
leaving $\theta$ as a \textit{pure location parameter} of the distribution
for $\hat{\theta}(\bi{x})$,
\begin{equation*}
 f(\hat{\theta}|\theta I) = \frac{1}{\sqrt{2\pi}\sigma(\hat{\theta})}
 \exp{\Biggl\{-\frac{(\hat{\theta}-\theta)^2}{2\sigma^2(\hat{\theta})}
 \Biggr\}} 
 \ ,
\end{equation*}
thus allowing to make a consistent inference about $\theta$.
The required number of collected events depends on the precision
required for the inference about the parameter and on the
form of the sampling distribution (see Figure\;\ref{fig:clt} with
two examples for the Weibull distribution).

\section{Inferring parameters of counting experiments}
\label{sec:counting}

Let an urn contain balls that are identical in every respect except
that some of the balls are coloured white and the remaining ones
are coloured red. Such an urn is usually referred to as the
Bernoulli urn and drawing from the urn is referred to as a
Bernoulli trial. We draw a ball from the urn blindfolded, observe
and record its colour, put it back into the urn and thoroughly
shake the urn in order to minimize any possible correlations
between successive draws. Then we repeat the process until
$n_0$ balls have been drawn, out of which $n$ have been recorded
to be white ($0\le n\le n_0$). If correlations between the
draws can be neglected, the probability of recording $n$ white balls
in $n_0$ draws is given by the binomial distribution (\cite{ken}, 
\S\,5.2-5.7, pp.\,163-168):
\begin{equation*}
 p(n|\theta n_0 I) = {n_0\choose n}\,\theta^{n}\,(1-\theta)^{n_0-n} \ ,
\end{equation*}
where the parameter $\theta$ of the distribution coincides with
the fraction of the white balls in the urn.

On the basis of known $n$ and $n_0$ we would like to make a
probabilistic inference about the value of the parameter $\theta$. 
Lacking any additional prior information, we write the pdf
for $\theta$ in accordance with \eqref{eq:bayprime}:
\begin{equation*}
 f(\theta|n n_0 I) = \frac{\pi(\theta)\,p(n|\theta n_0 I)}
 {\int_0^1\pi(\theta')\,p(n|\theta' n_0 I)\,d\theta'} =
 \frac{\pi(\theta)}{\eta(n,n_0)}\,p(n|\theta n_0 I) \ ,
\end{equation*}
where the form of the consistency factor $\pi(\theta)$ is yet 
to be determined.

We showed that we are able to uniquely determine the form of 
a consistency factor 
only in presence of invariance of the sampling
distribution under a continuous (Lie) group of transformations. 
But in the case of counting experiments, with a
discrete variate $n$ of the sampling distribution,
such an invariance is evidently absent. 
In this way we loose
ground to uniquely determine the form of the consistency factor simply
by following Cox-P\'{o}lya-Jaynes Desiderata. If we insist 
on making a 'probabilistic' inference, we are therefore restricted
to using 'consistency factors', also referred to as 
\textit{non-informative priors}, whose forms are chosen
on the basis of some \textit{ad hoc} criteria. But no matter how
carefully these criteria and non-informative priors
are specified, there is no guarantee that in this way our 
reasoning remains consistent. Then, without being protected
by the Desiderata, we stand at the mercy of all kinds of paradoxes,
stemming from inconsistencies that we may unintentionally
have committed. 

As an example, imagine a large number of urns, each containing
an unknown fraction of white balls. We make a series of Bernoulli 
trials by drawing a single ball from each of the urns, i.e. 
$n_0=1$ for each of the draws. The outcome of drawing
from the $i$-th urn can be a white ($n=1$) or a red ($n=0$) ball 
and the corresponding likelihood reads:
\begin{equation*}
 p(n|n_0\theta_iI) = 
 \begin{cases}
  \ \ \ \ \theta_i \ \ \ \ \, ; \ \ n=1 \\
  1-\theta_i \ \ ; \ \ n=0 
 \end{cases}
 \ ,
\end{equation*}
where $\theta_i$ is the (unknown) fraction of white balls in
the $i$-th urn. Let us try to make a 'probabilistic' inference
about the parameter by using a uniform non-informative prior distribution
of 'probability' for $\theta_i$:
\begin{equation*}
 \text{'}f(\theta|I)\text{'} = 1 \ .
\end{equation*}
Then, a 'pdf' can be assigned to $\theta_i$ simply by using Bayes' theorem
\eqref{eq:bayes}:
\begin{equation}
\text{'}f(\theta_i|n n_0 I)\text{'} = 
\frac{\text{'}f(\theta_i|I)\text{'}\,p(n|n_0\theta_i I)}
 {\int_0^1 \text{'}f(\theta_i|I)\text{'}\,p(n|n_0\theta_i I) \,d\theta_i} =
 \begin{cases}
  \ \ \ \ \ 2\theta_i \ \ \ \ \ \ \, ; \ \ n=1 \\
  2(1-\theta_i) \ \ ; \ \ n=0 
 \end{cases}
 \ .
\label{eq:noncon}
\end{equation}

\begin{figure}[tb]
\vskip -38mm
\hskip -33mm
\psfig{file=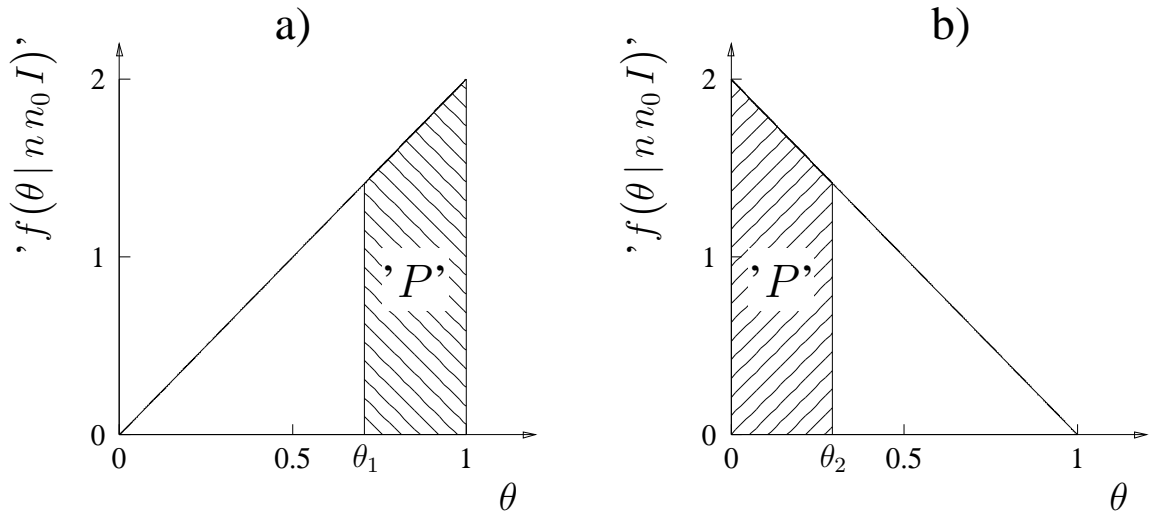,width=216mm}
\vskip -204mm
\label{fig:noncon}
\caption{'Probability density functions' assigned to the parameter of a
          binomial distribution when the number of draws is $n_0=1$
          and the result of drawing is either a) favourable ($n=1$)
          or b) non-favourable ($n=0$). The distributions are assigned 
          by choosing a uniform non-informative prior. The integrals of the 
          'densities', 'probabilities' '$P$' (hatched regions), on the
          intervals a) $(\theta_1=\sqrt{2}/2,\theta_2=1)$ and 
          b) $(\theta_1=0,\theta_2=1-\sqrt{2}/2)$ amount to 0.5}

\end{figure}

If \eqref{eq:noncon} is truly a pdf for $\theta$, then a 'probability'
\begin{equation*}
\text{'}P\bigl(\theta_i\in(\theta_1,\theta_2)|n n_0 I)\text{'} = 
\int_{\theta_1}^{\theta_2}
 \text{'}f(\theta_i|n n_0 I)\text{'} \,d\theta_i
\end{equation*}
should cover the true value of $\theta_i$ in 
$100\,\text{'}P\bigl(\theta_i\in(\theta_1,\theta_2)|n n_0 I)\text{'}$ 
per cent of the
inferences, but it is easy to see that such an inference is in 
general \textit{not} calibrated. Let us choose the shortest intervals 
with the 'probability' of containing $\theta_i$ being 50\% (see Figure
\ref{fig:noncon}):
\begin{equation*}
 (\theta_1,\theta_2) = 
 \begin{cases}
  \ \ \ \bigl(\frac{\sqrt{2}}{2},1\bigr) \ \ \ \ \ \, ; \ \ n = 1 \\
  \bigl(0,1-\frac{\sqrt{2}}{2}\bigr) \ \ ; \ \ n = 0 
 \end{cases}
 \ .
\end{equation*}
If, for example, the fraction of white balls in each of
the urns were exactly one-half, our intervals would \textit{never}
cover the true value, i.e. our inference is manifestly
non-calibrated. Non-informative priors of different forms,
for example
\begin{equation*}
 \text{'}f(\theta|I)\text{'} = \frac{1}{\theta(1-\theta)} \ \ \text{or}
 \ \ \text{'}f(\theta|I)\text{'} = \frac{1}{\sqrt{\theta(1-\theta)}}
\end{equation*}
(see \cite{jay}, \S\,12.4.3, p.\,384, eq.\,(12.50), and
\cite{har}, \S\,11.3, p.\,106, eq.\,(11.63)), are not immune to such
kind of problems, either.

As on two previous occasions, such an obstacle in the way of
consistent parameter inference can be overcome simply by collecting
more data, such that
\begin{equation}
\begin{split}
 n_0 - n &\gg 1 \ , \\ 
 n &\gg 1 \ .
\end{split}
\label{eq:dcond}
\end{equation}
When the above condition is fulfilled, the sampling distribution
for $x_n$,
\begin{equation*}
 x_n\equiv \frac{n}{n_0} \ ,
\end{equation*}
 converges to a Gaussian distribution $N(\mu,\sigma)$
(\cite{ken}, \S\,4.15, pp.\,138-140) with
\begin{equation*}
 \mu(\theta) = \theta \ \ \text{and} \ \ 
 \sigma(\theta) = \sqrt{\frac{\theta(1-\theta)}{n_0}}
\end{equation*}
in the sense that
\begin{equation*}
\begin{split}
 \sum_{n=n_1}^{n_2}p(n|n_0\theta I) &= \sum_{n=n_1}^{n_2}p(x_n|n_0\theta I) \\
 \simeq \int_{x_{n_1}-\frac{1}{2n_0}}^{x_{n_2}+\frac{1}{2n_0}} 
 f(x|\mu\sigma I)\,dx &=
 \int_{x_{n_1}-\frac{1}{2n_0}}^{x_{n_2}+\frac{1}{2n_0}}
 \frac{1}{\sqrt{2\pi\sigma}}\,
\exp\biggl\{-\frac{(x-\mu)^2}{2\sigma^2}\biggr\}\,dx\ .
\end{split}
\end{equation*}
The dispersion parameter $\sigma(\theta)$ of the limiting distribution
decreases with increasing number of draws from the urn and
for sufficiently large $n_0$ it can be approximated by
\begin{equation*}
 \sigma(\theta) \simeq \sigma(x_n) = \sqrt{\frac{x_n(1-x_n)}{n_0}} \ .
\end{equation*}
In this way, $\theta$ becomes a pure location parameter of a
Gaussian distribution with the corresponding uniform consistency
factor and with the corresponding pdf: 
\begin{equation*}
 f(\theta|x_nn_0I) = \frac{\pi(\theta)}{\eta(x_n)}\,p(x_n|n_0\theta I) =
 \frac{1}{\sqrt{2\pi}\sigma(x_n)}
 \,\exp\biggl\{-\frac{(x_n-\theta)^2}{2\sigma^2(x_n)}\biggr\} \ .
\end{equation*}

At this point we find it appropriate, for the sake of completeness, to make
a comment on the density of information $n n_0 I$ that probability 
$p(\theta|n n_0 I)$ is conditional upon. Recall that a continuity 
requirement concerning sets of different states of knowledge about
inferred propositions
was listed within the common sense
Desideratum\,\textit{II}. But with $I$ merely representing 
information about the type of counting sampling distribution,
and with $n_0=1$ being a fixed number of experiments performed,
the possible information about $\theta$ after the first trial consists
two \textit{atoms}, $n=0$ and $n=1$, that do \textit{not}
allow for such a requirement to be met. In this way the proof
of Cox's Theorem (see, for example, \cite{cox}, or
\cite{jay}, \S\,2.1-2.2, pp.\,24-45), stating that the basic
Desiderata necessarily imply (up to an isomorphic transformation)
the product rule \eqref{eq:product}, the sum rule \eqref{eq:sum},
and their corollary, Bayes' Theorem \eqref{eq:bayes}, 
misses an indispensable fact.
That is, in a situation like that there is no explicit reason
why an inference about a parameter of a sampling distribution
should be made in accordance with the Consistency Theorem
that is deduced by assuming, among other things, Bayes' Theorem to
be the only consistent way to update the probability distribution
for the inferred parameter
(recall Section\,\ref{sec:subjectivity}). However, in the 
dense limit \eqref{eq:dcond}, the continuity of information (i.e. 
of $n/n_0$) is recovered and the procedure of inference 
about $\theta$ becomes uniquely determined by the adopted
Desiderata.

Note that for a consistent and calibrated inference both conditions
of \eqref{eq:dcond} must be met. Suppose for a moment that only
\begin{equation*}
 n_0-n \gg 1
\end{equation*}
holds so that the binomial sampling distribution can be approximated by
the Poisson limit (\cite{ken}, \S\,5.8-5.9, pp.\,168-171):
\begin{equation*}
 p(n|\mu I) = \frac{\mu^n}{n!}\,e^{-\mu} \ ,
\end{equation*}
where the parameter $\mu$ of the distribution represents the
expected number of white balls drawn. Suppose that only
red balls are drawn, i.e. $n=0$, and that we want to make a
'probabilistic' inference about $\mu$ by choosing a uniform
non-informative prior. The corresponding 'pdf' for $\mu$
then reads:
\begin{equation*}
 \text{'}f(\mu|n=0\,I)\text{'}
 =\frac{\text{'}f(\mu|I)\text{'}\,p(n=0|\mu I)}
 {\int_0^{\infty}\text{'}f(\mu'|I)\text{'}\,p(n=0|\mu' I)\,d\mu'} 
 = e^{-\mu} \ ,
\end{equation*}
which implies a 'probability' for $\mu \le \mu_0$,
\begin{equation*}
 \text{'}P(\mu\le\mu_0|n I)\text{'} = \int_0^{\mu_0} 
\text{'}f(\mu|n=0\,I)\text{'}\,d\mu 
 = 1-e^{-\mu_0} \ ,
\end{equation*}
or, equivalently, implies a 'probability' for $\mu > \mu_0$,
\begin{equation*}
 \text{'}P(\mu > \mu_0|n I)\text{'} = 
 \int_{\mu_0}^{\infty}\text{'}f(\mu|n=0\,I)\text{'} \,d\mu 
 = e^{-\mu_0} \ ,
\end{equation*}
where $\mu_0$ takes an arbitrary positive value. For example,
for $\mu_0=3$, the two 'probabilities' equal $\sim 0.95$ and $\sim 0.05$, 
respectively.

Imagine now such draws from a number of urns $N$, all ending up
with $n_i = 0$. We can make use of the general sum rule \eqref{eq:gensum}
and calculate a 'probability', $\text{'}P_N\text{'}$, for at least one 
out of $N$ parameters $\mu_i$ being greater than $\mu_0$. Since the draws 
from different urns are not correlated, the 'probability' equals:
\begin{equation*}
 \text{'}P_N\text{'} 
  = 1 - \Bigl(1-\,\text{'}P(\mu > \mu_0|n I)\text{'}\Bigr)^N \ .
\end{equation*}
With increasing $N$, the value of $\text{'}P_N\text{'}$ approaches unity:
for a sufficiently large number of urns we can claim with 'certainty'
that at least for one of the urns the parameter $\mu_i$ is
greater than $\mu_0$, regardless of the chosen value of the latter.
But claiming the existence of white balls in the urns on the
basis of observing only red ones, is clearly a logically
unacceptable result, pointing to serious flaws of this kind
of inference. Note that it was \textit{not} the choice of the uniform 
non-informative prior that was decisive for the above result since 
every $\text{'}f(\mu|I)\text{'}$ with the existing integral
\begin{equation*}
 \int_0^{\infty}\text{'}f(\mu'|I)\text{'}\,p(n=0|\mu' I)\,d\mu'
\end{equation*}
would lead to the same kind of a logically unacceptable conclusion.
 
If an urn contains only red balls, the requirement
\begin{equation*}
 n \gg 1
\end{equation*}
can \textit{never} be met. Is there some kind of inference that
could be made in such cases? Let us perform two Bernoulli trials
by drawing $n_0$ and $2n_0$ balls from an urn, with all of the
drawn balls in both trials being red. The evidence against the 
presence of white balls in the urn that was obtained by the
second trial may be reasonably held to be stronger than the evidence 
from the first trial. Yet \textit{how much stronger}? It seems to
us that in such cases the degrees of belief, although still comparable,
cannot be expressed quantitatively (see also \cite{pol1}, Chapter\,XV, 
p.\,137), i.e. in order to avoid all sorts of paradoxes we should 
remain on a qualitative level.

\section{Historical digression}
\label{sec:history}

      In Section\,\ref{sec:subjectivity} we stressed that,
      contrary to the prior probability distribution in 
      Bayes' Theorem (\ref{eq:bayes}\,--\:\ref{eq:bayespdf}),
      the consistency factor in equations 
      (\ref{eq:bayprime}\,--\:\ref{eq:bayprime2}) 
      does \textit{not} represent any kind of probability
      distribution. Overlooking this fact led to perpetual
      \textit{philosophical} argument throughout the history
      of probabilistic reasoning, with far-reaching \textit{practical}
      consequences.

      A natural starting point for every sequential updating of
      information is a state of complete ignorance, e.g.
      complete prior ignorance about the value of an inferred
      parameter $\theta$ of a sampling distribution for $x$.
      The original sin is then committed in an attempt to
      make use of Bayes' Theorem, \eqref{eq:bayes} or
      \eqref{eq:bayespdf}, in such a situation. According
      to the Theorem, in order to obtain a probability distribution
      for the inferred parameter after observing $x_1$ both
      the likelihood, i.e. the probability for observing $x_1$ given
      a particular value of the parameter, and a prior probability 
      distribution for $\theta$, i.e. a distribution of our belief 
      in different values of the parameter prior to observing $x_1$,
      must be provided. With the known form of the sampling
      distribution, the calculation of the likelihood is a rather
      straightforward task. The problem arises when we try to
      formulate the distribution of our belief prior to the first recorded
      event, for up untill then we had been completely ignorant about 
      the possible 
      values of $\theta$. What we are trying to do is to assign
      a prior probability distribution based on ignorance,
      i.e. we are trying to establish the so-called \textit{ignorance}
      or \textit{non-informative probability distributions}. But
      according to the definition of probability adopted
      as a degree of reasonable belief that is based on relevant
      information at hand (see Section\,\ref{sec:intro}),
      a probability assigned on grounds of ignorance is simply
      a \textit{contradiction in terms}. A prior probability
      distribution that is based only on ignorance 
      thus \textit{cannot} be the realm of a consistent probability
      theory. We will see that apart from being self-contradicting
      on the conceptual level, the concept
      of ignorance or non-informative priors inevitably also produces 
      many practical inconsistencies and paradoxes.

      This delusion about probability distributions based 
      on ignorance has been present for more than two centuries.
      In a scholium to his essay \cite{bay} Bayes suggested that
      in the absence of all prior knowledge it is reasonable to 
      assume a uniform distribution for $p$, where $p$ stands
      for, for example, an unknown fraction of white balls in 
      an urn experiment as described in Section\,\ref{sec:counting}. 
      Laplace (\cite{lap1}, p.\,XVII) was also very explicit on the same
      subject:``When the probability of a single event is unknown we may
      suppose it equal to any value from zero to unity.'' 
      The above assumption is usually referred to as the Bayes principle
      (\cite{ken}, \S\,8.19, p.\,298), the Laplace principle
      of insufficient reason \cite{kas} or the principle of
      indifference (\cite{jay}, \S\,2.4, p.\,40). 
      When applied to estimation of 
      a general unknown parameter, the assumption would read:
      in the limit of complete prior ignorance about
      the value of an inferred parameter, the prior distribution
      should be uniform.

      We strongly disagree with such a rule and propose a very
      simple rule to replace it: \textit{when there is no (prior)
      information, no (prior) probabilities are to be assigned
      whatsoever}. Or to paraphrase de\,Finetti: \textit{prior
      probability does not exist}.
      Knowing that a uniform prior probability 
      distribution in the range $(\theta_a,\theta_b)$ 
      has been assigned to the value of a parameter $\theta$
      as a result of positive knowledge,
      and not knowing anything about $\theta$
      with the exception of its admissible range, are 
      \textit{two fundamentally different states of knowledge}. 

      In order to illustrate the difference,
      we return to drawing coloured balls from urns. Imagine having 
      a good number of urns, each containing red and white balls,
      with the fractions of white balls, $\theta_i$, in each of the
      urns being unknown. According to the first scenario,
      we \textit{possess information} that the fractions 
      are distributed uniformly in the range (0,1), while according
      to the second scenario we \textit{know nothing} about the 
      distribution of the values of $\theta_i$. It is easy to see that
      the two states of knowledge are completely different.

      In the first case it is possible to assign a probability
      for $\theta_i$ of the $i$-th urn which is in an arbitrary
      interval $(\theta_1,\theta_2) \subset [0,1]$ even before drawing
      the first ball from the urn simply by integrating
      the (uniform) prior probability distribution:
       \begin{equation}
         P\bigl(\theta_i\in(\theta_1,\theta_2)|I\bigr) =
         \int_{\theta_1}^{\theta_2}f(\theta|I)\,d\theta =
         \int_{\theta_1}^{\theta_2}d\theta =\theta_2-\theta_1 \ .
       \label{eq:0draw}
       \end{equation}    
      On making the first draw from the $i$-th urn, it can be either
      a white ($n_i=1$) or a red ($n_i=0$), implying the corresponding
      likelihood to be:
      \begin{equation*}
        p(n_i|\theta_i I) = 
        \begin{cases} \ \ \ \ \ \theta_i \ \ \ \ \ \, \ ;\ \  n_i=1 \\
                      (1-\theta_i) \ \ ;\ \  n_i = 0 \ ,
        \end{cases} 
      \end{equation*}
      The updated posterior probability therefore reads:
      \begin{equation}
       P\bigl(\theta_i\in(\theta_1,\theta_2)|n_iI\bigr)= 
       \begin{cases} \ \ \ \ \ \ \ \ \ \ \ \,
        \theta_2^2-\theta_1^2 \ \ \ \ \ \ \ \ \ \ \ \ \ ; \ \ n_i=1 \\
        (1-\theta_1)^2 - (1-\theta_2)^2 \ \ ; \ \ n_i=0 \ ,
       \end{cases}
      \label{eq:1draw}
      \end{equation}
     where the update was made by using Bayes' Theorem 
     \eqref{eq:bayes}. Note that both probabilities  
     are calibrated: when repeating the above assessments for all
     of the urns, intervals $(\theta_1,\theta_2)$ contain the true
     fractions of white balls in the each of the urns 
     in $100P\bigl(\theta_i\in(\theta_1,\theta_2)|I\bigr)$ percent of
     cases before any of the balls were drawn, while after
     drawing the coverage is exactly 
     $P\bigl(\theta_i\in(\theta_1,\theta_2)|n_iI\bigr)$.
         
     Within the second scenario, with complete prior ignorance about
     the distribution of $\theta_i$ within the urns, we \textit{cannot} 
     make a probabilistic inference about fraction of white balls 
     in a particular urn before any of the balls is drawn. The statement that 
     the probability for $\theta_i$ to be in the interval $(\theta_1,\theta_2)$
     equals the length of the interval (see eq. \eqref{eq:0draw}) 
     would in general be uncalibrated
     unless the distribution of $\theta_i$ within the urns is truly
     uniform - but this we do \textit{not} know and need \textit{not}
     be true. The same holds for the probability statement
     \eqref{eq:1draw} after drawing one ball from each of the urns.
     We saw in the previous section that in the limit of complete 
     prior ignorance, consistent and calibrated probability statements 
     about the parameters $\theta$ can be made only after drawing enough
     balls from each of the urns so that the Gaussian limit of the
     binomial sampling distribution can be applied.
 
     Many failed to recognize the fundamental difference between 
     \textit{knowing} the prior probability 
      distribution of an inferred parameter to be uniform
      and \textit{not knowing} anything about the value
      of the parameter. Harold Jeffreys, for example,
     wrote (\cite{jef}, \S\,1.22, p.\,29):
     ``If there is originally no ground to believe one of a set of
     alternatives rather than another, their prior probabilities are equal.''
     The same standpoint was persistently advocated also by Edwin
     Jaynes (\cite{jay}, \S\,18.11.1, p.\,573):
     ``Before we can use the principle of indifference to assign 
     numerical values of probabilities, there are two different
     conditions that must be satisfied: (1) we must be able
     to analyze the situation into mutually exclusive, exhaustive
     possibilities; (2) having done this, we must then find 
     available information gives us no reason to prefer any of
     the possibilities to any other. In practice, these conditions
     are hardly ever met unless there is some evident element
     of symmetry in the problem. But there are two entirely
     different ways in which condition (2) might be satisfied.
     It might be satisfied as a result of ignorance, or it might
     be satisfied as a result of positive knowledge about 
     the situation.''\footnote{A very similar idea expressed in 
     very similar words by Anthony O'Hagan can be found in 
     \cite{hag}, \S\,4.39, p.\,112.}

     We could not agree more with Jaynes if it were not for the
     last sentence. Take, for example, two different samples of 
     radionuclides. For the first sample we do not know anything about
     the isotopes except that the permissible range of their expected
     decay times is in an interval $(\tau_a,\tau_b)$, while for the 
     second sample the isotopes were chosen in such a way that the 
     distribution of their expected decay times can be well approximated 
     by a uniform distribution in the same interval.
     As a result of ignorance, we cannot make inferences about 
     the expected decay times $\tau_i$ of the isotopes in the first
     sample before measuring their actual decay times $t_i$.
     After the decay time $t_i$ of the $i$-th isotope is measured, 
     however, we can make a probabilistic statement about $\tau_i$ 
     according to \eqref{eq:bayprime1}:
     \begin{equation*}
     f(\tau_i|t_iI) = \frac{\pi(\tau_i)}{\eta(t_i)}\,f(t_i|\tau_iI)
                    = \frac{t_i}{\tau_i}\,f(t_i|\tau_iI) 
                    = \frac{t_i}{\tau^2_i}\,e^{-t_i/\tau_i} \ ,
     \end{equation*}
     where $I$ stands for prior ignorance about the expected 
     decay time of a particular radionuclide, while the consistency 
     factor $\pi(\tau_i)$ was chosen according \eqref{eq:cflsd}.
     From Section\,\ref{sec:unique} we recall that our inference 
     is consistent
     if the integral of the above pdf outside the admissible range
     for $\tau_i$ is small compared to the precision required.
     For the second sample of isotopes, as a result of positive knowledge,
     there is a pdf for the expected decay time of each of the isotopes
     at our disposal even prior to measuring the actual decay times:
     \begin{equation*}
      f(\tau_i|I') = 
      \begin{cases}
       (\tau_b-\tau_a)^{-1} \ \ ; \ \ \tau_a < \tau_i < \tau_b \\
       \ \ \ \ \ \ \ 0 \ \ \ \ \ \ \ \ \ \,\,\ ; \ \ \text{otherwise} \ \ ,
      \end{cases}
     \end{equation*}
     where $I'$ stands for positive prior knowledge. After $t_i$ is
     measured (and \textit{not} knowing the results of other possible 
     measurements of decay times from the same sample), 
     the distribution of our belief can be updated by means of Bayes'
     theorem \eqref{eq:bayes1}:
     \begin{equation*}
     f(\tau_i|t_iI') = \frac{f(\tau_i|I')\,f(t_i|\tau_iI')}
        {\int_{\tau_a}^{\tau_b}f(\tau'_i|I')\,f(t_i|\tau'_iI')\,d\tau'_i} \ .
     \end{equation*}
     Note that when appropriate prior pdf's exist, the inferences
     are always consistent and calibrated. Figure\,5
     shows examples of inferences with and without existing 
     prior probability distributions.
\begin{figure}[tb]
\vskip -38mm
\hskip -31mm
\psfig{file=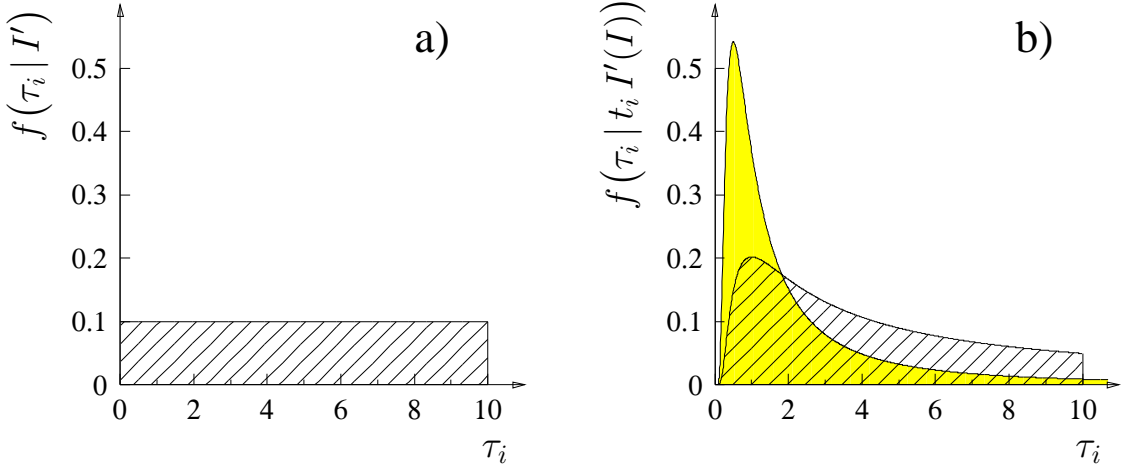,width=210mm}
\vskip -198mm
\caption{ Probability density functions for the expected decay time
           $\tau_i$ of the $i$-th isotope from a sample of isotopes
           with the expected decay times distributed approximately
           uniformly in an admissible range 
           $(\tau_a,\tau_b)=(0,10)$ (hatched areas),
           and the corresponding pdf for an isotope from a sample
           with the expected decay times in the same interval but
           with an unknown distribution of their actual values
           (shaded area). Figure a) displays the pdf before
           any of the actual decay times are measured, while Figure
           b) shows appropriate pdf's based on a measured
           value $t_i=1$ in both cases. All decay times are given
           in arbitrary but equal units. While the hatched histograms 
           are limited within the permissible range, the integral
           of the shaded pdf in the range $(10,\infty)$ is  
           approximately $\epsilon_2=0.095$.}
\end{figure}

      Apart from the problems with calibration, the Bayes postulate contains
      also some insurmountable inconsistencies. Suppose we are
      estimating a parameter $\theta$ with no prior information
      available. According to the postulate, the prior 'probability'
      distribution in such a case is given by a uniform 'pdf':
      \begin{equation*}
        \text{'}f(\theta|I)\text{'} = 1.
      \end{equation*}
      But, instead of $\theta$, we could have equally well chosen
      different parameterization, say $\lambda(\theta)$, where
      \begin{equation}
      \begin{CD}
       \theta @>>> \lambda
      \end{CD}
      \label{eq:bitl}
      \end{equation}
      is a bijective parameter transformation. Then, due 
      to the consistency Desideratum\,\textit{III.c},
      the appropriate prior 'pdf' for the transformed variate 
      $\lambda$, obtained according to \eqref{eq:vartr1}, reads:
      \begin{equation}
        \text{'}f(\lambda|I)\text{'} = 
        \text{'}f\bigl(\theta(\lambda)|I\bigr)\text{'}\,
        \Bigl|\frac{d\theta}{d\lambda}\Bigr| = 
        \Bigl|\frac{d\theta}{d\lambda}\Bigr| \ .
      \label{eq:noninfpt}
      \end{equation}
      In the case of a non-linear transformation \eqref{eq:bitl},
      the absolute value of the derivative in \eqref{eq:noninfpt}
      is not a constant, i.e. the prior 'pdf' for $\lambda$ is 
      \textit{not} uniform. But since a one-to-one mathematical
      transformation like \eqref{eq:bitl} does \textit{not}
      change the state of knowledge about the inferred parameters,
      we also remain completely ignorant about $\lambda$. The Bayes
      postulate would therefore imply a uniform prior distribution
      for $\lambda$, which obviously contradicts \eqref{eq:noninfpt}.
      That is, \textit{Bayes postulate in general contradicts
      consistency Desideratum\,III.c)}.

      A sophisticated version of the principle of insufficient
      reason is referred to as \textit{the principle of maximum
      entropy}. The term \textit{information entropy}
      was introduced by Claude Shannon (\cite{sha}, \S\,6) as
\begin{equation*}
 S\equiv -\sum_{i}p_i\,\ln{p_i} \ ,
\end{equation*}
where $p_i$ is the probability assigned to inferred parameter $\theta$
of taking a value within an interval $(\theta_i,\theta_i+d\theta_i)$,
and the sum covers the whole admissible range of $\theta$.
In order for it to be used for determination of non-informative 
prior distributions,
$p_i$ is interpreted as 
\begin{equation*}
 p_i\equiv \text{'}p(\theta_i|I)\text{'} = 
           \text{'}f(\theta_i|I)\text{'}\,d\theta_i \ ,
\end{equation*}
where $\text{'}f(\theta_i|I)\text{'}$ stands for the non-informative
prior 'pdf' for $\theta$.
The principle of maximum entropy then states (\cite{jay}, \S\,11.3, 
pp.\,350) that \textit{the function $\text{'}p(\theta_i|I)\text{'}$ 
which maximizes entropy
represents the most honest description of what we know about
the value of the inferred parameter}.

Suppose, for example, that we know only that $\theta\in(\theta_a,\theta_b)$ 
is a parameter of a specified sampling distribution, and that the
prior 'probability' distribution for $\theta$ is subject to 
the usual constraint
\begin{equation}
 \sum_i\text{'}p(\theta_i|I)\text{'}= \sum_i p_i = 1 \ .
\label{eq:constraint}
\end{equation}
Then, the principle of maximum entropy and the above constraint are
taken simultaneously into account if $p_j$
maximizes the function $H$,
\begin{equation*}
 H\equiv S - \alpha\Bigl(\sum_i p_i - 1\Bigr) = 
  -\sum_{i}p_i\,\ln{p_i}- \alpha\Bigl(\sum_i p_i - 1\Bigr) \ ,
\end{equation*}
that is, if
\begin{equation*}
 \frac{\partial H}{\partial p_j} = -(\ln{p_j}+1+\alpha) = 0 \ ,
\end{equation*}
where $\alpha$ is a Lagrange multiplier. It is therefore the constant
non-informative prior 'probability' distribution
\begin{equation*}
 p_j=\text{'}f(\theta_j|I)\text{'}\,d\theta_j = e^{-(1+\alpha)} = 
 \text{constant}
\end{equation*}
that maximizes the information entropy,
subject to the usual normalization condition \eqref{eq:constraint}.
If we choose intervals of uniform width
\begin{equation*}
 d\theta_i = \text{constant} \ ,
\end{equation*}
the principle of maximum entropy yields a uniform non-informative
prior 'pdf'. But a non-linear one-to-one re-parameterization
\begin{equation*}
\begin{CD}
 \theta  @>>> \lambda(\theta)  
\end{CD}
\end{equation*}
implies widths of intervals
\begin{equation*}
 d\lambda_i = 
 \Bigl|\frac{\partial \lambda}{\partial \theta}\Bigr|_{\theta=\theta_i}\,
 d\theta_i
\end{equation*}
and the corresponding 'pdf'
\begin{equation*}
 \text{'}f(\lambda_i|I)\text{'} = \text{'}f(\theta_i|I)\text{'}\,
 \Bigl|\frac{\partial \lambda}{\partial \theta}\Bigr|^{-1}_{\theta=\theta_i}
 \propto
 \Bigl|\frac{\partial \lambda}{\partial \theta}\Bigr|^{-1}_{\theta=\theta_i} 
 \ , 
\end{equation*}
neither of the two being uniform. And so an immediate question can be raised
which (and why) is the distinguished parameterization of the sampling
distribution with both the interval widths and the non-informative
prior 'pdf' being uniform. In other words, this means that the 
principle of maximum entropy \textit{cannot} solve
the ancient ambiguity of how to find the elusive non-informative 
'probability' distributions. 

      Jeffreys (\cite{jef}, \S\,3.1) proposed the following
      solution to the problem of non-informative priors. He suggested 
      that for parameters
      with the admissible range coinciding with the whole real
      axis we should keep to Bayes postulate, i.e. to the
      uniform priors, while for parameters known to be positive
      the proper way to express complete ignorance is to assign
      uniform prior probability to its logarithm, i.e. in the latter
      case the prior 'pdf' should be:
      \begin{equation}
       \text{'}f(\theta|I)\text{'} 
        \propto \frac{1}{\theta} \ \ ; \ \ \theta\in(0,\infty) \ .
      \label{eq:jeffpr}
      \end{equation}
      He tried to justify this on the grounds of invariance of the 
      former prior under translation,
      \begin{equation*}
       \lambda = \theta + b \ \ ; \ \ b\in(-\infty,\infty) \ ,
      \end{equation*}
      and on the grounds on invariance of the latter prior on
      raising $\theta$ to a power of $n$ or to scaling it by a
      positive constant $a$:
      \begin{equation*}
       \lambda = \theta^n \ \ \text{and} \ \ \lambda = a\theta \ .
      \end{equation*}

      Why invariance of the non-informative
      priors under these particular transformations? For the location, 
      scale and dispersion parameters the answer was extensively
      searched for by utilizing related transformation groups
      (see refs. \cite{vil}, \cite{kas}, \cite{har} and Chapter 12 
      in \cite{jay}). 

      For example, when we are 
      completely ignorant about location $\mu$ of a distribution for 
      $x$ prior to starting to collect data, a mere shift of location,
      \begin{equation}
       \begin{CD}
        \mu @>>> \mu' = \mu + b \ ,
       \end{CD}
      \label{eq:deftransloc}
      \end{equation}
      could not change our state of knowledge. That is,
      according to \textit{the principle of group invariance}, 
      the non-informative prior 'probability' distribution
      for $\mu$ should be invariant under a group of transformations
      \eqref{eq:deftransloc}. It is easy to see that the uniform prior is 
      the only one that satisfies this condition. 

      Similarly, when we are completely ignorant about
      the scale $\tau$ of a distribution for $t$, the prior 
      'probability' distribution for $\tau$
      should be invariant under a scale transformation,
      \begin{equation}
      \begin{CD}
       \tau @>>> \tau' = a\tau \ .
      \end{CD}
      \label{eq:deftranssc}
      \end{equation}
      The only non-informative prior 'probability' distribution satisfying
      the above condition, is expressed by Jeffreys' prior 
      \eqref{eq:jeffpr}. To see that, we should realize that the required
      invariance implies equality for prior 'probabilities'
\begin{equation*}
 \text{'}p(\tau|I)\text{'} = \text{'}f(\tau|I)\text{'}\,d\tau =
        \phi(\tau)\,d\tau
\end{equation*}
and
\begin{equation*}
 \text{'}p(\tau'|I)\text{'} = \text{'}f(\tau'|I)\text{'}\,d\tau' =
        \widetilde{\phi}(\tau')\,d\tau' = \widetilde{\phi}(a\tau)\,a\,d\tau \ ,
\end{equation*}
or, equivalently,
\begin{equation*}
 \phi(\tau) = \widetilde{\phi}(a\tau)\,a \ .
\end{equation*}
If this is to be true for any $a\in(0,\infty)$, it must also be true
for $a = \tau^{-1}$. Hence:
\begin{equation*}
 \text{'}f(\tau'|I)\text{'} = \phi(\tau) 
  = \widetilde{\phi}(1)\,\frac{1}{\tau} \ ,
\end{equation*}
and the above statement about Jeffreys' prior is proved.

      In the same spirit, a non-informative prior 'probability'
      distribution 
      representing complete ignorance about both location $\mu$ and scale
      $\sigma$ should be invariant under simultaneous location and scale 
      transformations:
      \begin{equation}
      \begin{split}
       \begin{CD}
       &\mu @>>> \mu' = a\mu + b \ , \\
       &\sigma @>>> \hskip -7mm \sigma' = a\sigma \ .
       \end{CD}
      \end{split}
      \label{eq:deftranslocsc}
      \end{equation}
      Following the idea of the proof from the preceding example leads to 
      a non-informative 'pdf' of the form:
      \begin{equation}
       \text{'}f(\mu\sigma|I)\text{'} = \frac{1}{\sigma^2} \ .
      \label{eq:jeffpr1}
      \end{equation}

      However, strong objections can be raised against
      the form of \eqref{eq:jeffpr1}. For example, 
      non-informative prior \eqref{eq:jeffpr1} leads to a 
      marginalization paradox (see, for example, references
      \cite{sto} and 
      \cite{daw}, and Chapter\,10, pp.\,81-94 of reference \cite{har}).
      Let $\mu$ be a location
      parameter with known fixed value, i.e. we are inferring
      only a dispersion parameter $\sigma$. By inserting the non-informative
      Jeffreys' prior \eqref{eq:jeffpr} in the Bayes' Theorem
      \eqref{eq:bayespdf} we obtain:
      \begin{equation}
      \text{'}f(\sigma|\mu\bar{x}s I)\text{'} =
      \frac{\text{'}f(\sigma|I)\text{'}\,f(\bar{x}s|\mu\sigma I)}
           {\int_0^{\infty}
           \,\text{'}f(\sigma'|I)\text{'}\,f(\bar{x}s|\mu\sigma' I)\,d\sigma'} 
      \propto\frac{1}{\sigma}\,f(\bar{x}s|\mu\sigma I) \ .
      \label{eq:marpar1}
      \end{equation}
      Note that the result is identical to the one obtained by
      inserting the appropriate consistency factor \eqref{eq:cflsd} 
      into the Consistency Theorem \eqref{eq:bayprime1}.
      But we can also obtain
      a 'pdf' for $\sigma$ given $\mu$, $\bar{x}$ and $s$ in
      another way. We start with a simultaneous inference
      about unknown parameters $\mu$ and $\sigma$:
      \begin{equation*}
      \text{'}f(\mu\sigma|\bar{x}s I)\text{'} =
      \frac{\text{'}f(\mu\sigma|I)\text{'}\,f(\bar{x}s|\mu\sigma I)}
           {\int_{-\infty}^{\infty}d\mu'\int_0^{\infty}d\sigma'\;
           \text{'}f(\mu'\sigma'|I)\text{'}\,f(\bar{x}s|\mu'\sigma' I)} 
      \propto\frac{1}{\sigma^2}\,f(\bar{x}s|\mu\sigma I) \ .
      \end{equation*}
      Then we make use of the product rule \eqref{eq:product1} and
      rewrite the above expression as:
      \begin{equation*}
       \text{'}f(\mu\sigma|\bar{x}s I)\text{'} = 
      \text{'}f(\sigma|\mu\bar{x}s I)\text{'}\:
       \text{'}f(\mu|\bar{x}sI)\text{'} \ ,
      \end{equation*}
      where $\text{'}f(\mu|\bar{x}sI)\text{'}$ is marginalized 'pdf' 
      (see eq.\,\eqref{eq:marginal}):
      \begin{equation*}
       \text{'}f(\mu|\bar{x}sI)\text{'} = 
       \int_{0}^{\infty}\text{'}f(\mu\sigma'|\bar{x}s I)\text{'}
       \,d\sigma' \ .
      \end{equation*}
      By combining the last three equations 
      we can therefore write down the 'pdf' for $\sigma$ given $\mu$
      and observed $\bar{x}$ and $s$ also as:
      \begin{equation}
      \text{'}f(\sigma|\mu\bar{x}s I)\text{'} =
      \frac{\text{'}f(\mu\sigma|\bar{x}s I)\text{'}}
      {\text{'}f(\mu|\bar{x}sI)\text{'}} 
      \propto\frac{1}{\sigma^2}\,f(\bar{x}s|\mu\sigma I) \ .
      \label{eq:marpar2}
      \end{equation}
      The pdf's \eqref{eq:marpar1} and \eqref{eq:marpar2} are in an
      obvious contradiction, i.e. the non-informative priors
      \eqref{eq:jeffpr} and \eqref{eq:jeffpr1} are inconsistent. For 
      further discussion about the marginalization paradox see Appendix\,B.

      When recognizing this fundamental difficulty, some
      authors claim the procedure of marginalization \eqref{eq:marginal}
      to be illegitimate (see, for example, ref.\;\cite{har}, 
      \S\,10, pp.\,81-94 and \S\,17.2, pp.\,163-164), despite the
      fact that - according to Cox's Theorem - 
      the procedure is implied by the basic Desiderata. Others
      apply transformations different from \eqref{eq:deftranslocsc}
      in attempts to solve the problem (see, for example, \cite{jay},
      \S\,12.4, p.\,378, equation (12.18)). But for the latter,
      since the modified transformations do \textit{not} correspond to 
      a simultaneous translation and scale transformation, the original 
      motivation to relate the complete prior ignorance about location, 
      scale and dispersion parameters to invariance of the corresponding 
      non-informative priors with respect to the transformations
      (\ref{eq:deftransloc}-\ref{eq:deftranslocsc}), is definitely lost.

Note that the formalism related to the principle of group invariance
of non-informative priors is remarkably similar to the formalism
applied for determination of consistency factors. 
But no matter how strong this similarity may appear at first glance,
there is a fundamental difference between the two methods, leading
to substantially different results. Group invariance of non-informative
prior probabilities is imposed as a principle, additional to the 
basic Desiderata,
while we made use of invariance of likelihoods (being
well defined probabilities) as a necessary condition for
equivalence of two states of knowledge that led to functional
equation \eqref{eq:funeqcon} for consistency factors. In the 
latter case it is the ratio of $\pi(\theta)\,d\theta$ and 
$\eta(x)\,dx$ that is to be invariant
under simultaneous transformations $\bar{g}_a(\theta)\in\bar{\mathcal{G}}$
and $g_a(x)\in\mathcal{G}$ of the inferred parameter $\theta$
and the sampling variate $x$, while $\pi(\theta)\,d\theta$ itself
need \textit{not} be invariant under $\bar{\mathcal{G}}$ since 
it is, by definition, uniquely determined only up to a multiplicative 
constant;
and it is due to this degree of freedom that marginalization
paradoxes, stemming from strict applications of the principle
of group invariance, are avoided.

In practice, the difference between the probability distribution
assigned simultaneously to a location and a dispersion parameter
by multiplying the appropriate likelihood by the corresponding
consistency factor \eqref{eq:cfmusigma}, and the 'probability' 
distribution assigned by utilizing the non-informative prior
\eqref{eq:jeffpr1}, will fade away with increasing number of collected 
events. So, from a pragmatic standpoint, arguments about which function
correctly expresses a state of complete prior ignorance might 
amount to quibbling over pretty small peanuts (\cite{jay}, \S\,6.15,
p.\,183). But, from a standpoint of principle, this is definitely
not true, for the convergence of the limiting distributions by itself 
certainly does not guarantee either of the two to be correct, i.e. we might
have been completely wrong in both cases. Fundamental difficulties 
with non-informative priors introduced very serious consequences
for inductive reasoning. For example, applications of the
Laplace principle of insufficient reason that led to logically 
unacceptable results, provided P\'{o}lya (\cite{pol1}, Chapter\;XV,
\S\,6, pp.\,133-136) with the main reason to persist in a qualitative level
when linking induction with probability. Others (see, for example,
references \cite{fis1} and \cite{edw}) refused even a qualitative 
correspondence. As noted already by Jeffreys (\cite{jef}, 
\S\,3.1, p.\,120), ``a succession of authors have said that the prior 
probability is nonsense and therefore that the principle of 
inverse probability, which cannot work without it, is nonsense too.''

In this way two, at first glance fundamentally distinct, schools
of inductive reasoning emerged. The first one, usually referred to
as the \textit{Bayesian school} due to the central role of the Bayes'
theorem in the process of inference, recognizes probability
as a degree of reasonable belief and applies probability theory
in the course of inductive reasoning. The second one, usually
referred to as the \textit{frequentist school} due to its strict
frequency interpretation of probability, advocates the usage
of the calculus of probability only for treatment of 
so-called \textit{random phenomena}. The aim of the frequentist school 
is to avoid the supposed mistakes and inconsistencies of the probabilistic
inductive inference,
so they relegate the problems of inductive inference, e.g. estimations
of distribution parameters, to a new field, \textit{statistical inference}.
Lacking applications of probability theory may, however, represent a serious
drawback when making inductive inferences. For example, the Fisher-Behrens
problem, introduced in Section\,\ref{sec:locscale}, may become an 
insurmountable obstacle outside the probabilistic parameter inference
(see, for example, \cite{ken2}, \S\,19.47, Example\,19.10, pp.\,160-162 
 and \S\,26.28-26.29, pp.\,441-442). In particular, difficulties stem from
the illegitimacy of the marginalization procedure within the theory 
of statistical inference.

Some of the substitutes for the calculus of probability that are proposed
within the framework of statistical inference, are put forward as solutions of
specific problems, such as the principles of least squares and minimum
chi squared. The \textit{principle of Maximum Likelihood} \cite{fis1},
however, is usually advocated as one of general application
(see \cite{ken}, \S\,8.22-8.27, pp.\,300-304, and \cite{ken2},
Chapter\,18, pp.\,46-104). The principle states that, when confronted
with a choice of the values of a parameter $\theta\in(\theta_a,\theta_b)$,
we choose the particular value $\hat{\theta}$ which maximizes the 
corresponding likelihood,
$p(x|\theta I)$, for the observed data $x$. In general, the principle
\textit{contradicts} our basic Desiderata. Imagine a problem of 
inference when there is, apart from $x$, some additional information
$I$ at hand that would allow for an assignment of 
probability distribution for $\theta$ on its own. Neglecting this 
information directly contradicts Desideratum \textit{III.b}. For special
cases, when the pdf for $\theta$ based on prior information $I$, 
$f(\theta|I)$, is uniform in a wide interval around $\hat{\theta}$, 
the contradiction is removed and the principle can no longer 
be disputed. 

When there is no prior knowledge about the value of a particular
inferred parameter, according to the consistency Desiderata,
the inference should be made on the basis of both the likelihood
containing the information about $\theta$ from the measurement $x$,
and the consistency factor containing information about $\theta$
coming from the known form of the sampling distribution for $x$.
Ignoring the latter in general implies our reasoning to be
inconsistent, i.e. to be in a direct contradiction with Desideratum
\textit{III.c}. Imagine, for example, two persons inferring average
decay times $\tau_i$ of unknown particles on the basis of single decay time
measurements $t_i$. Mr.\,A proceeds according to the Maximum Likelihood 
principle by extracting values
\begin{equation*}
 \hat{\tau}_i^{\rm A} = t_i 
\end{equation*}
of the parameter that maximizes the likelihoods
\begin{equation*}
 p(t_i|\tau_i I) = f(t_i|\tau_i I)\,dt
                 = \frac{1}{\tau_i}\,e^{-t_i/\tau_i}\,dt
\end{equation*}
for observing measured decay times in an interval $(t_i,t_i+dt)$,
given particular $\tau_i$'s. In accordance with the adhered principle,
$\hat{\tau}_i^{\rm A}$ should be the value of the parameter being indicated
by the datum $t_i$ as the strongest candidate, i.e. in the long term
the fraction of Mr.\,A's confidence intervals, ($\hat{\tau}_i^{\rm A}$, 
$\hat{\tau}_i^{\rm A}+d\tau$), covering the true values of inferred
parameters, should be larger than the corresponding fractions of any
other interval of the same width, $d\tau$. Mr.\,B, on the other hand,
chooses the consistent procedure: he extracts the values 
\begin{equation*}
 \hat{\tau}_i^{\rm B} = \frac{t_i}{2} 
\end{equation*}
of the parameters that maximize the pdf's for $\tau_i$
(see equations \eqref{eq:bayprime1} and \eqref{eq:cflsd}), 
\begin{equation*}
 f(\tau_i|t_i I) = \frac{\pi(\tau_i)}{\eta(t_i)}\,f(t_i|\tau_i I) =
 \frac{t_i}{\tau^2_i}\,e^{-t_i/\tau_i} \ ,
\end{equation*}
taking this way into account both information from the immediate data $t_i$,
and information $I$ about the form of the sampling distribution that is
contained in the consistency factor $\pi(\tau_i)$. It is easy to
see that, contrary to the claims of Mr.\,A, the coverage of Mr.\,B's
confidence intervals for $\tau_i$, ($\hat{\tau}_i^{\rm B}$,
$\hat{\tau}_i^{\rm B}+d\tau$), surpasses that of Mr.\,A's intervals,
($\hat{\tau}_i^{\rm A}$, $\hat{\tau}_i^{\rm A}+d\tau$), by a factor
of $4\,e^{-1} \simeq 1.47$.

Again, for special cases when inferring explicit
location parameters, the consistency of the principle of Maximum 
Likelihood is restored.

      Last but not least, in order to avoid unnecessary though frequent
      misunderstandings, we would like to clarify the following. Most 
      of the advocates of the so-called subjective Bayesian school of 
      thought
      find both the existence of complete prior ignorance
      about the inferred parameter (\cite{hag}, \S\,4.15, p.\,102), 
      and the existence of an exact equivalence of information 
      possessed by two different persons inferring the same parameter
      (\cite{hag}, \S\,1.16, p.\,11), impossible and thus irrelevant
      to a theory of inductive reasoning. We believe that such 
      statements are as poorly grounded as it would have been absurd, 
      for example,
      rejecting the use of right angled and similar triangles when constructing
      images within geometrical optics just due to the fact that
      no real triangle is exactly right angled and that no two real
      triangles are exactly similar.
      In Section\,\ref{sec:intro} we demonstrated that a particular state 
      of knowledge is \textit{exactly the same} as the state after an
      arbitrary one-to-one variate transformation, while
      in Section\,\ref{sec:objectivity} we stressed 
      that ignorance is just a
      limiting state of knowledge and the natural starting
      point in any actual inference, just as zero is the natural
      starting point in adding a column of numbers.

      Similarly, it might also have been argued for experiments like
      drawing balls from urns, tossing dice, or collecting the decay 
      times of unstable particles, are oversimplified and therefore
      not adequate for calibration of methods derived for 
      the real scientific inferences about unknown parameters. But such 
      simple (\textit{not} oversimplified) experiments usually serve as
      a paradigm for all experiments with well known sampling
      distributions of the collected data, and with well controlled
      experimental conditions, both features being among the basic
      assumptions of our theory (see 
      Sections\,\ref{sec:ignorance}\:and\:\ref{sec:objectivity}).
 
\section{On the probability of general hypotheses}
\label{sec:probhyp}
\hfill \parbox{0.48\linewidth}{\small
\noindent
Theories are nets: only he who casts will catch.
\\ \vskip -2mm
\hfill
Friedrich von Hardenburg (Novalis)
} \\ \vskip 2mm

Inference about a parameter, say $\theta$, of a distribution for a 
certain sampling variate, say $x$, is, by definition,
always conditional on a specified model (recall Section\,\ref{sec:ignorance}).
The value of a parameter is always estimated under the assumption
that within the specified family of distributions, with each member of the
family being completely determined by the value of the inferred parameter,
there is a distribution, specified by the so-called \textit{true} value of 
$\theta$, that corresponds to the actual sampling distribution of $x$.
How well can such an assumption be justified: what is the probability
for the specified model to be true? Or, putting it in a wider framework,
what is the probability for a general hypothesis \dA to be true? For example:
what is the probability $p(A_2|I)$ of Newton's law of universal 
gravitation \cite{new}, here denoted by $A_2$, judged in the light of 
the facts $I$ collected in the first edition of the Principia?

In order to answer such a question, we must be able
to analyze the situation into mutually exclusive, exhaustive
possibilities ${A_1, A_2, ...}$ in order to allow for the usual 
normalization 
\begin{equation*}
 \sum_iP(A_i|I) = 1 \ .
\end{equation*}
Recall that the unity in the above expression is only a matter of convention
within Desideratum\,\textit{I}, but a normalization of the probability for 
an exhaustive set to an arbitrary (positive) constant value is necessary
in order to define the scale of the assigned probabilities. 
For example, without
such a normalization it would have been impossible to say whether a certain 
probability, say $p(A_2|I)=0.75$, is either high or low.

Unfortunately, the normalization represents
a task that \textit{cannot be consistently accomplished}. First of all,
the absolute status of a hypothesis embedded in the universe of \textit{all}
conceivable theories \textit{cannot} be stated. That is, its probability within
the class of all conceivable theories is neither large or small; it is
simply undefined because the class of all conceivable theories
is undefined (\cite{jay}, \S\,9.16.1, p.310).

Consequently, it would only be possible to express the plausibility of 
a hypothesis
within a class of well--defined alternatives. But what is the criterion that 
a hypothesis should satisfy in order to take a place within such a class? 
Let us, for example, choose a class of alternatives $A_n$ 
to Newton's law of universal gravitation such that the gravitational 
attraction between two planets could be 
inversely proportional to the $n$-th power of the distance $r$ between the 
planets, where $n$ is limited to positive integers. Since we are not able
to justify the restriction of $n$ to positive integers only -- these have
been chosen completely arbitrarily -- we extend the class by allowing $n$ to
take any real value. Then, since there is no obvious reason that the 
alternatives should be limited to power--laws, we add hypotheses 
from the exponential family to our class. And so on and so forth,
into an infinite regress that brings us to the conclusion that \textit{there
is no such thing as a well--defined class of alternatives that would 
allow for expressing the quantitative plausibility of a hypothesis}.

Scientific theories can never be \textit{justified}, or \textit{verified}.
Under certain circumstances a hypothesis \dA can be trusted more than
a hypothesis \dB -- perhaps because \dB is contradicted by certain results
of observations, and therefore \textit{falsified} by them, whereas \dA
is not falsified; or perhaps because a greater number of predictions
can be derived with the help of \dA than with the help of \dB \cite{pop3}; or
because the likelihood for observing measured data $x$, given \dA
is correct, $p(x|AI)$, is higher than the corresponding likelihood 
$p(x|BI)$, conditional on \dB being true. The best we can say about
an hypothesis is that up to now it has been able to show its worth,
and that it has been more successful than other hypotheses although,
in principle, it can never be justified or verified; we saw that 
\textit{we cannot even state its probability} for being true.

Some inductive arguments are stronger than others, and
some are very strong. But how much stronger or how strong we
cannot express \cite{key}. If predictions made by a theory are borne
out by future observations, then we become more confident of the
hypotheses that led to them; and if the predictions never fail
in vast number of tests, we come eventually to call them \textit{physical
laws}. On the other hand, if the predictions prove to be wrong,
we have learned that our hypotheses are wrong or incomplete, and
from the nature of the error we may get a clue as to how they might
be improved (\cite{jay}, \S\,9.16.1, p.311). But there is absolutely 
no guarantee for the corrected theory to be correct.

After all has been said and done, it becomes evident that
there is nothing \textit{absolute} about the theory of consistent 
inference about 
parameters of sampling distributions. The theory does not rest
upon a rock--bottom: an inference about a parameter is necessarily conditional 
on a specified model (i.e. on a specified family
of sampling distributions) whose truth we are \textit{never} able to prove.
The whole structure of the theory rises, as it were, above a swamp.
It is like a building erected on piles. The piles are driven down from 
above into the swamp, but not down to any \textit{natural} or 
\textit{given} base; and when we cease our attempts to drive our piles
into a deeper layer, it is not because we have reached firm ground.
We simply stop when we are satisfied that they are firm enough to 
carry the structure, at least for the time being (\cite{pop},
\S\,30, p.\,111).

\section{Conclusions}
\label{sec:conclusions}

      This article presents an attempt to formulate a consistent 
      theory for inferring parameters of sampling probability distributions.
      The theory is developed by following general rules, referred to as the 
      Cox-P\'{o}lya-Jaynes Desiderata. We extended the existing
      applications of the Desiderata in order to allow for consistent
      inferences in the limit of complete prior ignorance about the
      values of the parameters. 

      Starting from that limit, the Consistency 
      Theorem\;(\ref{eq:bayprime}, \ref{eq:bayprime2}) 
      is to be used for \textit{assigning} probabilities
      on the basis of the collected data. The form of the Theorem
      is very similar to the form of Bayes' Theorem \eqref{eq:bayes}
      that is used for \textit{updating} the assigned probability
      distributions, but we stressed an important difference between the
      two. While in Bayes' Theorem the prior probability $f(\theta|I)$
      represents a distribution of credibility of different values
      of the parameter $\theta$ that is based on information $I$
      prior to the inclusion of data $x$ in our inference, $\pi(\theta)$
      in Consistency Theorem is just a consistency factor that 
      depends on the form of the
      sampling distribution and is determined in a way that ensures
      consistent reasoning. Contrary to prior and posterior
      probabilities for an inferred parameter $\theta$, $p(\theta|I)$
      and $p(\theta|xI)$, to the corresponding pdf's, $f(\theta|I)$
      and $f(\theta|xI)$, and to the likelihood for an observed $x$ given
      $\theta$, $p(x|\theta I)$, the consistency
      factors \textit{by no means} represent any kind of probability
      distributions and should \textit{not} be confused with the
      ill-defined non-informative priors. That is, no probabilistic inference 
      is ever to be made on the basis of the form of the consistency
      factor alone. When this is recognized, there is absolutely
      no need for the factors to possess (or not to possess) any of
      the properties of the pdf's: we find arguments for the factors to
      be either normalizable (see, for example, references \cite{daw},
      \cite{hag}, \S\;3.27-3.29, pp.\;77-78, and \cite{jay}, \S\;15.12, 
      p.\;488) or non-normalizable (see, for example, \cite{jef}, 
      \S\;3.1, p.\;121, or \cite{jay}, \S\,15.10, p.\,485) 
      completely irrelevant.

      The developed theory is only an effective one.
      We met several examples where the prior information and the 
      collected data were too meagre
      to permit a consistent parameter inference. We saw that, under
      very general conditions, the remedy is just to collect more data
      relevant to the estimated parameter. Probability theory does not
      guarantee in advance that it will lead us to a consistent answer
      to every conceivable question.
      But, on second thoughts, this shuld not be too big a surprise,
      since we are accustomed to effective theories in all
      branches of science outside pure mathematics. Formulating
      an \textit{absolute theory} of inductive reasoning, based on
      imperfect information, might just turn out to be an overambitious 
      task. 

      By giving up the idea of the existence of (self-contradicting)
      non-informative probability distributions, and the illusion
      of an absolute theory, all paradoxes and inconsistencies, 
      extensively discussed in the preceding sections, are
      solved and we arrive at a position where we can write down a logically
      consistent quantitative theory of inference about parameters.
      
      The theory is operational in the sense that it is verifiable
      from long range consequences. 
      We saw that all the predictions
      were automatically calibrated, i.e. that the predicted long range
      relative frequencies coincided with the actual frequencies
      of occurrence. This is a very important feature that allows
      for a reconciliation between the frequentist and the Bayesian
      approaches to statistics, probably the same kind of reconciliation
      that Maurice Kendal \cite{ken3} had in mind:
      ``Neither party can avoid ideas of the other in order to 
        set up and justify a comprehensive theory.'' 
      In this way the distinction between the \textit{theory of probability}
      and that of \textit{statistical inference} might be removed, leaving
      a logical unity and simplicity. \\

\vskip 3mm
\noindent
{\large\bf Acknowledgment}
\\
\vskip 0mm
\noindent
Prof.\;Gabrijel Kernel, dr.\;Igor Mandi\'{c}, prof.\;Ale\v s Stanovnik and 
prof.\;Peter \v Semrl read a preliminary draft of this paper. 
We wish to thank
them for their kindness and for their help in correcting several
inaccuracies. They are not responsible, of course, for any errors
that remain as to the opinion expressed concerning the nature of
consistent plausible reasoning.

\appendix
{\Large{\bf Appendix}}
\section{Cox's Theorem}
\label{sec:cox}

In what follows we present, for the sake of completeness, a proof of 
Cox's Theorem, borrowing mainly from the original proof of 
Richard Cox \cite{cox} and from the proofs of Edwin Jaynes 
(\cite{jay}, \S\,2.1-2.2, pp.\,24-35) and Kevin Van\,Horn \cite{vho}, 
but first we prove the following Lemma:
\begin{lemm}
Suppose that plausibilities $(A|B)$, $(B|AI)$, 
$(B|I)$ and $(A|BI)$
can be assigned\footnote{Throughout the proof of
Cox's Theorem it is always assumed that all considered
degrees of plausibility \textit{can} be assigned.} and that they
completely and uniquely determine the plausibility $(AB|I)$ for the logical
product $AB$ to be true. Then $(AB|I)$ is to be a function
either of $(A|B)$ and $(B|AI)$ or of $(B|I)$ and $(A|BI)$ only.
\end{lemm}
{\bf Proof.} \;According to the above assumptions, there are fifteen 
different combinations of plausibilities
$(A|B)$, $(B|AI)$, $(B|I)$ and $(A|BI)$,
corresponding to fifteen different subsets of arguments of a function $H$
from which we might compute $(AB|I)$:
\begin{equation}
\begin{split}
 t &= H(x) = H(u) \ ,\hskip 16mm \\
 t &= H(y) = H(v) \ ,
\end{split}
\label{eq:lemm1}
\end{equation}
\begin{equation}
 t=H(x,u) \ , \hskip 27mm
\label{eq:lemm1f}
\end{equation}
\begin{equation}
 t=H(y,v) \ , \hskip 27mm
\label{eq:lemm1a}
\end{equation}
\begin{equation}
 t = H(x,y) = H(u,v)\ , \hskip 8mm
\label{eq:lemm1g}
\end{equation}
\begin{equation}
 t = H(x,v) = H(u,y)\ , \hskip 8mm
\label{eq:lemm1b}
\end{equation}
\begin{equation}
 t = H(x,y,v) = H(u,v,y) \ ,
\label{eq:lemm1c}
\end{equation}
\begin{equation}
 t=H(x,y,u)=H(u,v,x) \hskip 2mm
\label{eq:lemm1e}
\end{equation}
and
\begin{equation}
 t = H(x,y,u,v) \ , \hskip 20mm
\label{eq:lemm1d}
\end{equation}
where we used abbreviations
\begin{equation*}
 x\equiv(A|I),\ \ y\equiv(B|AI),\ \ u\equiv(B|I),\ \ v\equiv(A|BI)\ \ 
 \text{and}\ \ t\equiv(AB|I) \ .
\end{equation*}
Interchangeability of $x$ and $u$, as well as of $y$ and $v$, in 
the above expressions is implied by commutativity of the logical
product $AB$, $AB = BA$. We \textit{always assume 
that $H$ is continuous, as well as differentiable and non-decreasing 
in all of its arguments}. Moreover, in the case where none of the
arguments equals $\mathsf{F}$, $H$ should be \textit{strictly increasing
in all of its arguments}.

The Lemma is then proved by the method of trying out all possible subsets
of arguments and demonstrating that all but two of them inevitably lead
to conclusions that contradict the basic Desiderata.

\begin{description}
\item[\textit{a)}] In four out of the fifteen possible 
cases, $H$ is a function of a single variable.
Let, for example, in the case of $t = H(x)$, $A$ being true, 
e.g. a tautology. Then $AB$ is equivalent to $B$ (i.e. $t$ is 
equivalent to $u$), so we have
\begin{equation*}
 (B|I) = H(1) = \text{constant} \ ,
\end{equation*}
regardless the proposition $B$ and the state of information $I$.
By letting $B$ be a tautology and a false proposition,
respectively, the above equation implies $1=\mathsf{F}$,
which evidently contradicts the requirements of 
Desideratum\;\textit{I.} Similarly, an assumption 
$A=B$ in the case of $t = H(y)$ leads to the same kind of contradiction,
so we conclude that the plausibility $(AB|I)$ \textit{cannot} be expressed
as a function of a single variable.

\item[\textit{b)}] In a very similar way we can also rule out the possibility
\eqref{eq:lemm1a}.
Namely, by choosing $A=B$ we obtain
\begin{equation*}
 t=H(1,1)= \text{constant} \ .
\end{equation*}
In the case of \eqref{eq:lemm1b},
identical contradictions are obtained by choosing 
either \dA or \dB to be tautologies.
\item[\textit{c)}] According to the first one of the possibilities 
\eqref{eq:lemm1c}, 
\begin{equation}
 t=H(x,y,v) \ ,
\label{eq:lemm1cprime}
\end{equation}
the plausibility $\bigl(A(BC)|I\bigr)$ is expressible as
\begin{equation}
 \bigl(A(BC)|I\bigr) = H\bigl[x,(BC|AI),(A|BCI)\bigr] \ ,
\label{eq:lemm2}
\end{equation}
where $C$, as long as it is consistent with both $I$ and $B$, is 
a completely arbitrary proposition. We can therefore choose $C=A$, implying
\begin{equation*}
 \bigl(A(BC)|I\bigr)=t, \ \ (BC|AI)=y, \ \ (BC|I)=t 
\ \ \text{and} \ \ (A|BCI)=1 \ ,
\end{equation*}
so that \eqref{eq:lemm2} reduces to
\begin{equation*}
 t = H(x,y,1) \ .
\end{equation*}
which is incompatible with the original assumption \eqref{eq:lemm1cprime}
about the value of $t$ being dependent on \textit{three} independent variables 
$x$, $y$ and $v$.
\item[\textit{d)}] In the case \eqref{eq:lemm1d} where $H$ is to be 
a function of \textit{four independent} variables $x$, $y$, $u$ and $v$, 
the plausibility $\bigl(A(BC)|I\bigr)$ is expressible as
\begin{equation}
 \bigl(A(BC)|I\bigr) = H\bigl[x,(BC|AI),(BC|I),(A|BCI)\bigr] \ .
\label{eq:lemm2a}
\end{equation}
By setting $C=A$, \eqref{eq:lemm2a} reduces to
\begin{equation}
 t=H(x,y,t,1) \ ,
\label{eq:lemm3}
\end{equation}
or, equivalently, to
\begin{equation}
 H(x,y,u,v) = H\bigl[x,y,H(x,y,u,v),1\bigr] \ .
\label{eq:lemm4}
\end{equation}
Note that due to the interchangeability of $x$ and $u$ and of $y$ and $v$,
\eqref{eq:lemm3} can be rewritten as
\begin{equation}
 t=H(t,1,x,y) \ .
\label{eq:lemm5}
\end{equation}

Differentiation of \eqref{eq:lemm4} with respect to
$x$, $y$, $u$ and $v$, respectively, yields a system of four
differential equations:
\begin{equation}
\begin{split}
 H_1(x,y,u,v) &= H_3\bigl[x,y,H(x,y,u,v),1\bigr]\,H_1(x,y,u,v)
               + H_1\bigl[x,y,H(x,y,u,v),1\bigr] \ , \\
 H_2(x,y,u,v) &= H_3\bigl[x,y,H(x,y,u,v),1\bigr]\,H_2(x,y,u,v)
               + H_2\bigl[x,y,H(x,y,u,v),1\bigr] \ , \\
 H_3(x,y,u,v) &= H_3\bigl[x,y,H(x,y,u,v),1\bigr]\,H_3(x,y,u,v) \ , \\
 H_4(x,y,u,v) &= H_3\bigl[x,y,H(x,y,u,v),1\bigr]\,H_4(x,y,u,v) \ . 
\end{split}
\label{eq:lemmdif}
\end{equation}
Since $H(x,y,u,v)$ is to be increasing in all of its arguments,
the derivatives \linebreak $H_3(x,y,u,v)$ and $H_4(x,y,u,v)$ must be 
positive (i.e. different from zero) in the case of 
$x$, $y$, $u$ and $v$ all being different from $\mathsf{F}$. Then,
according to the latter two of the four equations, the derivative
$H_3\bigl[x,y,H(x,y,u,v),1\bigr]$
should equal unity,
\begin{equation}
 H_3\bigl(x,y,t,1) = 1 \ .
\label{eq:contrd0}
\end{equation}
This, when inserted to the first 
equation of \eqref{eq:lemmdif}, further implies
\begin{equation}
 H_1\bigl[x,y,H(x,y,u,v),1\bigr]=H_1(x,y,t,1) = 0 \ .
\label{eq:contrd1}
\end{equation}
Since both \eqref{eq:contrd0} and \eqref{eq:contrd1} must hold for arbitrary
propositions \dA and \dB, they must also hold for \dB being a tautology,
implying
\begin{equation*}
 t = (AB|I) = (A|I) = x \ \ \text{and} \ \ y = (B|AI) = 1 \ .
\end{equation*}
In this case \eqref{eq:contrd0} and \eqref{eq:contrd1} read
\begin{equation}
 H_1(x,1,x,1) = 0 \ \ \ \text{and} \ \ \ H_3(x,1,x,1) = 1 \ .
\label{eq:contrd2}
\end{equation}

The same kind of reasoning as above, however, applied to equation 
\eqref{eq:lemm5}, leads to
\begin{equation*}
 H_1(x,1,x,1) = 1 \ \ \ \text{and} \ \ \ H_3(x,1,x,1) = 0 \ ,
\end{equation*}
which, when compared to \eqref{eq:contrd2}, is \textit{an evident
inconsistency}. In this way
the possibility for $t$ to be a function of four independent 
variables $x$, $y$, $u$ and $v$, is finally rejected.

Note that three additional possibilities for $H$, \eqref{eq:lemm1e}
and \eqref{eq:lemm1f},
can be excluded by following the same patterns of reasoning
as in the case of $H(x,y,u,v)$.
\end{description}
After trying out thirteen different possibilities we thus end up with 
only two admissible subsets of arguments for $H$:
$x$ and $y$, or $u$ and $v$, the latter being a consequence of
the aforementioned interchangeability of variables.
All other subsets, \eqref{eq:lemm1}, \eqref{eq:lemm1f}, \eqref{eq:lemm1a},
\eqref{eq:lemm1b}, \eqref{eq:lemm1c}, \eqref{eq:lemm1e} and \eqref{eq:lemm1d},
 have been ruled out as incompatible
with the basic Desiderata. Therefore, if $(AB|I)$ is truly to be uniquely and 
completely determined by the plausibilities $(A|B)$, $(B|AI)$, 
$(B|I)$ and $(A|BI)$, there must exist a function $H$ \eqref{eq:lemm1g}, such
that
\begin{equation*}
 (AB|I)= H\bigl[(A|I),(B|AI)\bigr]= H\bigl[(B|I),(A|BI)\bigr] \ ,
\end{equation*}
which completes the proof of the Lemma.

\vskip 3mm \noindent
In summary, existence of the function $H$ \eqref{eq:lemm1g}
thus represents the starting point of the proof of Cox's Theorem.
Recall that Desideratum\;\textit{II.} requires $H$ to be strictly 
increasing and twice differentiable in both of its arguments, so we have
\begin{equation*}
 H_1(x,y) \ge 0 \ \ \ \text{and} \ \ \ H_2(x,y) \ge 0 \ ,
\end{equation*}
with equalities if and only if $y$ and $x$ represent impossibilities,
respectively.

Suppose now we try to find the plausibility $(ABC|I)$ that three propositions,
\dA, \dB and $C$, would be true simultaneously. Because of the fact
that Boolean algebra is associative, $ABC=(AB)C = A(BC)$, we can 
express the plausibility that we are searching for in two different ways,
\begin{equation*}
 \bigl((AB)C|I\bigr) = H\bigl[(AB|I),(C|ABI)\bigr] = 
 H\bigl[H(x,y),z\bigr] 
\end{equation*}
and
\begin{equation*}
 \bigl(A(BC)|I\bigr) = H\bigl[(A|I),(BC|AI)\bigr] = \ H\bigl[x,H(y,z)] \ , 
\end{equation*}
where another abbreviation, $z\equiv (C|ABI)$, was used. 
According to Desideratum\,\textit{III.a}, the two ways must 
lead to the same result, i.e. if our reasoning is to be 
consistent, function $H$ must solve the Associativity Equation:
\begin{equation}
 H\bigl[H(x,y),z\bigr] = H\bigl[x,H(y,z)\bigr] 
\label{eq:assoc}
\end{equation}
(see, for example, reference\,\cite{acz}, \S\,6.2, pp.\,253-273, and
\S\,7.2.2, pp.\,327-330, and references quoted therein).

In order to solve it, we first differentiate the Associativity Equation
with respect to $x$ and $y$, obtaining in this way:
\begin{equation}
 H_1\bigl[H(x,y),z\bigr]\,H_1(x,y) = H_1\bigl[x,H(y,z)\bigr] 
\label{eq:difassoc1}
\end{equation}
and
\begin{equation}
 H_1\bigl[H(x,y),z\bigr]\,H_2(x,y) = H_2\bigl[x,H(y,z)\bigr]\,H_1(y,z) \ .
\label{eq:difassoc2}
\end{equation}
Dividing \eqref{eq:difassoc2} by \eqref{eq:difassoc1} yields:
\begin{equation}
 K(x,y)=K\bigl[x,H(y,z)\bigr]\,H_1(y,z) \ ,
\label{eq:K1}
\end{equation}
where
\begin{equation*}
 K(x,y) \equiv \frac{H_2(x,y)}{H_1(x,y)} \ .
\end{equation*}
Equation \eqref{eq:K1} can also be rewritten as:
\begin{equation}
 K(x,y)\,K(y,z) = K\bigl[x,H(y,z)\bigr]\,H_2(y,z) \ .
\label{eq:K2}
\end{equation}
The right side of \eqref{eq:K1} is independent of $z$, therefore
when differentiated with respect to $z$, it vanishes for all $x$ and $y$
and $z$:
\begin{equation}
 K_2\bigl[x,H(y,z)\bigr]\,H_1(y,z)\,H_2(y,z)+
 K\bigl[x,H(y,z)\bigr]\,H_{1,2}(y,z) = 0 \ .
\label{eq:difz}
\end{equation}
The right side of \eqref{eq:K2}, when differentiated with respect
to $y$, equals \eqref{eq:difz}, and must therefore vanish
for all $x$, $y$ and $z$, too. It is then evident that both the left and the
right side of eq.\,\eqref{eq:K2} must be independent of $y$, i.e.
this means that
\begin{equation*}
 \frac{d}{dy}\ln{\bigl(K(x,y)\,K(y,z)\bigr)} = 
 \frac{d}{dy}\ln{K(x,y)} + \frac{d}{dy}\ln{K(y,z)} = 0
\end{equation*}
or, equivalently,
\begin{equation*}
 \frac{d}{dy}\ln{K(x,y)} = -\frac{d}{dy}\ln{K(y,z)} \ .
\end{equation*}
This further implies both the right and the left side of the above
equation to be independent of either $x$ or $z$:
\begin{equation}
 \frac{d}{dy}\ln{K(x,y)} = -\frac{d}{dy}\ln{K(y,z)} \equiv 
\frac{d}{dy}\ln{h'(y)} \ ,
\label{eq:hy}
\end{equation}
where
\begin{equation*}
 h'(y) \equiv \frac{d}{dy}h(y)
\end{equation*}
is a strictly positive function of $y$.
Permutation of the variables results in an expression very
much like \eqref{eq:hy}:
\begin{equation}
 \frac{d}{dx}\ln{K(z,x)} = -\frac{d}{dx}\ln{K(x,y)} \equiv 
\frac{d}{dx}\ln{h'(x)} \ .
\label{eq:hx}
\end{equation}
By subtracting equation \eqref{eq:hx}, multiplied by $dx$,
from equation \eqref{eq:hy}, multiplied by $dy$, we obtain:
\begin{equation*}
 d\ln{K(x,y)} = d\ln{\biggl\{\frac{h'(y)}{h'(x)}\biggr\}} \ ,
\end{equation*}
or, when integrated,
\begin{equation}
 \frac{H_2(x,y)}{H_1(x,y)} = \frac{1}{a}\frac{h'(x)}{h'(y)} \ ,
\label{eq:H2overH1}
\end{equation}
where $a$ is an arbitrary integration constant. By introducing
\begin{equation*}
 G(x,y)\equiv ah(x)+h(y) \ ,
\end{equation*}
equation \eqref{eq:H2overH1} can be rewritten as:
\begin{equation*}
 H_1(x,y)G_2(x,y)-H_2(x,y)G_1(x,y) = 0 \ .
\end{equation*}
We saw on two previous occasions (recall Section\,\ref{sec:invariance},
eqns.\,(\ref{eq:eqF1G2}-\ref{eq:cdfxth}), and the repeated situation
in Section\,\ref{sec:calibration}), that the general solution of such
a functional equation reads:
\begin{equation}
 H(x,y)=k\bigl(G(x,y)\bigr) = k\bigl(ah(x)+h(y)\bigr) \ ,
\label{eq:funck}
\end{equation}
where $k$ is an arbitrary function of a single variable $G(x,y)$. 
When inserted into the Associativity Equation, the solution \eqref{eq:funck}
yields:
\begin{equation}
 k\biggl\{ah\Bigl[k\bigl(ah(x)+h(y)\bigr)\Bigr]+h(z)\biggr\} =
 k\biggl\{ah(x)+ h\Bigl[k\bigl(ah(y)+h(z)\bigr)\Bigr]\biggr\} \ .
\label{eq:assocrewrtn}
\end{equation}
If the above equality is to be true for every $z$, then
the function $h\Bigl[k\bigl(ah(y)+h(z)\bigr)\Bigr]$ must take the
form:
\begin{equation*}
 h\Bigl[k\bigl(ah(y)+h(z)\bigr)\Bigr] = l(y)+h(z) \ ,
\end{equation*}
or, equivalently, then the function $k\bigl(ah(y)+h(z)\bigr)$
must take the form:
\begin{equation}
 k\bigl(ah(y)+h(z)\bigr)=h^{-1}\bigl(l(y)+h(z)\bigr) \ ,
\label{eq:funck1}
\end{equation}
where $h^{-1}$ is an inverse function of $h$, while
$l(y)$ is a function of $y$ whose form we are about to determine.
In order to do that, we first differentiate equation \eqref{eq:funck1}
with respect to $z$ and obtain:
\begin{equation*}
 k'\bigl(ah(y)+h(z)\bigr)\,h'(z) = (h^{-1})'\bigl(l(y)+h(z)\bigr)\,h'(z) \ ,
\end{equation*}
thus implying equality between $k'$ and $(h^{-1})'$:
\begin{equation*}
 k'\bigl(ah(y)+h(z)\bigr) = (h^{-1})'\bigl(l(y)+h(z)\bigr)\ .
\end{equation*}
Taking this equality into account, we also differentiate \eqref{eq:funck1}
with respect to $y$ and obtain a differential equation:
\begin{equation*}
 ah'(y) = l'(y) \ ,
\end{equation*}
whose integral reads:
\begin{equation}
 l(y) = ah(y) + b \ ,
\label{eq:funcl}
\end{equation}
where $b$ is an integration constant. This then implies the form of $H(x,y)$
to be:
\begin{equation}
 H(x,y)=h^{-1}\bigl(ah(x)+h(y)+b\bigr) \ .
\label{eq:solhxy}
\end{equation}

To determine the value of the constant $a$, let us insert the 
solution \eqref{eq:solhxy} into \eqref{eq:assocrewrtn}, obtaining
in this way an equation:
\begin{equation*}
 a\bigl(ah(x)+h(y)+b\bigr) + h(z) = ah(x) + ah(y) +h(z) + b \ ,
\end{equation*}
or
\begin{equation*}
 \bigl(ah(x)+b\bigr)\,(a-1) = 0 \ , 
\end{equation*}
with two possible solutions: $a=b=0$ or $a=1$. Since the former
violates the requirement for monotonicity of $H(x,y)$, 
$H_1(x,y)>0$, the only acceptable solution of the Associativity Equation,
reads:
\begin{equation*}
 H(x,y)=h^{-1}\bigl(h(x)+h(y)+b\bigr) \ ,
\end{equation*}
or, written in terms of plausibilities:
\begin{equation}
 h\bigl[(AB|I)\bigr]=h\bigl[(A|I)\bigr]+h\bigl[(B|AI)\bigr]+b \ .
\label{eq:solassoc}
\end{equation}
By exponentiation, the solution takes the form
\begin{equation}
 w(AB|I)=w(A|I)\,w(B|AI)\,e^b \ ,
\label{solassoc1}
\end{equation}
with
\begin{equation*}
 w(A|I)= w\bigl[(A|I)\bigr]\equiv \exp\bigl\{h\bigl[(A|I)\bigr]\bigr\}
\end{equation*}
being a function of plausibilities that is by construction both
positive and strictly increasing with respect to its argument.

Suppose now that given information $I$, a proposition $A$ is certain,
i.e. true beyond any reasonable doubt, and that $B$ is another proposition.
Then, the state of knowledge about the propositions $A$ and $B$ being
simultaneously true, $AB$, is the same as the state of knowledge
about only $B$ being true, which can be expressed by a simple equation
of Boolean algebra as:
\begin{equation*}
 AB=B \ .
\end{equation*} 
Therefore, by Desideratum\,\textit{III.c}, 
we must assign equal plausibilities for $AB$ and $B$, 
\begin{equation*}
 (AB|I)=(B|I) \ ,
\end{equation*}
and we also will have 
\begin{equation*}
 (A|BI) = (A|I)
\end{equation*}
because if $A$ is already certain given $I$, then given any other
information $B$ not contradicting $I$, it remains certain.
In this case the \textit{product rule} reads:
\begin{equation*}
 w(B|I)=w(A|I)\,w(B|I)\,e^{b} \ ,
\end{equation*}
so our function $w(x)$ must have a property that for
certain events \dA
\begin{equation*}
 w(A|I) = e^{-b} \ .
\end{equation*} 
With no loss of generality we may choose the value of the constant
$b$ to be zero, i.e. the value of function $w(A|I)$ for a certain
event $A$ to be one. In other words, any continuous positive 
strictly increasing function $w$ with upper bound $e^{-b}$
can be renormalized by being multiplied with $e^{b}$ so that its upper
bound equals unity. Note that the \textit{compositum} of plausibility
assignment $(A|I)$ and the renormalized functions $w$,
$w(A|I)$ meets all
the requirements for plausibilities, i.e. renormalized functions
$w(A|I)$ are also plausibilities by themselves.
Then, our general product rule takes the form:
\begin{equation}
 w(AB|I) = w(A|I)\,w(B|AI) = w(B|I)\,w(A|BI) \ ,
\label{eq:finalproduct}
\end{equation}
where in the case that, apart from $(A|I)$ and $(B|AI)$, also 
plausibilities $(B|I)$ and $(A|BI)$ can be assigned, the last equality 
is due to Desideratum\,\textit{III.a}. Evidently, the product rule 
can also be rewritten as
\begin{equation}
 w^a(AB|I) = w^a(A|I)\,w^a(B|AI) = w^a(B|I)\,w^a(A|BI) \ ,
\label{eq:finalproduct1}
\end{equation}
where $a$ is a non-zero but otherwise arbitrary constant.

Now suppose that \dA is impossible, given $I$. Then, the proposition
$AB$ is also impossible given $I$:
\begin{equation*}
 w(AB|I) = w(A|I) \ ,
\end{equation*}
and if \dA is already impossible given $I$, then, given any further
information \dB which does not contradict $I$, \dA would still
be impossible:
\begin{equation*}
 w(A|BI) = w(A|I) \ .
\end{equation*}
In this case, the product rule reduces to
\begin{equation*}
 w(A|I) = w(A|I)\,w(B|I) \ ,
\end{equation*}
which must hold regardless of plausibility for \dB, given $I$. There
are three possible values of $w(A|I)$ that could satisfy this condition:
it could be either $-\infty$, $\infty$ or zero. The choice $-\infty$ is
ruled out due to the requirement for plausibilities to take
non-negative values, while $\infty$ contradicts the requirement for
plausibilities to be monotonically increasing, both thus implying
the plausibility for impossible events uniquely to be zero.

Now, in order to derive the sum rule, we suppose that the plausibility
for proposition \dA to be false, given information $I$, $w(\bar{A}|I)$,
must depend in some way on the plausibility $w(A|I)$ that it is true, i.e.
there must exist some functional relation
\begin{equation*}
 w(\bar{A}|I) = S\bigl[w(A|I)\bigr] \ .
\end{equation*}
Qualitative correspondence with common sense requires that 
$S\bigl(w(A|I)\bigr)$ be a continuous, twice differentiable, 
strictly decreasing function with the extreme values 
\begin{equation}
 S(0)=1 \ \ \ \text{and}\ \ \ S(1)=0 \ .
\label{eq:boundary}
\end{equation}
But it cannot be just any function with these properties, for
it must be fully consistent with the product rule. We make use
of the latter and express the plausibility for \dA and \dB
simultaneously to be true as:
\begin{equation*}
 w(AB|I) = w(A|I)\,w(B|AI) = w(A|I)\,S\bigl[w(\bar{B}|AI)\bigr]
         = w(A|I)\,S\biggl[\frac{w(A\bar{B}|AI)}{w(A|I)}\biggr] \ .
\end{equation*}
We also invoke consistency of the product rule, so that:
\begin{equation}
 w(A|I)\,S\biggl[\frac{w(A\bar{B}|AI)}{w(A|I)}\biggr] =
 w(B|I)\,S\biggl[\frac{w(\bar{A}B|AI)}{w(B|I)}\biggr] \ .
\label{eq:conssum}
\end{equation}
Since this must hold for every $A$ and $B$, given $I$, it must
also hold when
\begin{equation}
 B = \bar{A}+\bar{C} \ ,
\label{eq:specialb}
\end{equation}
i.e. when
\begin{equation*}
 \bar{B} = AC \ ,
\end{equation*}
where \dC is any new proposition. But then, according to
simple results of Boolean algebra:
\begin{equation*}
 A\bar{B} = \bar{B} \ \ \ \text{and} \ \ \ \bar{A}B=\bar{A} \ ,
\end{equation*}
and by using the abbreviations
\begin{equation*}
 x\equiv w(A|I) \ \ \ \text{and} \ \ \ y\equiv w(B|I) \ ,
\end{equation*}
\eqref{eq:conssum} becomes a functional equation
\begin{equation*}
 x\,S\biggl[\frac{S(y)}{x}\biggr] = y\,S\biggl[\frac{S(x)}{y}\biggr] \ .
\end{equation*}
By defining new variables,
\begin{equation*}
 u\equiv \frac{S(y)}{x} \ \ \ \text{and} \ \ \ v\equiv \frac{S(x)}{y} \ ,
\end{equation*}
the functional equation is further reduced to
\begin{equation}
 x\,S(u) = y\,S(v) \ .
\label{eq:xyuv}
\end{equation}
On the way to solution, we differentiate \eqref{eq:xyuv} with respect to 
$x$, $y$, and $x$ and $y$, respectively, obtaining in this way:
\begin{equation}
 S(u)-u\,S'(u) = S'(u)\,S'(x) \ ,
\label{eq:xyuv1}
\end{equation}
\begin{equation}
 S'(u)\,S'(y) = S(v) - v\,S'(v)\ ,
\label{eq:xyuv2}
\end{equation}
and
\begin{equation}
 S''(u)\,S'(y)\,\frac{u}{x} = S''(v)\,S'(x)\,\frac{v}{y} \ .
\label{eq:xyuv3}
\end{equation}
In order to eliminate $x$ and $y$, we multiply 
equations \eqref{eq:xyuv} and \eqref{eq:xyuv3},
\begin{equation*}
 S''(u)\,S(u)\,S'(y)\,u = S''(v)\,S(v)\,S'(x)\,v \ ,
\end{equation*}
and express $S'(x)$ and $S'(y)$ from \eqref{eq:xyuv1} and \eqref{eq:xyuv2},
arriving in this way to:
\begin{equation*}
 \frac{S''(u)\,S(u)\,u}{S'(u)\bigl[S(u)-u\,S'(u)\bigr]} = 
 \frac{S''(v)\,S(v)\,v}{S'(v)\bigl[S(v)-v\,S'(v)\bigr]} \ .
\end{equation*}
Then, evidently both sides of the above equation must equal a constant,
say $a-1$, and the above equation splits into two identical
differential equations of the form:
\begin{equation*}
 \frac{dS'}{S'} = (a-1)\,\Bigl(\frac{du}{u}+\frac{dS}{S}\Bigr) \ .
\end{equation*}
The solution that is obtained by two successive integrations and
that satisfies the boundary conditions \eqref{eq:boundary}, reads:
\begin{equation}
 S(u) = \bigl(1-u^a\bigr)^{\frac{1}{a}} \ .
\label{eq:su}
\end{equation}
In this way we obtained the so-called \textit{sum rule}:
\begin{equation}
 w^a(\bar{A}|I) + w^a(A|I) = 1 \ .
\label{eq:finalsum}
\end{equation}

Since our derivation of the functional equation \eqref{eq:xyuv}
used the special choice \eqref{eq:specialb} for \dB, \eqref{eq:su}
is a necessary condition to satisfy the general consistency 
requirement \eqref{eq:conssum}. To check for its sufficiency, 
we substitute \eqref{eq:su} in \eqref{eq:conssum} and obtain
an evident equality (c.f. eq. \eqref{eq:finalproduct}):
\begin{equation*}
 w(A|I)\,w(B|AI) = w(B|I)\,w(A|BI) \ .
\end{equation*}
Therefore, equation \eqref{eq:su} is the necessary and sufficient
condition on $S(x)$ for consistency in the sense \eqref{eq:conssum}.

Out of all possible plausibility functions $w(A|I)$ we then choose 
the \textit{probability} $P(A|I)$,
\begin{equation*}
 P(A|I)\equiv w^a(A|I) \ ,
\end{equation*}
for which the product and the sum rule evidently take the forms:
\begin{equation*}
 P(AB|I) = P(A|I)\,P(B|AI) = P(B|I)\,P(A|BI)
\end{equation*}
and
\begin{equation*}
 P(A|I) + P(\bar{A}|I) = 1 \ .
\end{equation*}
This completes the proof of Cox's Theorem.

\section{More on the marginalization paradox}
\label{sec:dsz}

In 1972, Stone and Dawid
published an article \cite{sto} containing two examples
of the so-called \textit{marginalization paradox}. The examples
refer to inferences that could be made in two different ways leading
to two different results, despite the two possible ways
being completely equivalent. Since at least one of the two
chosen ways involves the procedure of marginalization \eqref{eq:marginal},
such inconsistencies are usually referred to as the marginalization 
paradox. Stone and Dawid judged the usage of improper (i.e. non-integrable) 
non-informative priors as the cause of the paradox. 

If correct, these arguments would seriously threaten the consistency
of probability theory, for we saw that our consistency factors
for location, scale and dispersion parameters, \eqref{eq:cflsd}
and \eqref{eq:cfmusigma}, despite being conceptually different
from non-informative priors, are all non-integrable over their
particular (infinite) domains. But since the form of the consistency
factors is uniquely determined by the Cox-P\'{o}lya-Jaynes Desiderata
(see Section\,\ref{sec:unique}), the arguments of Stone and Dawid, 
if correct, would
imply that it is \textit{impossible} to construct a consistent and 
calibrated theory of qualitative inference about parameters of sampling
distributions.

If the appropriate consistency factors \eqref{eq:cflsd}
and \eqref{eq:cfmusigma} are used, there is \textit{no} paradox
in the first of the two aforementioned examples. The origin of the
paradox in Example \#2, however, is better camouflaged
and is discussed in the following.

Let the data $\bi{x}=(x_1,x_2,...,x_n)$ consist of $n\ge 2$ observations
from the normal sampling distribution $N(\mu,\sigma)$. Suppose that 
before the data $\bi{x}$ were collected we had been completely
ignorant about the values of the parameters and that we are not
interested in either of the two parameters separately but only in
their ratio $\theta$: 
\begin{equation*}
 \theta \equiv \frac{\mu}{\sigma} \ .
\end{equation*}

Let the inference be made by two persons, say Mr.\,A
and Mr.\,B, each choosing his own way to obtain the probability
distribution for $\theta$. Mr.\,A strictly obeys the rules
of probability as developed in the present paper and makes use 
of Consistency Theorem \eqref{eq:bayprime1}, of the consistency factor
\eqref{eq:cfmusigma}, and of the sequential use of Bayes' Theorem
\eqref{eq:bayes1}, in order to obtain first a two-dimensional pdf for
$\mu$ and $\sigma$:
\begin{equation}
\begin{split}
 f(\mu\sigma|\bi{x} I) & \propto \pi(\mu,\sigma)\,
 f(\bi{x}|\mu\sigma I) 
 =\frac{1}{\sigma}\prod_{i=1}^{n} f(x_i|\mu\sigma I) \\
 & \propto \frac{1}{\sigma^{n+1}}\,\exp{\biggl\{ - \frac{n}{2\sigma^2}
   \bigl( (\bar{x}-\mu)^2 + s^2\bigr)\biggr\} } \\
 & \propto \frac{1}{\sigma^{n+1}}\,\exp{\biggl\{-\frac{n}{2\sigma^2}\, 
 \Bigl(\mu^2+\frac{R^2}{n}-\frac{2rR\mu}{n}\Bigr)\biggr\} } \\
 & \propto \frac{1}{\sigma^{n+1}}\,\exp{\biggl\{-\frac{n\theta^2}{2}
   -\frac{R^2}{2\sigma^2}+\frac{rR\theta}{\sigma}\biggr\} } \ ,
\end{split} 
\label{eq:modela}
\end{equation}
where:
\begin{equation*}
 \bar{x} \equiv \frac{1}{n}\sum_{i=1}^{n}x_i \ \ , \ \ 
 s^2 \equiv \frac{1}{n}\sum_{i=1}^{n}(x_i-\bar{x})^2 \ ,  
\end{equation*}
and
\begin{equation*}
 R^2 \equiv n(\bar{x}^2+s^2) \ \ , \ \
 r \equiv \frac{n\bar{x}}{R} \ .
\end{equation*}
The admissible range of variables $\bar{x}$ is the entire
real axis, for $s^2$ and $R=\sqrt{R^2}$ it is only its
positive half, while the range of $r$ is the interval
$(-\sqrt{n},\sqrt{n})$.

By multiplying the above pdf by the appropriate Jacobian,
\begin{equation*}
 |J|=\biggl|\frac{\partial(\mu,\sigma)}{\partial(\theta,\sigma)}\biggr|
    = \sigma \ ,
\end{equation*}
Mr.\,A obtains the pdf for $\theta$ and $\sigma$,
\begin{equation*}
 f(\theta\sigma|\bi{x}I)\propto 
 \frac{1}{\sigma^{n}}\,\exp{\biggl\{-\frac{n\theta^2}{2}
   -\frac{R^2}{2\sigma^2}+\frac{rR\theta}{\sigma}\biggr\} } \ .
\end{equation*}
Since he is interested only in $\theta$ but not in $\sigma$,
he integrates out the latter in order to obtain the marginal
pdf for $\theta$,
\begin{equation*}
 f(\theta|\bi{x}I) = \int_0^{\infty}f(\theta\sigma|\bi{x}I)\,d\sigma
 \propto e^{-n\theta^2/2}\,H_{n-2}(r,\theta) \ ,
\end{equation*}
where function $H_n(r,\theta)$ is introduced as:
\begin{equation}
 H_n(r,\theta)\equiv \int_0^{\infty} u^n\, \exp{\Bigl\{
              - \frac{u^2}{2} + r\theta u \Bigr\} }\,du \ .
\label{eq:defhn}
\end{equation}
Evidently, Mr.\,A's pdf for $\theta$ is a function of
statistics $r$ alone and, when also appropriately normalized, it reads:
\begin{equation}
 f(\theta|\bi{x}I) = \frac{e^{-n\theta^2/2}}{\eta(r)}\,H_{n-2}(r,\theta) \ ,
\label{eq:mraspdf}
\end{equation}
where $\eta(r)$ is the usual normalization factor.

Seeing Mr.\,A's result, Mr.\,B decides to simplify the calculation.
His intuitive judgment is that he ought to be able equivalently but
more easily derive the (marginal) pdf for $\theta$ by 
a direct application of Consistency Theorem in a reduced model,
i.e. that $\theta$ should be inferred directly
from the sampling distribution for $r$ only, following in this
way the so-called \textit{reduction principle} 
\cite{daw}. He starts with the appropriate pdf for $\bar{x}$
and $s^2$,
\begin{equation*}
 f(\bar{x}s^2|\theta\sigma I') \propto \sigma^{-n}\,s^{n-3}\,
 \exp{\biggl\{-\frac{n}{2\sigma^2}\bigl[(\bar{x}-\mu)^2 + s^2\bigr]\biggl\}} 
 \ ,
\end{equation*}
transforms it into the pdf for $r$ and $R$,  
\begin{equation}
\begin{split}
 f(rR|\theta\sigma I') &= f(\bar{x}s^2|\theta\sigma I')\,
 \biggl|\frac{\partial(r,R)}{\partial(\bar{x},s^2)}\biggr|^{-1} \\ 
 &\propto \frac{R^{n-1}}{\sigma^n}\,
 \bigl(n-r^2\bigr)^{\frac{n-3}{2}}\,\exp{\biggl\{-\frac{n\theta^2}{2}
 - \frac{R^2}{2\sigma^2} + \frac{rR\theta}{\sigma}\biggr\} } \ ,
\end{split}
\label{eq:pdfrr}
\end{equation}
and then reduces it by means of marginalization to a pdf for
$r$ only:
\begin{equation}
 f(r|\theta\sigma I') = \int_0^{\infty} f(rR|\theta\sigma I')\,dR
 \propto\bigl(n-r^{2}\bigr)^{\frac{n-3}{2}}\,e^{-n\theta^2/2}\,H_{n-1}(r,\theta)
 \ .
\label{eq:pdfr}
\end{equation} 
Note that the pdf for $r$ is independent of parameter $\sigma$, i.e.
$f(r|\theta\sigma I') = f(r|\theta I')$. 
Then, according to Consistency Theorem, Mr.\,B's pdf for $\theta$ reads:
\begin{equation}
 f(\theta|rI') = \frac{\widetilde{\pi}(\theta)}{\widetilde{\eta}(r)}\,
                f(r|\theta I') = 
                \frac{\widetilde{\pi}(\theta)}{\widetilde{\eta}(r)}\,
                e^{-n\theta^2/2}\,H_{n-1}(r,\theta) \ ,
\label{eq:mrbspdf}
\end{equation}
with $\widetilde{\pi}(\theta)$ and $\widetilde{\eta}(r)$ being
the appropriate consistency and normalization factors, the latter
also containing the factor $(n-r^2)^{\frac{n-3}{2}}$ from equation
\eqref{eq:pdfr}.

Now, if the two marginal pdf's for $\theta$ of Mr.\,A and Mr.\,B,
\eqref{eq:mraspdf} and \eqref{eq:mrbspdf}, are to be equal, there
must exist a consistency factor $\widetilde{\pi}(\theta)$ such
that
\begin{equation}
 h(r)\,\widetilde{\pi}(\theta)\,H_{n-1}(r,\theta) = H_{n-2}(r,\theta) 
\label{eq:nonsolvbl}
\end{equation}
for all real $r,\theta\in(-\infty,\infty)$ and for every integer $n\ge 2$, 
where
\begin{equation*}
 h(r)\equiv \frac{\eta(r)}{\widetilde{\eta}(r)} \ .
\end{equation*}
But it is easy to demonstrate by \textit{reductio ad absurdum}
that a general solution $\widetilde{\pi}$ of equation
\eqref{eq:nonsolvbl} does \textit{not} exist. For a moment
we therefore suppose that such a solution indeed exists.  
By differentiating logarithm of \eqref{eq:nonsolvbl} with respect
to $r$ (with respect to $\theta$) we obtain
\begin{equation}
 \frac{h'(r)}{h(r)} = \theta\,\biggl(
 \frac{H_{n-1}(r,\theta)}{H_{n-2}(r,\theta)} -
 \frac{H_{n}(r,\theta)}{H_{n-1}(r,\theta)}\biggr) 
\label{eq:difhr}
\end{equation}
and
\begin{equation}
 \frac{\widetilde{\pi}'(\theta)}{\widetilde{\pi}(\theta)} 
 = r\,\biggl(
 \frac{H_{n-1}(r,\theta)}{H_{n-2}(r,\theta)} -
 \frac{H_{n}(r,\theta)}{H_{n-1}(r,\theta)}\biggr) \ , 
\label{eq:difht}
\end{equation}
where we made use of the following properties of functions $H_n$:
\begin{equation*}
 \frac{\partial}{\partial r}H_{n}(r,\theta) = \theta\,H_{n+1}(r,\theta) 
 \ \ \ \text{and} \ \ \ 
 \frac{\partial}{\partial\theta}H_{n}(r,\theta) = r\,H_{n+1}(r,\theta) \ .
\end{equation*}
When divided, equations \eqref{eq:difhr} and \eqref{eq:difht} yield
a functional equation: 
\begin{equation*}
 \frac{h'(r)}{h(r)}\,r = 
 \frac{\widetilde{\pi}'(\theta)}{\widetilde{\pi}(\theta)}\,\theta \ ,
\end{equation*}
whose general solution are functions $h(r)$ and $\widetilde{\pi}(\theta)$
of the form
\begin{equation*}
 h(r) \propto r^{c} \ \ \ \text{and} \ \ \ 
 \widetilde{\pi}(\theta) \propto \theta^{c} \ ,
\end{equation*}
with $c$ being a real constant that could be either positive,
negative or exactly zero. When inserted in equation \eqref{eq:nonsolvbl},
functions $h(r)$ and $\widetilde{\pi}(\theta)$ imply:
\begin{eqnarray}
 H_{n-2}(r,\theta) &=& C\,r^c\,\theta^{c}\,H_{n-1}(r,\theta) 
\ \ \ \ \ \ ; c > 0 \ , 
\label{eq:nonsolvbl1} \\
 H_{n-1}(r,\theta) &=& C\,r^{|c|}\,\theta^{|c|}\,H_{n-2}(r,\theta) 
\ \ \ ; c < 0 \ , 
\label{eq:nonsolvbl2} \\
 H_{n-2}(r,\theta) &=& C\,H_{n-1}(r,\theta) \hskip 1.3cm \,\ ; c = 0 \ ,
\label{eq:nonsolvbl3}
\end{eqnarray}
where C is an arbitrary constant. 

If this is to be true for any value of $\theta$ within its permissible
range, it must also be true for $\theta = 0$. But then, for positive
$c$, this implies $H_{n-2}(r,\theta=0)=0$, which, when compared to
$H_{n-2}(r,\theta=0)=2^{\frac{n-3}{2}}\,\Gamma\bigl(\frac{n-1}{2}\bigr)$
that follows directly from the definition of $H_n$ \eqref{eq:defhn},
is a clear contradiction. For negative $c$ we derive a contradiction
in an identical way. For $r\neq 0$, the result of differentiation of equation 
\eqref{eq:nonsolvbl3} that corresponds to the value of
$c$ being exactly zero, reads:
\begin{equation}
  H_{n-1}(r\neq 0,\theta) = C\,H_{n}(r\neq 0,\theta)\ .
\label{eq:nonsolvbl4}
\end{equation}
Dividing \eqref{eq:nonsolvbl3} with \eqref{eq:nonsolvbl4} and
again setting $\theta =0 $ yields:
\begin{equation*}
 \frac{n-1}{2}=\frac{\Gamma^2\Bigl(\frac{n}{2}\Bigr)}
                    {\Gamma^2\Bigl(\frac{n-1}{2}\Bigr)} \ ,
\end{equation*}
which is, since it is to be valid for any integer $n\ge 2$, an
evident contradiction. Hence, the proof is completed.

Stone and Dawid in \cite{sto}, as well as Dawid, Stone and Zidek 
in \cite{daw}, considered it obvious that Mr.\,A and Mr.\,B
made their inferences about $\theta$ from the same information
(i.e. from the measured value of $r$) and should therefore indeed
come to the same conclusions: the pdf's for $\theta$, 
\eqref{eq:mraspdf} and \eqref{eq:mrbspdf} ought to be identical.
The fact that Mr.\,B cannot reproduce Mr.\,A's result whatever
consistency factor $\widetilde{\pi}(\theta)$ he chooses, served them
as a proof that either Mr.\,A of Mr.\,B must be guilty of some
transgression. According to their reasoning, the blame is to be put
on Mr.\,A for using a non-integrable consistency factor
$\pi(\mu,\sigma)$ \eqref{eq:cfmusigma}.

From the standpoint that we advocate in the present paper, it is
evident that such reasoning is unjustified. First, as we have stressed
in many places and especially in Sections\,\ref{sec:invariance}
and \ref{sec:objectivity}, in the limit of complete prior
ignorance about the value of an inferred parameter, information
about the parameter of a sampling distribution after recording
events from that particular distribution consists of \textit{two}
pieces: of the value of the likelihood and of the form of the
sampling distribution (i.e. of the specified model). But the model
$I$ \eqref{eq:modela} of Mr.\,A is different from the model $I'$ 
\eqref{eq:pdfr} of Mr.\,B. In particular, Mr.\,B obtained 
the reduced model \eqref{eq:pdfr} by marginalization of the
sampling distribution \eqref{eq:pdfrr}. Since every marginalization
represents an irreversible process (a reverse transformation from the
reduced model to the original one does not exist), Mr.\,B is
deliberately throwing away available information, acting in this way
against the consistency Desideratum \textit{III.b}. 
Therefore, Mr.\,A and  Mr.\,B \textit{do not} infer parameter $\theta$
from the same information so it is natural that they come
to different conclusions (see also ref.\,\cite{jay},
\S\,15.8, pp.\,472-473). The fact that the inferences of Mr.\,A and Mr.\,B
will \textit{never} coincide is thus an unequivocal proof that some of the 
cogent information was lost during Mr.\,B's reduction of his model.
Moreover, since there is \textit{one and only one} logically
consistent procedure (i.e. a procedure that is in a complete
accordance with the basic Desiderata) to infer $\theta$ and this is 
the procedure applied by Mr.\,A,
and since the reduction of the sampling distribution
put Mr.\,B in a position where he cannot find the appropriate
consistency factor that would reproduce Mr.\,A's result, 
Mr.\,B is no longer able to make
any kind of consistent inference about the parameter $\theta$. 

Second, Mr.\,A is perfectly capable of making \textit{predictions}
about the value of the inferred parameter that are \textit{verifiable} 
at long-range consequences.
We performed numerous Monte Carlo experiments and inferred the
value $\theta$ of the ratio of parameters $\mu$ and $\sigma$ 
of the generator of normally distributed random numbers. As expected,
confidence intervals based on Mr.\,A's marginal distribution
\eqref{eq:mraspdf} for $\theta$ covered the true value (i.e. the
ratio of the parameters that were actually used by the generator)
exactly in the percentage $\delta$ of cases that was predicted according
to \eqref{eq:prconfint}. The coverage was exact regardless of
the number $n$ of observations that we based our inference upon
each time, the chosen value $\delta$, and the type
of chosen confidence interval (it could have equally well been 
the shortest of all possible intervals, the central interval, 
the lower-most or the upper-most interval, or any other
interval with the chosen probability \eqref{eq:prconfint} that
equals $\delta$). That is, the predictions of Mr.\,A 
are calibrated. On the other hand, the
predictions of Mr.\,B who is not able to reproduce the predictions
of Mr.\,A, will thus necessarily be \textit{non-calibrated}.

Third, the inference of Mr.\,A was based only on steps that are
all (including the use of the consistency factor \eqref{eq:cfmusigma}
and the procedure of marginalization of the pdf 
$f(\theta\sigma|\bi{x}I)$) \textit{deduced directly} (i.e. without any
other assumptions) from the basic Desiderata. In particular, absolutely
no assumption was ever made about the existence of an integral of
the consistency factor over its domain: it \textit{need not} exist,
\textit{neither} is it \textit{forbidden} to exist. That is, for Mr.\,A
the existence of the integral is completely irrelevant and we therefore
see no reason whatsoever why the eventual non-integrability of 
the consistency factor \eqref{eq:cfmusigma} could be any kind of
transgression.

After all is said and done, there are only two possibilities left
at Mr.\,B's disposal. He can either correct his intuitive reasoning
and abandon the principle of reduction, or he can
develop a completely new theory of inference on his own by incorporating
the reduction principle in his set of basic rules. But, as exhibited above,
such a theory would necessarily be logically inconsistent (e.g. it would
allow, among other things, some of the available information 
that is relevant to a particular inference to be ignored), as
well as non-operational, and thus of no practical importance.

The basic Desiderata and their direct applications such as the Cox,
Bayes and Consistency Theorems, are adequate for conducting
inference so they must always take precedence over intuitive
\textit{ad hoc} devices like the above principle of reduction.
We agree with Edwin Jaynes (\cite{jay}, \S\,15.7, p.\,469) that in order
to avoid inconsistencies, the rules of inference must be obeyed strictly,
in every detail. Intuitive shortcuts that violate those rules might,
by a coincidence, lead to correct results in some very special
cases, but will in general lead to inconsistent and non-calibrated
inferences.

\end{document}